\documentclass[twocolumn]{aastex631}
\usepackage{amsmath,amsfonts,amssymb,graphicx,chngcntr,multirow, float,booktabs, afterpage}
\usepackage[]{hyperref}
\hypersetup{colorlinks=true}

\received{}
\revised{}
\accepted{}

\shorttitle{Warped accretion flows in tidal disruption events }

\begin{document}

\title{ Observational signatures of warped accretion flows in tidal disruption events }
\author{Andrew Mummery} 
\affiliation{School of Natural Sciences, Institute for Advanced Study, 1 Einstein Drive, Princeton, NJ 08540, USA}
\email{amummery@ias.edu}
\author{Eliot Quataert}
\affiliation{Department of Astrophysical Sciences, Princeton University, Princeton, NJ 08544, USA}

\date{\today}


\begin{abstract}
    We analyze the possible observational signatures and evolution of global warps in the accretion flows produced in tidal disruption events. These warps are driven by Lense-Thirring torques resulting from the misalignment of the disk's rotation axis and the black hole's spin axis. In a previous paper we argued that these global warps should be generically present in TDE accretion flows as they enter a late-time thin disk phase of evolution.  We isolate three possible observational signatures of warps: 
    (i) anomalous (reddened) late time optical colors, resulting from irradiative heating of the warped outer disk by the inner flow, (ii) delayed X-ray rises, resulting from the inner disk being slowly revealed to a distant observer as the warped flow gradually aligns with the spin axis, and (iii) reverberation of turbulent X-ray flares into optical/UV frequencies as the inner disk heats the outer disk during a flaring state.  We present preliminary observational evidence that all three effects may have already been observed in TDE systems. Observations of TDEs may well offer the cleanest astronomical signatures of the strong-gravity effects which cause disk warping, and provide unique constraints on the poorly understood evolution of warped accretion disks. 
\end{abstract}
\section{Introduction}

Every tidal disruption event (TDE) should produce a warped accretion disk.
The stars which are ultimately tidally disrupted are scattered toward the supermassive black hole from the surrounding
nuclear cluster on an orbit essentially uncorrelated with the hole's spin,
so the debris disk that forms is generically misaligned with the spin
equatorial plane. A misaligned flow experiences \cite{LenseThirring1918} torques that then should twist it into a warp
\citep{Bardeen1975}. In a companion paper \citep{PaperI} we argued that
this misaligned state is not a transient curiosity of the early, super-Eddington
flow but the natural configuration of the long-lived, thin-disk phase: the disk does not, in general,
align with the spin during a super-Eddington accretion phase, so that by the time
the flow settles into a thin disk its outer angular momentum still points
close to the original stellar orbit. Warping should therefore be the rule,
not the exception, for TDE disks.

Warped disks are, however, considerably less well understood than their
flat counterparts. For planar disks we possess a detailed and well-tested
framework, from the classical \citet{SS73} solution through its
relativistic generalizations \citep{NovikovThorne73, PageThorne74} to
fully time-dependent relativistic treatments \citep{Mummery23a} which describe their
structure, spectra and evolution. The physics of \emph{warped} disks ---
whose orbital plane varies with radius --- has been studied for decades
\citep{Papaloizou1983, Pringle1992, Ogilvie1999}, and direct hydrodynamic
and (general-)relativistic magnetohydrodynamic simulations have begun to
probe warp propagation, precession, alignment and disk tearing
\citep[e.g.,][]{Fragile2007, Sorathia2013b, Nealon2015, Liska2021}. Such calculations remain computationally demanding, limited in dynamic
range and duration, and often far from the parameter space relevant for the rapidly evolving disks expected in
 TDEs. Much of the non-linear, time-dependent behavior --- and especially
its imprint on the emergent light --- therefore remains comparatively
poorly understood. We believe that this is an opportunity. A generic, and
plausibly observable, disk configuration exists in the regime
where accretion theory is least well understood.

Wide-field optical (e.g., ZTF \citealt{Hammerstein22, Yao2023} and soon
Rubin/LSST) and X-ray (e.g., \emph{eROSITA} \citealt{Grotova25}, \emph{Einstein Probe} \citealt{Yuan15}) surveys are
now delivering TDEs in large, multi-wavelength samples
\citep[e.g.,][]{Yao2023, Guolo24, Grotova25,MummeryVV25}. Setting aside the early
optical/UV flare --- whose origin remains debated, and which may not be
directly tied to the accretion flow --- a growing body of evidence points
to a long-lived accretion disk which dominates the multi-wavelength emission at late times. The observation of super-soft, disk-like X-ray
spectra \citep{Mummery_Wevers_23, Guolo24, Guolo26}, and the late-time UV ``plateaus'' now seen in most events are
naturally explained as emission from a settled disk
\citep[e.g.,][]{vanVelzen19,MumBalb20a, Mummery24, Guolo26}. It is this disk phase that our
warp model describes, and in which an evolving warp should in principle leave its imprint.

Several specific, and growing, observational puzzles are really questions
about the geometry of this disk. A growing number of TDEs brighten in X-rays
months to years \emph{after} the optical peak \citep[e.g.,][]{Gezari17},
posing the question of whether the delay reflects the slow {formation}
of the disk or a change in its {visibility} or structure. Many events
show large-amplitude, erratic X-ray variability
\citep{Mummery25Variability, Chakraborty26}, sometimes accompanied by
variability in the UV. Further, in their disk-dominated phase some TDEs display
optical/UV colors that depart from the multi-color blackbody expected of a
flat disk --- in places reddening transiently before recovering --- a
behavior that is hard to reconcile with a planar flow
\citep[e.g.,][]{Mummery_et_al_2024}. A global warp is a plausible common link
between all of these observations: we shall show that a warp reddens the outer-disk colors through irradiative
heating, modulates the observed spectral hardness with viewing angle, hides and then
reveals the inner (X-ray) disk as it relaxes toward the equatorial plane,
and reprocesses inner-disk variability into a delayed optical/UV response.
Moreover, unlike the quasi-steady disks of active galactic nuclei --- where
warp physics can only be inferred from a static snapshot --- each TDE lets
us watch a disk form, warp, and relax in real time.

A particular appeal of this framework is its connection to black hole spin.
The warp exists only because the metric is that of a {spinning} black hole hole.
The Lense--Thirring torque $\tau_{\rm LT} \propto a$, and hence the warp radius, scale with spin
($r_{\rm warp}\propto a^{2/3}$ in the canonical estimate). A non-rotating hole
would produce none of the signatures we discuss. The presence of a warp
signature is therefore, in itself, evidence of a spinning black hole  and a misaligned
disk system, and the scale of the effect will encodes the spin to some degree. This is valuable
because spin is notoriously difficult to measure: current
constraints in TDEs rely on continuum-fitting of the disk-dominated spectrum
\citep[e.g.,][]{Wen20, Guolo26}, or on Hills-mass arguments at the highest
black hole masses \citep{Mummery24}, all of which are model-dependent and often degenerate with other effects.
Warp signatures offer an independent, geometric handle --- most powerfully
at the population level, where the incidence and timing of warped disk signatures will carry information about the
spin distribution of the black holes that host them. 

Earlier studies of misaligned TDE disks focused largely on solid-body
Lense--Thirring precession during the super-Eddington phase, and its
potential imprint on the early light curve
\citep[e.g.,][]{StoneLoeb2012, Franchini2016}. Our emphasis here is
different. 
We assume that the thin-disk-phase flow is globally
warped (based on arguments put forward in \citealt{PaperI}, which we review later) and we construct an explicit model which takes physical
parameters of the warp --- its amplitude $\theta_{\rm warp}$,
characteristic radius $r_{\rm warp}$, outer edge $R_{\rm out}$, and the
observer's viewing angle --- and produces emergent spectral energy
distributions and their time evolution. We use this model to isolate three
observational signatures that we argue may already be present in the TDE
literature: (i) anomalous (reddened) optical/UV colors, produced when the hot
inner disk irradiatively heats the tilted outer disk; (ii) delayed X-ray
rises, produced as the inner disk is gradually revealed to a distant
observer while the warp relaxes; and (iii) reverberation, in which a
turbulent X-ray flare is reprocessed into a delayed optical/UV response by
the warped outer disk. Throughout, we treat the misaligned disk as a
single, smoothly-warped surface, and set aside the alternative possibility
of disk tearing (Section~2). Our longer-term aim is to invert this problem
--- to use TDE observations to constrain the warp, and through it the
more poorly understood physics of warped disks themselves. 

The paper is organized as follows. In Section~2 we review the basic
physics of Lense--Thirring-driven warping, estimate the warp radius, argue
that TDE disks should generically be warped, and set out the disk geometry.
Section~3 develops the self-irradiation of a warped disk and the resulting
temperature and spectral structure. Section~4 presents the three
observational signatures for a disk of fixed warp. Section~5 follows their
evolution in time. Section~6 discusses candidate
signatures already observed in three TDEs, namely AT2019dsg, AT2022dsb and AT2018fyk. We conclude in Section~7. Some technical results are presented in two Appendices.

\section{Basic physics of warped accretion flows and their geometry}

\subsection{Lense--Thirring precession and disk warping}
\label{sec:LT_warping}
A spinning black hole drags the spacetime  surrounding in the direction of its rotation, a result known as the Lense--Thirring effect
\citep{LenseThirring1918}. One important impact of this physics is that any orbit whose angular-momentum vector $\mathbf{\hat l}$ is misaligned with the black hole spin axis $\mathbf{\hat{s}}$ is forced to precess about $\mathbf{\hat{s}}$ at a rate given by the mismatch of its azimuthal and vertical frequencies
\begin{equation}
  \Omega_{\rm LT}(r) =\Omega_{\phi}(r,a) - \Omega_z(r,a)\approx  \frac{2GJ}{c^2 r^3},
  \label{eq:omega_LT}
\end{equation}
where $J = aGM^2/c$ is the black hole angular momentum and $a$ is the dimensionless spin parameter, and the last expression is the leading order (Newtonian limit) contribution.  The steep $r^{-3}$ radial dependence of this torque is the ultimate origin of a disk warp. Inner annuli want to precess far more rapidly than outer annuli, so if a misaligned disk were composed of non-interacting rings, each ring would precess independently and the disk would shear apart.

A real disk resists this differential precession through internal stresses (pressure, turbulent stresses, magnetic fields, etc.) which communicate angular momentum between neighboring annuli.  The competition between the Lense--Thirring torque (which tries to precess each annulus independently) and the internal torques (which try to keep neighboring annuli coplanar) can in the right conditions produce a warp. By ``a warp'' we mean a smooth, radially varying inclination of the disk midplane and the line of nodes of the flow.

In the simplest viscous hydrodynamic theories of misaligned disks, how the warp
propagates depends on the relative strength of two quantities, namely the
\cite{SS73} viscosity-parameter $\alpha$ that acts on vertical shear (the oscillatory
sloshing motions driven by pressure gradients perpendicular to the warped
midplane), and the disk aspect ratio $H/R$ \citep{Papaloizou1983,Ogilvie1999}.
When $\alpha > H/R$ the sloshing motions are overdamped and warps propagate
diffusively, with the warp amplitude subsequently decaying smoothly over a
viscous timescale \citep{Papaloizou1983,Pringle1992}.  When $\alpha < H/R$ the
sloshing motions are underdamped and warps propagate as bending waves at
approximately half the local sound speed \citep{Papaloizou1983,LubowOgilvie2002}.
Whether a given disk is in the diffusive or wavelike regime is therefore
controlled by how thin it is relative to $\alpha$.

In the diffusive regime the viscous torques dissipate energy in the warped
region, preferentially damping the misalignment of the innermost
annuli where the Lense--Thirring torque is strongest.  Over time,
the inner disk settles into the black hole's equatorial plane while
the outer disk retains its original orientation.  The resulting
quasi-steady-state structure --- a flat inner disk connected to a tilted outer
disk through a smooth warp --- is known as the Bardeen--Petterson effect
\citep{Bardeen1975}.

\subsection{The warp radius}
\label{sec:rwarp_estimate}
We now attempt a first estimate of the warp radius $r_{\rm warp}$. In doing so we
step onto noticeably less certain theoretical ground: as emphasised above, the
physics of warped disks is far less settled than that of their planar
counterparts, and the estimate that follows is intended to be an
order-of-magnitude guide rather than a precise prediction.

With time the inner disk is expected to align with the equatorial plane of the
black hole, while the outer disk will not yet have had time to do so. The
transition radius $r_{\rm warp}$ between the aligned inner disk and the tilted
outer disk can be estimated by balancing the Lense--Thirring precession
timescale against the warp communication timescale.

In the diffusive regime, in hydrodynamic theories of disk warping, the warp
diffusion coefficient is given by
\citep{Papaloizou1983}
\begin{equation}
  D_w = \frac{c_s^2}{2\alpha\,\Omega}
      = \frac{(H/R)^2\,r^2\,\Omega}{2\alpha},
  \label{eq:Dw}
\end{equation}
where $c_s = (H/R)\,r\,\Omega$ is the midplane sound speed
evaluated at radius $r$.  Equation~(\ref{eq:Dw}) embodies the
counterintuitive result that warp diffusion is less
efficient at higher viscosity, $D_w \propto 1/\alpha$.  This
arises because the warp is communicated by pressure-driven
oscillatory flows (``sloshing'') within the disk, and viscosity
damps these flows rather than driving them. In this linear isotropic theory
the same result is often written as an effective vertical viscosity
$\alpha_2\simeq 1/(2\alpha)$ in the reduced warp equations, so that
$D_w=\alpha_2(H/R)^2 r^2\Omega$ \citep{Ogilvie1999}. This ``$\alpha_2$'' is an
effective coefficient, not a local turbulent stress $T^{z\phi}\neq\alpha_2 P$.

The warp diffuses over a radial scale $r$ on a timescale
$t_{\rm diff} \sim r^2/D_w$, while Lense--Thirring precession
operates on a timescale
$t_{\rm LT} \sim 1/\Omega_{\rm LT} \propto r^3$.
Equating the two relevant timescales  $t_{\rm diff}(r_{\rm warp}) = t_{\rm LT}(r_{\rm warp})$
\begin{equation}
  \frac{2\alpha\,r_{\rm warp}^{3/2}}{(H/R)^2}
  = \frac{r_{\rm warp}^3}{2a},
  \label{eq:rwarp_balance}
\end{equation}
leads to
\begin{equation}
  \frac{r_{\rm warp}}{r_g}
  \approx \left(\frac{4a\alpha}{(H/R)^2}\right)^{2/3}
  = (4a\alpha)^{2/3}\left(\frac{R}{H}\right)^{4/3}.
  \label{eq:rwarp}
\end{equation}
The numerical coefficient is order-unity and in reality depends on the
detailed disk structure (surface-density profile, boundary
conditions, etc.); a full steady-state solution is given by
\citet{Scheuer1996}.

The scalings are physically sensible.  Larger spin $a$
produces a stronger Lense--Thirring torque, extending alignment
to larger radii.  Larger $\alpha$ damps the sloshing motions
that carry the warp, weakening the outer disk's ability to
resist alignment.  Thinner disks (smaller $H/R$) have weaker
pressure-driven warp communication, allowing the
Lense--Thirring torque to dominate further out.  For $a = 0.5$,
$\alpha = 0.1$, and $H/R = 0.1$,
equation~(\ref{eq:rwarp}) gives
$r_{\rm warp} \approx 7\,r_g$. 
One notes that as the disk cools and thins ($H/R\to 0$),
eventually the entire disk should align ($r_{\rm warp}\to \infty$).

It is worth reflecting on the coefficient $\alpha_2$.  In writing
equation~(\ref{eq:rwarp}) in terms of the usual $\alpha$ we have used
the linear isotropic closure $\alpha_2\simeq 1/(2\alpha)$, so that
$D_w$ is not an independent diffusivity \citep{Ogilvie1999}.  More
generally, the 1D diffusive description of a warp involves a second
phenomenological ``viscosity'' $\nu_2=\alpha_2 c_s H$, distinct from the
$\alpha$ of standard thin-disk theory, that can be related to the
underlying turbulent stress only within an assumed closure and in
limiting regimes, and whose applicability to genuine MHD turbulence
in a warped geometry is far from established.  That this extra
coefficient is needed at all is a measure of how incompletely the
non-linear physics of warped disks is understood.  We regard this
less as a caveat and more as an opportunity: the uncertain
microphysics is imprinted on $D_w$, and hence on an (in principle)
observable length scale $r_{\rm warp}$.  A robust mapping between
warp models and TDE data could therefore constrain the warp-transport physics that  $\alpha$, $\alpha_2$, (etc.) 
are invented to represent.

A caveat on our central assumption is worth mentioning. Throughout this work we
treat the misaligned disk as a single, continuously warped surface that
joins the aligned inner region smoothly onto the tilted outer disk, as in
the classical Bardeen--Petterson picture \citep{Bardeen1975, Scheuer1996}.
A qualitatively different outcome has been proposed: if the differential
Lense--Thirring precession across the warp overwhelms the disk's ability to
communicate the warp internally, the disk may instead break into discrete,
near-independently precessing annuli --- so-called disk tearing
\citep{Nixon2012, Nealon2015, Liska2021, Kaaz23}. Whether, and how commonly, real
accretion flows tear rather than warp smoothly remains an open question.
The disk tearing process is intrinsically non-linear, time-dependent, and sensitive to
the (poorly constrained) internal stresses of a warped disk, and is not
readily captured by the single-surface, semi-analytic treatment we adopt. 
We therefore regard disk tearing, and its distinct observational
signatures, as beyond the scope of the present work and defer it to a
future study: all results here should be understood as applying to the
smoothly-warped regime.

\subsection{Why TDE disks should generically be warped}
\label{sec:why_warp}

In a tidal disruption event the star that is disrupted approaches the black hole on an orbit whose angular-momentum direction $\mathbf{\hat{l}}_\infty$ 
is generically misaligned with the black hole spin axis $\mathbf{\hat{s}}$.
Since the disruption geometry is set by the stellar orbit in the sphere of influence of the black hole --- which is not expected to bear any correlation with the spin of the central
black hole --- the star-black hole misalignment angle
\begin{equation}
    \theta_{\rm star} \equiv \arccos(\mathbf{\hat{s}}\cdot\mathbf{\hat{l}}_\infty)
\end{equation}
is expected to be isotropically distributed
(in other words $P(\theta_{\rm star})\propto\sin\theta_{\rm star}$, for $\theta_{\rm star}\in[0^\circ,90^\circ]$, with a typical mismatch $\theta_{\rm star}\sim 60^\circ$). 

The typical TDE therefore results in, at least initially, a generically misaligned flow of debris with respect to the black hole's spin axis. It had previously been argued that during a super-Eddington phase of accretion (if such a phase occurs, which is not obvious) then the TDE disk would precess as a solid body for a period of time, before aligning with the equatorial plane relatively rapidly \citep[e.g.,][]{StoneLoeb2012,Franchini2016}.

However, in our previous paper we argued that in fact (i) coherent precession is not expected to be present other than for a small number  ($\sim 2$) of cycles for fine tuned regions of parameter space, and (ii) the disk {\it will not} generically align with the black hole spin during a super-Eddington phase. 

The same physical mechanism underlies both results, namely that in the regime required for solid body precession (a thick $H/R \sim {\cal O}(1)$ disk), one naturally has a short ``viscous'' time. This short viscous time results in material rapidly being pushed out to large radii (to soak up the angular momentum of material moving inwards to accrete), leading to a very large lever arm for the torque to act over. As the Lense-Thirring torque drops rapidly with distance ($\sim 1/r^3$), this enlarged disk is unable to precess coherently as a solid body (there is too much angular momentum at large radii to precess as a solid body on an observable time frame). 

The same radial disk expansion makes it very difficult for the disk to align with the equatorial plane, at least initially. Again, there is a large amount of angular momentum situated at large radii (which is pointing along $\sim \mathbf{\hat l_\infty}$ as it was pushed out almost immediately) which would require a large torque for it to be re-aligned with $\mathbf{\hat s}$. The torque at these large radii is small $(\tau_{\rm LT} \sim 1/r^3)$ where gravity is quasi-Newtonian, and there is simply not enough time in the early phases of TDE disk evolution to pull the material into the equatorial plane. What this means is that for the majority of TDE phase space, by the time the flow enters a thin-disk phase of evolution, the asymptotic angular momentum of this thin disk will be $\sim \mathbf{\hat l_\infty}$, and so 
\begin{equation}
    \theta_{\rm warp} \approx \theta_{\rm star}.
\end{equation}

\subsection{Coordinate system}
\label{sec:coords}
\begin{figure*}
    \centering
    \includegraphics[width=0.49\linewidth]{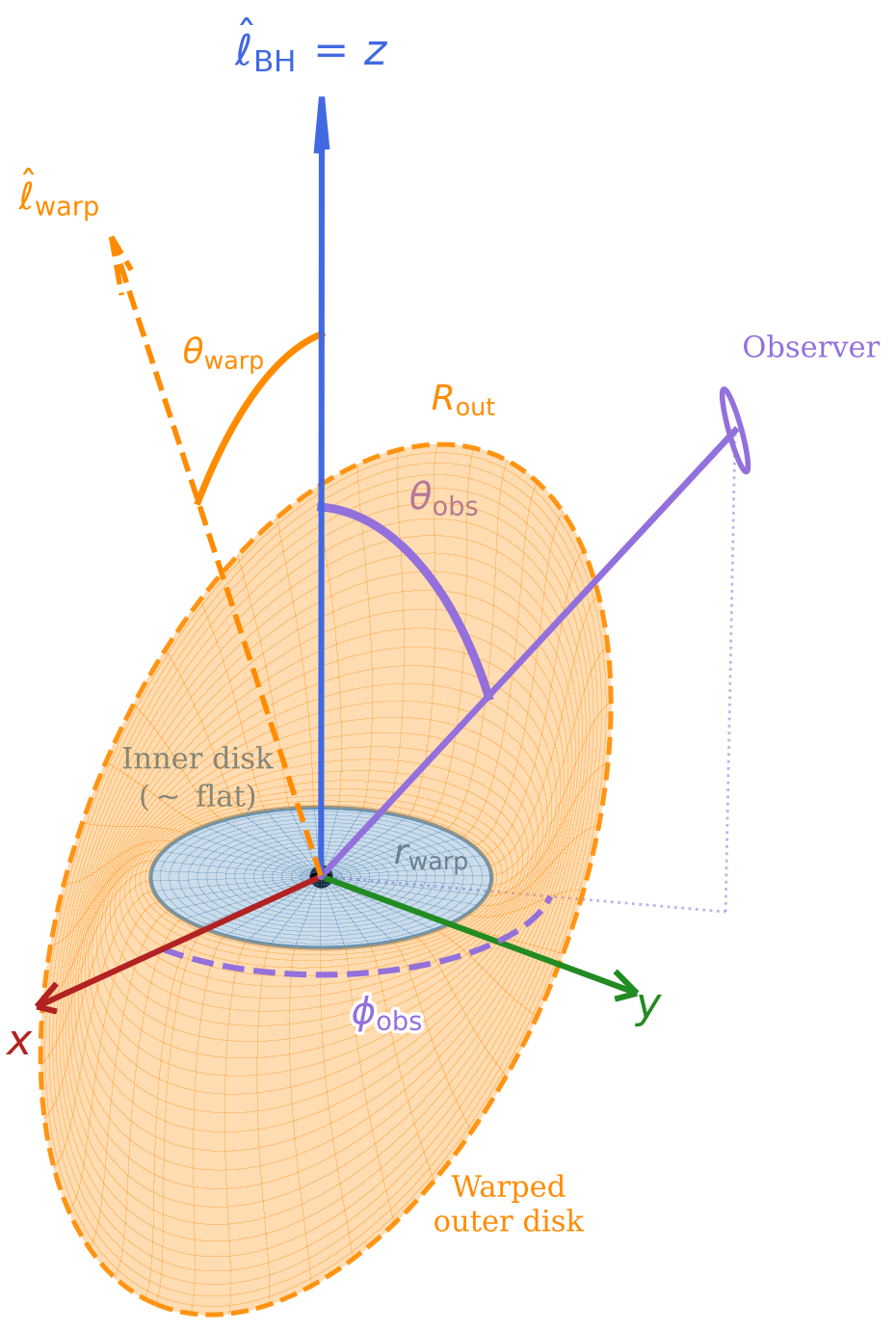}
    \includegraphics[width=0.49\linewidth]{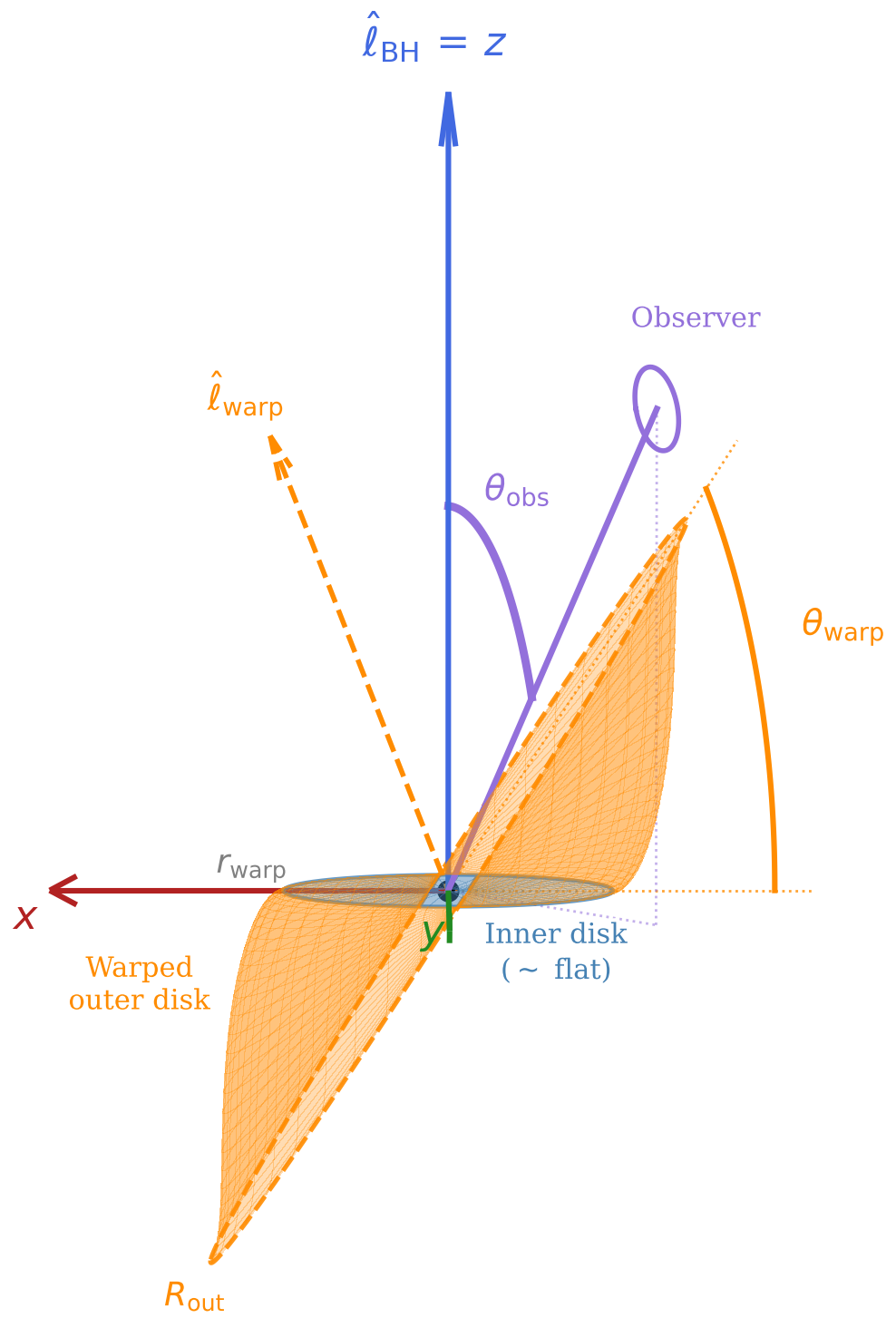}
    \caption{A schematic of the warped disk geometry considered in this work, and the definition of relevant angles and the coordinate system. The inner part of the disk is $\sim$ flat and in the equatorial plane (the radial extent of this flatness is exaggerated in this schematic), with the disk asymptotically approaching a misalignment of angle $\theta_{\rm warp}$ from the spin axis at large radii. The asymptotic line of nodes of the disk is oriented so that an observer along the $\mathbf{\hat x}$ axis sees the warp face on, with the general observer then described by two angles, $\phi_{\rm obs}$ (the azimuthal angle from the $\mathbf{\hat x}$ axis), and $\theta_{\rm obs}$ (the polar angle down from the black holes spin axis $\mathbf{\hat z}$).    }
    \label{fig:schematic}
\end{figure*}

Our coordinate system is shown in Figure \ref{fig:schematic}, which depicts a schematic of the warped disk setup. 

We work in a right-handed Cartesian frame centered on the black hole,
with the spin axis $\mathbf{\hat{s}} = \mathbf{\hat{z}}$ defining the polar
direction.  The inner disk therefore lies approximately in the equatorial plane
$z = 0$.  We orient the remaining axes so that the line of nodes
--- the intersection of the inner and outer disk planes --- lies
along $\mathbf{\hat{y}}$, and the warp peak (maximum $z$-height as $r \to \infty$) lies along $-\mathbf{\hat{x}}$ \citep[this is the original][coordinate convention, which differs from some of the more recent literature, e.g., \citealt{Speicher2025}]{Scheuer1996}.  The asymptotic
angular-momentum vector $\mathbf{\hat{l}}_\infty$ then lies in the
$xz$-plane 
\begin{equation}
  \mathbf{\hat{l}}_\infty
  = (\sin\theta_{\rm warp},0,\;\cos\theta_{\rm warp}).
  \label{eq:l_infty}
\end{equation}

An observer at infinity is specified by two angles: an inclination
$\theta_{\rm obs}$ from the spin axis ($\theta_{\rm obs} = 0^\circ$ is face-on, $\theta_{\rm obs} = 90^\circ$ is
edge-on) and an azimuth $\varphi_{\rm obs}$ around it, so that
\begin{equation}
  \mathbf{\hat{o}}
  = (\sin \theta_{\rm obs}\cos\varphi_{\rm obs},\;
     \sin \theta_{\rm obs}\sin\varphi_{\rm obs},\;
     \cos \theta_{\rm obs}).
  \label{eq:observer}
\end{equation}
The convention is that $\varphi_{\rm obs} = 180^\circ$ places the
observer on the $-\mathbf{\hat{x}}$ side of the disk (i.e., behind the warp direction as $r\to \infty$).

\subsection{The angular momentum profile}
\label{sec:beta}

The Bardeen--Petterson steady-state solution
\citep{Scheuer1996,Speicher2025} specifies the angular-momentum unit
vector $\mathbf{\hat{l}}(r)$ of each disk annulus at radius $r$.  In
our coordinate system,
\begin{align}
  l_x(r) &= \sin(\theta_{\rm warp})\,\cos(\gamma(r))\,
             e^{-\gamma(r)},  \label{eq:lx} \\
  l_y(r) &= \sin(\theta_{\rm warp})\,\sin(\gamma(r))\,
             e^{-\gamma(r)},  \label{eq:ly} \\
  l_z(r) &= \sqrt{1 - l_x^2 - l_y^2},  \label{eq:lz}
\end{align}
where $\gamma(r) \equiv 2\sqrt{r_{\rm warp}/r}$.  At large radii
($\gamma\to 0$) these reduce to
$\mathbf{\hat{l}} \to (\sin\theta_{\rm warp}, 0, \cos\theta_{\rm warp})
= \mathbf{\hat{l}}_\infty$, as required.

The
solution has the property that the horizontal component of
$\mathbf{\hat{l}}$ decays exponentially with a characteristic rate
\begin{equation}
\gamma(r) \equiv 2\sqrt{r_{\rm warp}/r},
\end{equation}
namely
\begin{equation}
  \mathbf{\hat l_\perp}^2 \equiv  l_x^2 + l_y^2
  = \sin^2(\theta_{\rm warp})\,\exp\bigl(-2\gamma(r)\bigr),
  \label{eq:lperp}
\end{equation}
so that
\begin{equation}
  l_z(r) = \sqrt{1 - \sin^2\!\theta_{\rm warp}\,
             \exp\!\bigl(-2\gamma(r)\bigr)}.
  \label{eq:lz_explicit}
\end{equation}
The same function $\gamma(r)$ simultaneously controls the Bardeen--Petterson twist (Section~\ref{sec:twist}).

\subsection{The geometry of the disk surface}
In Figure \ref{fig:schematic} we show a schematic view of a warped disk surface, which also defines the coordinate system and relevant angles in the problem. Figure \ref{fig:examples} shows three dimensional renderings of the inner regions of an actual warped disk geometry specified by $\mathbf{\hat l}$. 

The inner part of the disk is $\sim$ flat and in the equatorial plane (this flatness is exaggerated in this schematic), with the disk angular momentum asymptotically approaching a misalignment of angle $\theta_{\rm warp}$ from the spin axis at large radii. The asymptotic line of nodes of the disk is oriented so that an observer along the $\mathbf{\hat x}$ axis sees the warp face on, with the general observer then described by two angles, $\phi_{\rm obs}$ (the azimuthal angle from the $\mathbf{\hat x}$ axis), and $\theta_{\rm obs}$ (the polar angle down from the black holes spin axis $\mathbf{\hat z}$). 

There are three radii of interest in the problem, $R_{\rm in}$ (likely around or interior to the innermost stable circular orbit), $r_{\rm warp}$ (the canonical scale of the radius at which the disk starts to noticeably deviate from the equatorial plane) and $R_{\rm out}$ (the size of the accretion disk, starting compact for a TDE and then expanding). The ratios of all of these quantities ($r_{\rm warp}/R_{\rm in}, R_{\rm out}/R_{\rm in}$ and $R_{\rm out}/r_{\rm warp}$) all set interesting physical scales in the problem, which we will examine throughout this work. 


Mathematically, the disk surface itself is defined by the equation $h = \vec{ \mathbf{r}}\cdot \mathbf{\hat l} = 0$, where $\vec{\mathbf{r}}$ is a vector from the origin, and $\mathbf{\hat l}$ is the disk normal. The reason for this is that the disk normal depends only on the amplitude of the spherical radius $r \equiv |\vec{\mathbf{r}}|$, and so each disk annulus is simply a twisted and tilted great circle about the origin. Vectors that lie in the plane of the disk surface at radius $r$ therefore pass through the origin. The physics of this result is that the torque itself only depends on spherical radius $r$. Surface plots of the solutions of $h=0$ are shown in Figure \ref{fig:examples}. 


For computing (e.g.) the  luminosity emitted by/the irradiative flux received by a given disk element at $(r, \theta, \phi)$ one must take into account that the area of a disk element is not the simple ${\rm d}A_{\rm flat} = r{\rm d}r {\rm d}\phi$, with $\phi$ the usual cylindrical angle, once the disk has been twisted and tilted. As the warped disk is made up of a sequence of tilted rings, each with individual ring area ${\rm d}A_{\rm ring} = 2\pi r{\rm d}r$, the area  of a disk element can still be parameterized by an internal angle $\psi$, such that ${\rm d}A_{\rm ring} = r{\rm d}r {\rm d}\psi$. This can be related to the usual spherical cordinate system via 
\begin{equation}
\label{eq:dA_ring_phi}
\mathrm{d}A_{\rm ring}
   = r\,\mathrm{d}r\,\mathrm{d}\psi
   = \frac{r\cos\beta}{\sin^2\!\phi + \cos^2\!\phi\,\cos^2\!\beta.}\;\mathrm{d}r\,\mathrm{d}\phi.
\end{equation}
The derivation of this is somewhat non-trivial, and we discuss the geometric properties of this warped profile in detail in Appendix \ref{app:geo}. 


The final questions relevant for both observability and irradiation are (i) can disk element $A$ ``see'' disk element $B$?, and (ii) can the observer ``see'' disk element $A$?  As each disk annulus is a twisted and tilted great circle about the origin, both of these questions can be posed as simple ``$h=0$'' type problems.

Both problems are effectively just the question of whether or not a straight vector (we ignore light bending for computational ease, this is a bad approximation in the very inner regions) from a point $A$ passes through an intermediate disk element $P$ before reaching its destination $B$. Expressing this trajectory vector as $\vec{\mathbf{d}}$ (with unit vector $\mathbf{\hat d}$; this is either $\mathbf{\hat o}$ for the light ray trying to reach the observer, or $\vec{\mathbf{B}}-\vec{\mathbf{A}}$ for the two disk area elements case), then the the ray is blocked if 
\begin{equation}
    h = (\mathbf{r}_0 + t \mathbf{\hat d})\cdot \mathbf{\hat l}(|(\mathbf{r}_0 + t \mathbf{\hat d})|) = 0,
\end{equation}
for any $t$ along the trajectory. This equation, when solved for $t$, is approached numerically when required. 

Finally, we note that the \citep{Scheuer1996} warped disk profile has a point symmetry, where the two annuli at fixed radius and with azimuthal angles $\phi \to \phi+180^\circ$ (i.e., opposite sides of the disk) have formally identical physics. This cuts down computation of (e.g.,) the irradiation by a factor 2.  

The above discussion of disk geometry is of course determined by the properties of the unit vector $\mathbf{\hat l}$, for which we now discuss the individual physical components. 


\subsection{The Bardeen--Petterson tilt profile $\beta(r)$}
The local tilt angle $\beta(r)$ --- the inclination of the disk plane at radius $r$ relative to the black hole equatorial plane ---
is given by $\beta(r) = \arccos(l_z(r))$, or  
\begin{equation}
  \beta(r) = \arccos\!\left[\sqrt{1 - \sin^2(\theta_{\rm warp})\,
             \exp\!\left(-4\sqrt{r_{\rm warp}/r}\right)}\right].
  \label{eq:beta}
\end{equation}
The limits of $\beta(r)$ are the following
\begin{itemize}
  \item At the inner edge ($r \ll r_{\rm warp}$):
        $\gamma\to\infty$, the exponential vanishes, $l_z\to 1$,
        and $\beta\to 0$.  The inner disk lies flat in the equatorial
        plane, aligned with the black hole spin.
  \item At large radii ($r \gg r_{\rm warp}$): $\gamma\to 0$,
        the exponential approaches unity, and $\beta\to\theta_{\rm warp}$.
        The disk asymptotes to the orbital plane of the disrupted star.
  \item At any finite outer radius $R_{\rm out}$,
        $\beta(R_{\rm out}) < \theta_{\rm warp}$ because the
        exponential suppression is not yet negligible.  The effective
        wall presented by the outer disk to an observer is therefore
        less steep than $\theta_{\rm warp}$ alone would suggest.
\end{itemize}

The functional form of $\beta(r)$ is nearly universal when expressed in
dimensionless units: $\beta(r)/\theta_{\rm warp}$ as a function of
$r/r_{\rm warp}$ is approximately independent of $\theta_{\rm warp}$.
This universality is exact in the small-angle limit
($\theta_{\rm warp}\ll 1$), where
$\beta(r)\approx\theta_{\rm warp}\,\exp(-2\sqrt{r_{\rm warp}/r})$,
and remains an excellent approximation to within a few per cent
for $\theta_{\rm warp}\lesssim 60^\circ$.

\subsection{The Bardeen--Petterson twist $\gamma(r)$}
\label{sec:twist}

The function $\gamma(r)$ plays a dual role in the Bardeen--Petterson
solution. It sets both the exponential growth rate of the tilt
(equation~\ref{eq:lperp}) and the azimuthal twist of
$\mathbf{\hat{l}}(r)$.  The line of nodes at radius $r$ is rotated by
\begin{equation}
  \gamma(r) = 2\sqrt{r_{\rm warp}/r}
  \label{eq:gamma}
\end{equation}
relative to the asymptotic ($r\to\infty$) direction
(which is determined by $\mathbf{\hat{l}}_\infty \to \mathbf{\hat{x}}$).  At $r = r_{\rm warp}$ the twist angle
is $\gamma = 2$~rad $\approx 115^\circ$.
At large radii, $\gamma\to 0$ and the twist vanishes and 
the outer disk has its line of nodes aligned with $\mathbf{\hat{y}}$, as
defined by our coordinate choice. 

Physically, the twist arises because each annulus wants to precess about
the spin axis at the local Lense--Thirring rate, which scales as
$r^{-3}$.  The inner rings therefore accumulate more precession
angle, winding the line of nodes into a spiral pattern.

The twist has important consequences for the observational
appearance of the disk, specifically which lines of sight can see the inner disk.  Although it does not change the height profile of the disk (which is determined by $\beta$), it shifts the azimuthal position at which an observer sees the
maximum warp height, and thereby modifies the occultation geometry (Section~\ref{sec:blocking}).




\subsection{Sensitivity to $r_{\rm warp}$ and $\theta_{\rm max}$}
\label{sec:sensitivity}

The warp profile~(\ref{eq:beta}) depends on two parameters:
$r_{\rm warp}$ and $\theta_{\rm warp}$.  Their effects are
distinct 

\begin{itemize}
  \item 
        Increasing $\theta_{\rm warp}$ increases the tilt of the disk at every radius,
        increasing the height above the equatorial plane $z_{\rm max}(r) = r\sin\beta(r)$.  This increases (at fixed everything else) the irradiation received by the outer disk, and the probability of the inner disk being blocked from view. 
  \item 
        The transition from aligned to misaligned occurs at
        $r \sim r_{\rm warp}$.  A larger warp radius pushes the
        steep part of the warp to larger radii, reducing the ratio
        $R_{\rm out}/r_{\rm warp}$ and thereby reducing $\beta(R_{\rm out})$.
        Conversely, a smaller $r_{\rm warp}$ concentrates the warp at
        small radii, allowing $\beta$ to reach near-asymptotic values
        at the outer edge. As the warp radius enters as a square root within an exponential, even relatively small changes in  $r_{\rm warp}$ can have moderate impacts on the observed properties of these disks. 
\end{itemize}


\begin{figure*}
    \centering
    \includegraphics[width=0.49\linewidth]{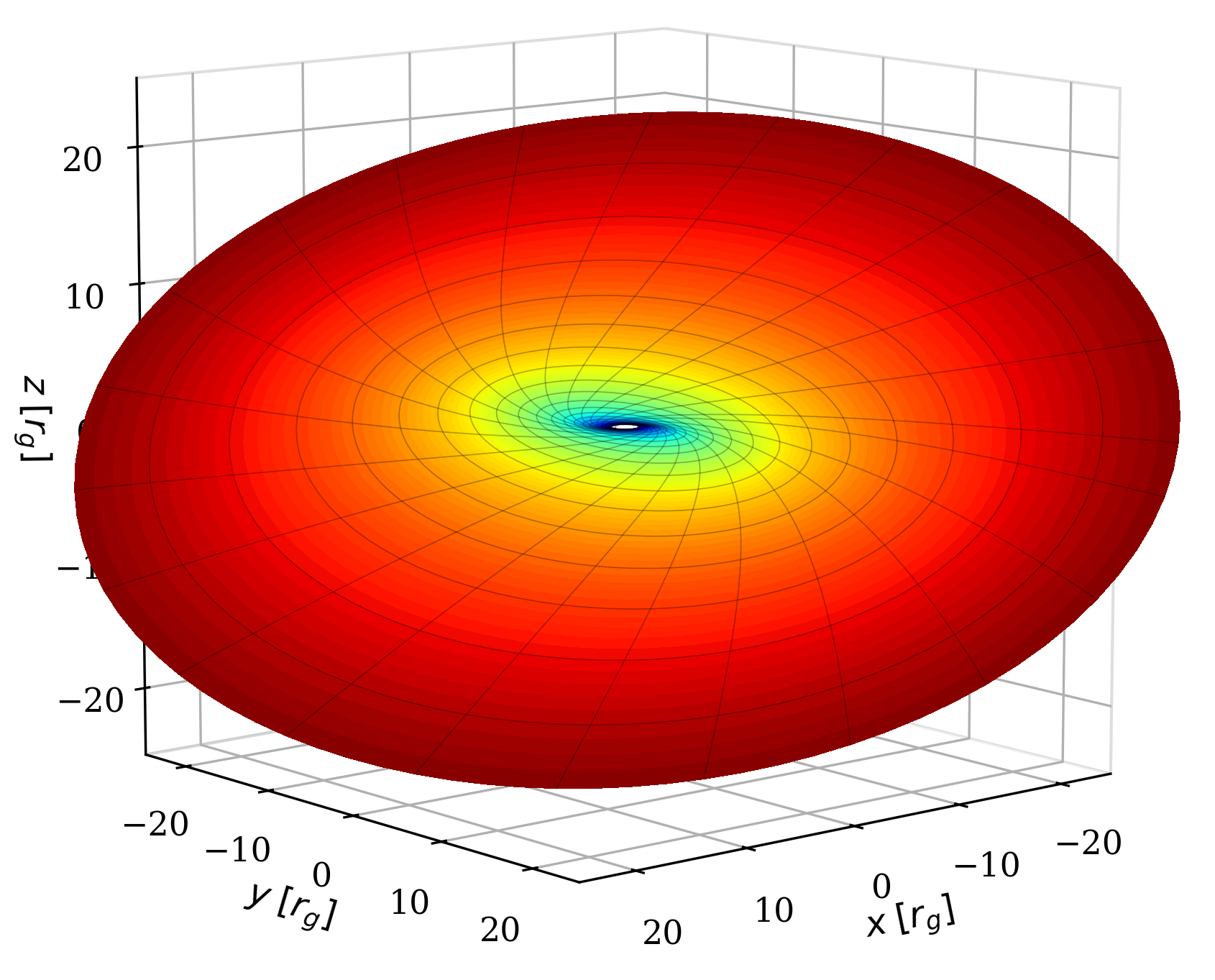}
    \includegraphics[width=0.49\linewidth]{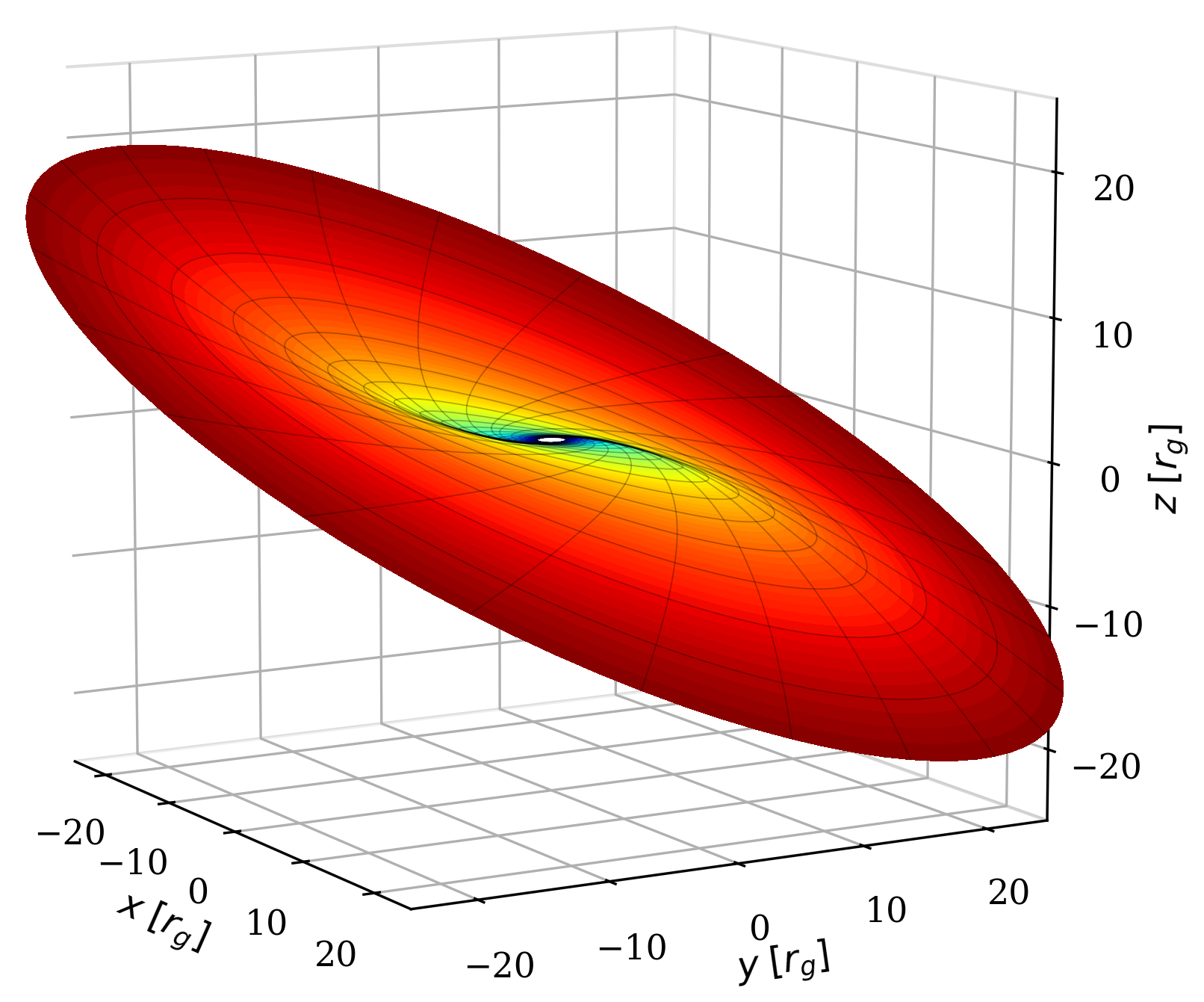}
    \includegraphics[width=0.49\linewidth]{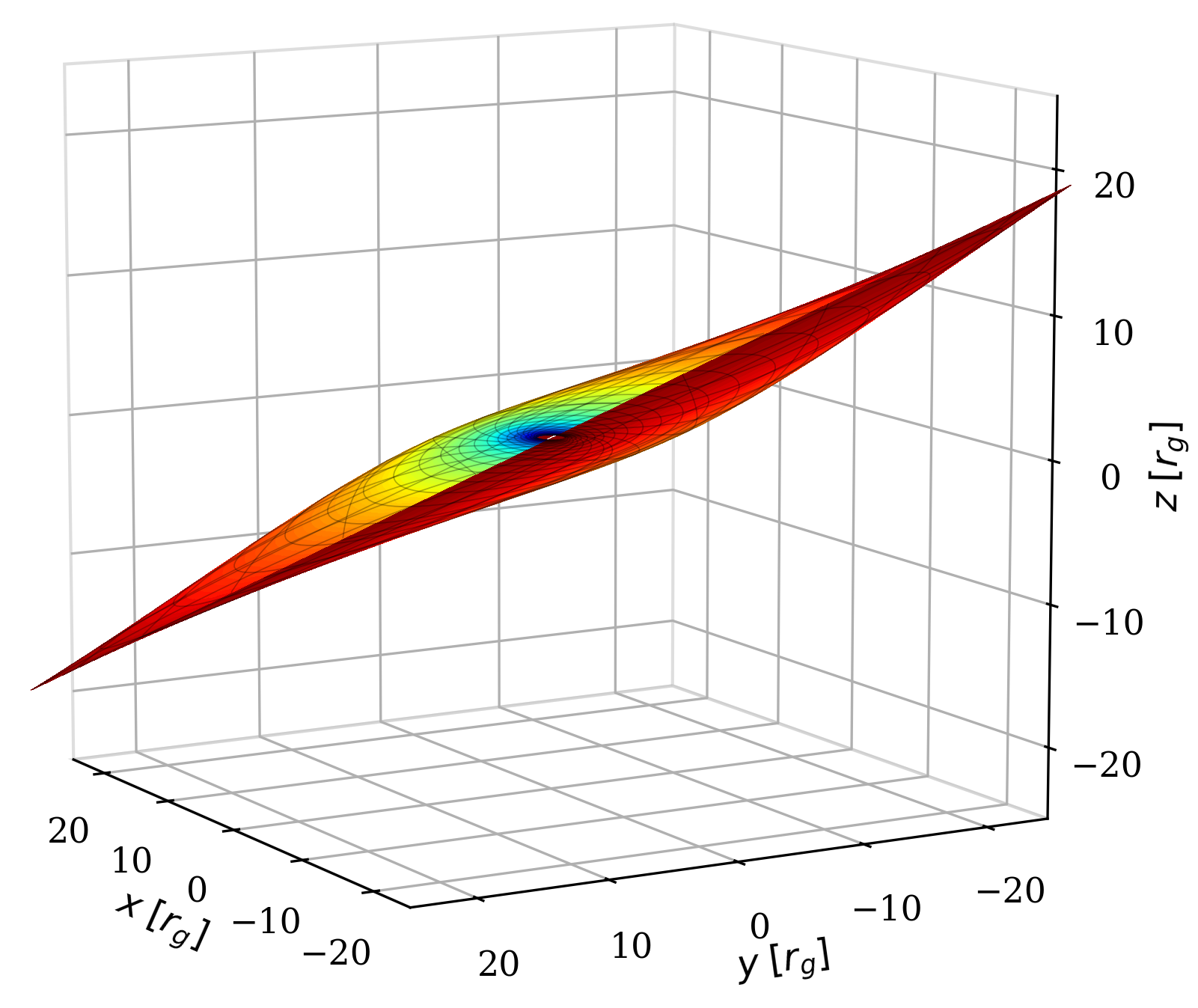}
    \includegraphics[width=0.49\linewidth]{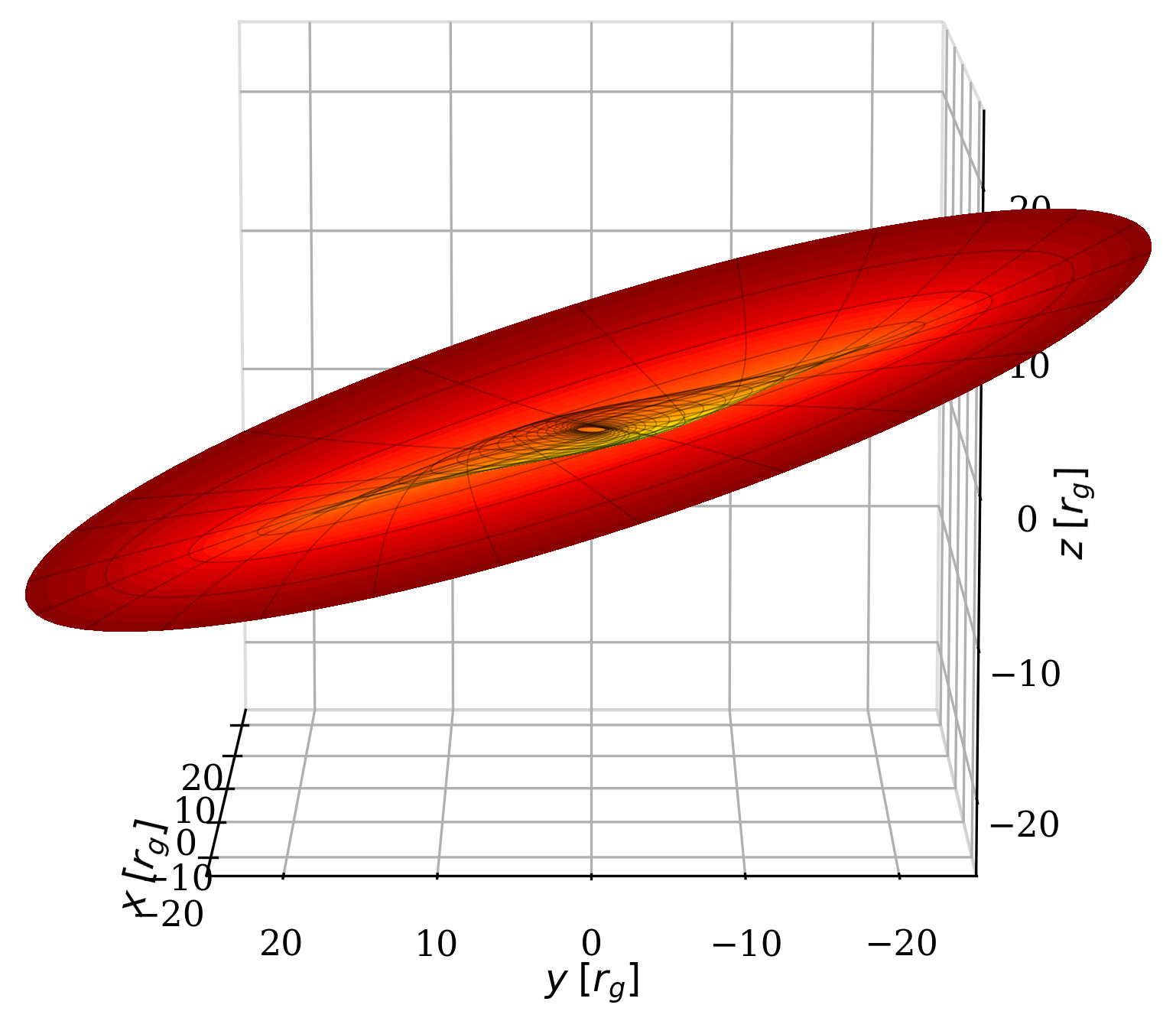}
    \caption{Different views of the actual \citealt{Scheuer1996} warped disk geometry, for $r_{\rm warp} = 5r_g$ and $\theta_{\rm warp} = 70^\circ$, observed at $\theta_{\rm obs}=80^\circ$ for different $\phi_{\rm obs}$. We have plotted only the inner $\sim 20$ gravitational radii for clarity. These views are chosen to highlight some of the interesting observer-dependent physics available in a highly warped disk system. The color displayed on the plot is simply the radius, to help visualize the warp structure itself. The observer in the top left panel sees a relatively ``normal'' disk system ($\phi_{\rm obs} = 50^\circ$), the upper right observer ($\phi_{\rm obs} = 330^\circ$) sees an inner disk which is highly inclined from their point of view, but a relatively $\sim$ flat outer disk. This would change the balance between harder (inner disk) and softer (outer disk) emission components.  The lower two panels highlight possibly geometric blocking effects, either concealing $\sim$ half the inner disk (lower left, $\phi_{\rm obs}=150^\circ$) or all the inner disk (lower right, $\phi_{\rm obs} = 180^\circ$). This lower right disk would be observed to be completely X-ray dark, independent of the actual X-ray emissivity of the inner disk.    }
    \label{fig:examples}
\end{figure*}

\begin{figure*}
    \centering
    \includegraphics[width=0.49\linewidth]{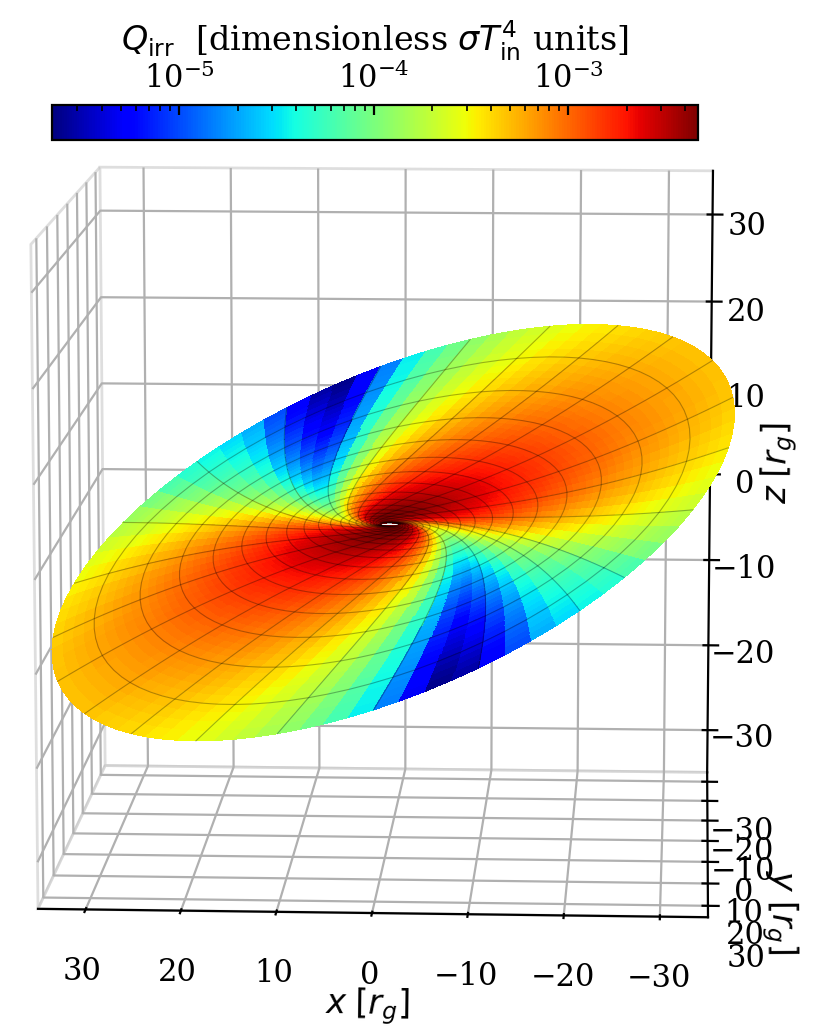}
    \includegraphics[width=0.48\linewidth]{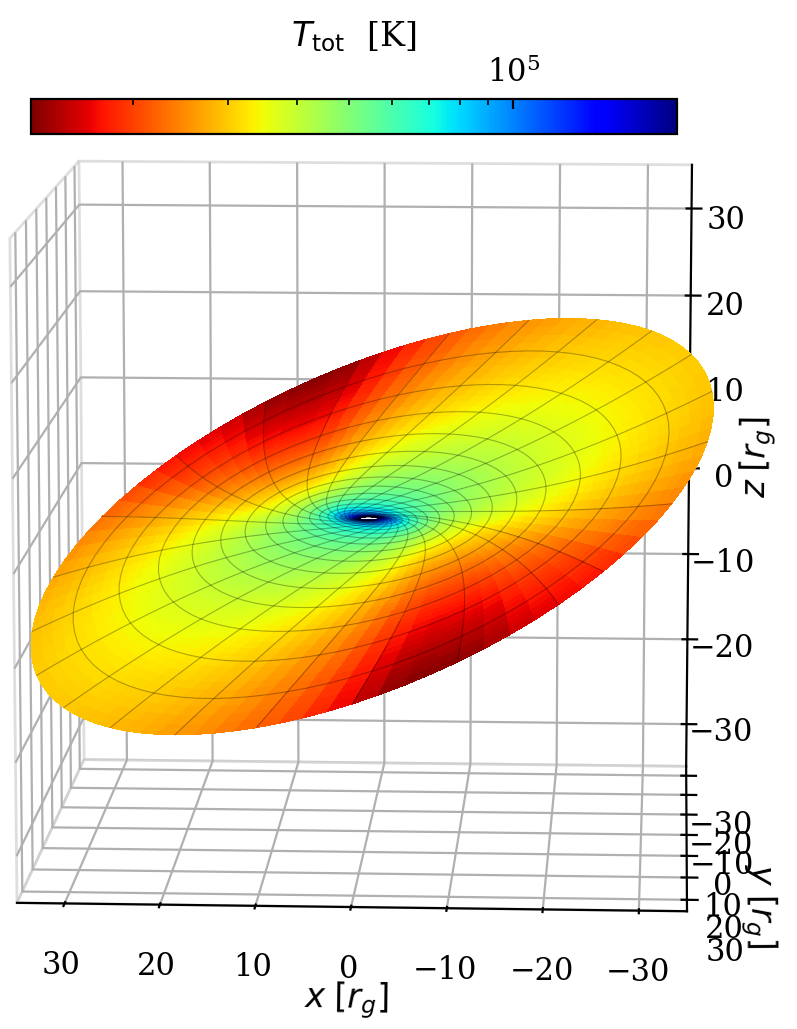}
    \caption{The dimensionless irradiation (in units of $\sigma T_{\rm in}^4$, left panel) profile of a warped disk ($r_{\rm warp} = 5r_g$ and $\theta_{\rm warp} = 70^\circ$) and the resulting two dimensional temperature profile (right panel). The irradiation profile directly tracks the warp, and makes the twist clear (the twisting blue, low irradiation, region are parts of the disk hidden behind the twist). The temperature profile shows a clear two component behavior, with the temperature along the steepest part of the warp falling off much more slowly with radius than the part of the disk behind the twist.   }
    \label{fig:irr_temp}
\end{figure*}

\begin{figure*}
    \centering
    \includegraphics[width=0.49\linewidth]{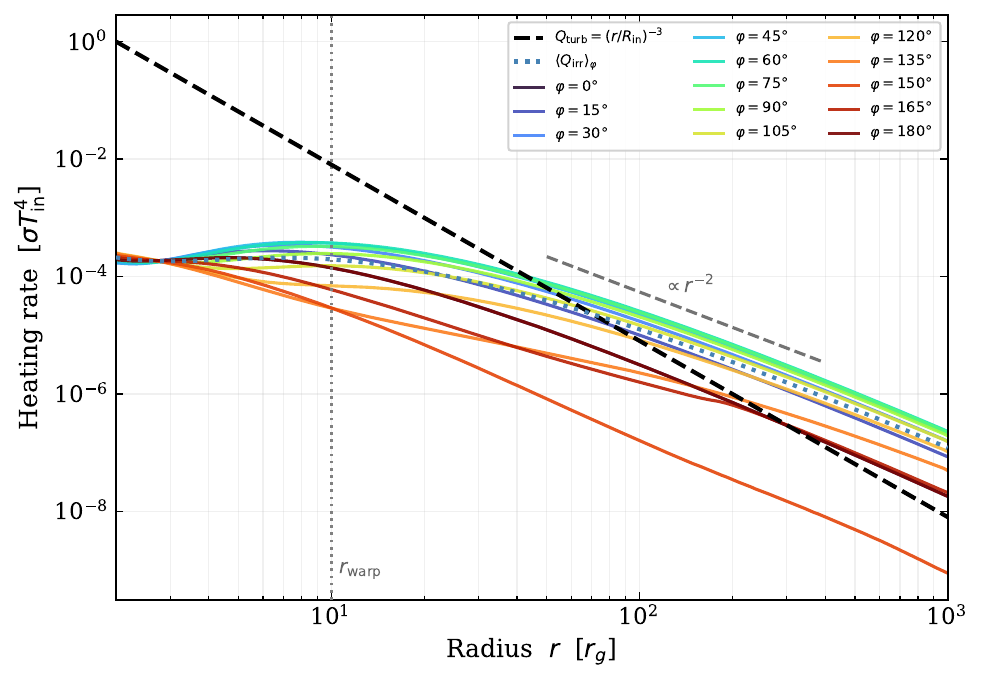}
    \includegraphics[width=0.49\linewidth]{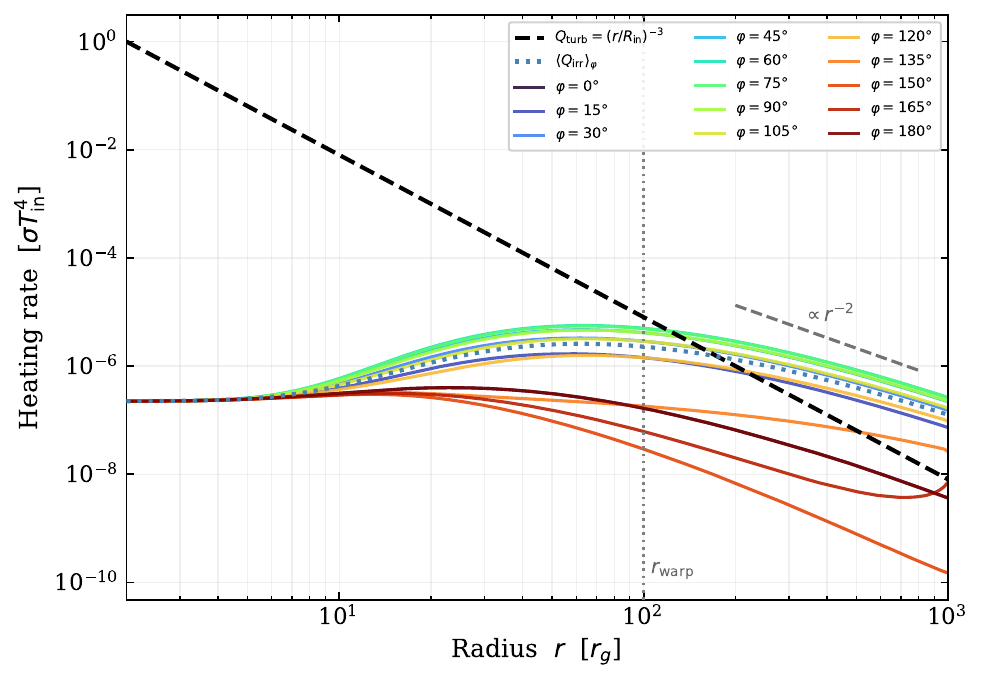}
    \includegraphics[width=0.49\linewidth]{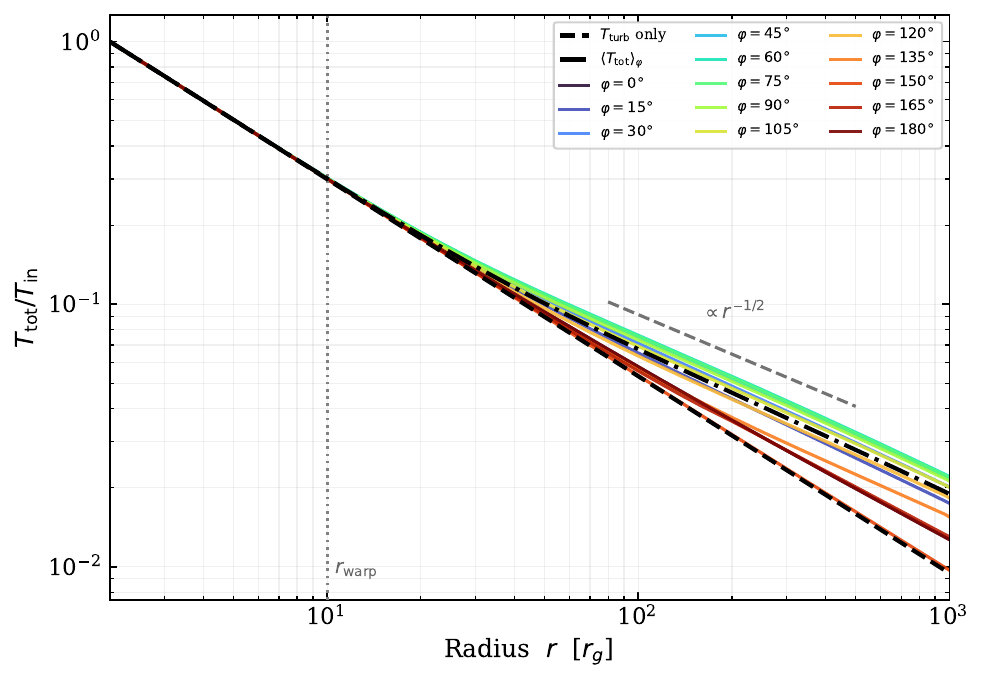}
    \includegraphics[width=0.49\linewidth]{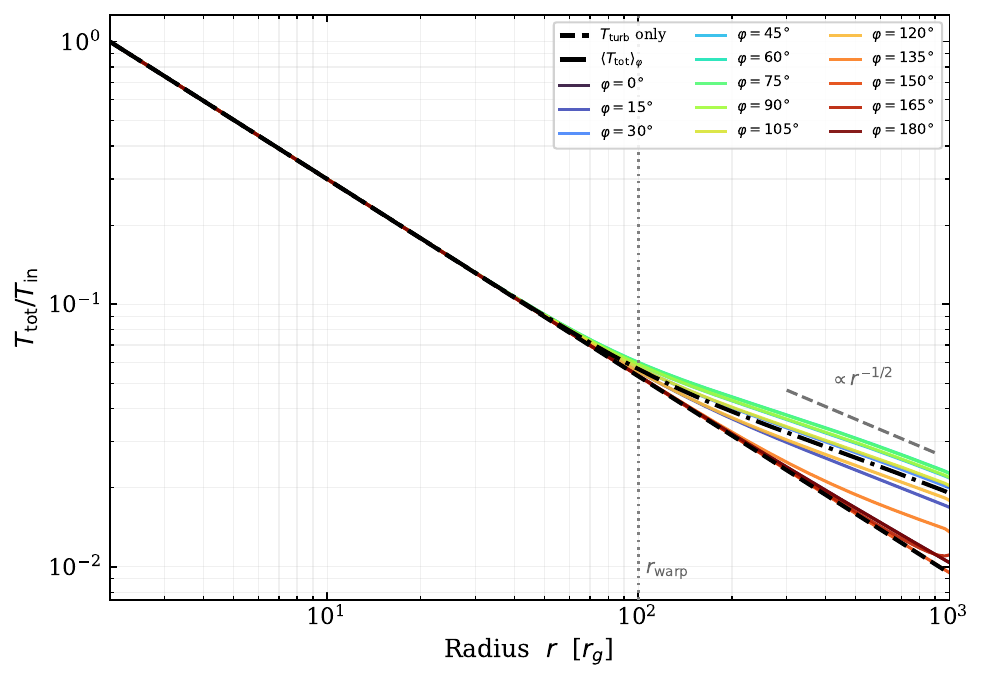}
    \caption{Dimensionless irradiative heating rates (upper, plotted in units of $\sigma T_{\rm in}^4$) and temperature profiles (lower, plotted in units of $T_{\rm in}$) plotted against radius for different azimuthal angles $\phi$ of two different warped disks, both with $\theta_{\rm warp}=50^\circ$ and with $r_{\rm warp}=10r_g$ (left column) and $r_{\rm warp} = 100 r_g$ (right column). The black dashed lines in all plots are the flat disk turbulence-only heating rates and temperatures. Starting at roughly the warp radius (where the disk gains non-trivial elevation) irradiation starts to dominate over turbulent heating for some azimuthal angles, approximately following the $r^{-2}$ profile expected of an inner point source (a reasonable approximation to an accretion flow which has luminosity dominated by $\sim R_{\rm in}$). This cause some regions of the disk to follow a $T \sim r^{-1/2}$  profile, which will have observational implications. Averaging over azimuthal angles gives the black dotted (upper) and dot-dashed (lower) profiles, which deviate noticeably from a flat disk. The non-monotonic behavior at some azimuthal angles results from the inner disk twist, which leads to non-trivial blocking of inner disk radiation from reaching outer regions (see e.g., Figure \ref{fig:irr_temp}).    }
    \label{fig:irr_temp_1d}
\end{figure*}


\section{The irradiation of warped disks}
\subsection{Self-irradiation in a warped geometry}
\label{sec:irr_kernel}

In a flat accretion disk, each annulus radiates into the half-space above (and below) the disk plane. In Newtonian gravity (where photons follow straight paths) no flat-disk annulus can irradiate any other, because all surface normals are parallel and no annulus subtends a finite solid angle as seen from any other. This is of course not true in a relativistic disk, as photons can be bent up and over the black hole, leading to irradiation even in the flat disk limit \citep[e.g.,][]{Thorne1974}. This effect, stemming from strong ray bending, does not typically modify the disk temperature profile to a strong degree \citep[e.g.,][]{Li05}. For the remainder of this paper we will work in the limit where photons follow straight lines, where we note this is an approximation which is poor in the innermost regions. 

A warped disk is fundamentally different. The tilted outer regions present their surfaces to the hot inner disk, intercepting a fraction of the inner-disk luminosity.  This self-irradiation can substantially modify the temperature profile at intermediate and large radii, with direct consequences for the observed spectral energy distribution \citep[e.g.,][]{Speicher2025}.

The radiative power exchanged between a source element ${\rm d}A_S$ at
position $\vec{\mathbf{S}}$ and a receiver element ${\rm d}A_P$ at position
$\vec{\mathbf{P}}$ is given by the standard Lambertian view-factor formula
\begin{equation}
  {\rm d}P_{S\to P}
  = I(r_S)\,
    \frac{|\mathbf{\hat{d}}\cdot\mathbf{\hat{n}}_S|\;|\mathbf{\hat{d}}\cdot\mathbf{\hat{n}}_P|}
         {|\vec {\mathbf{S}} - \vec{\mathbf{P}}|^2}\,{\rm d}A_S\,{\rm d}A_P,
  \label{eq:view_factor}
\end{equation}
where $\mathbf{\hat{d}} = (\vec{\mathbf{ S}} - \vec{\mathbf{P}})/|\vec{\mathbf{S}} - \vec{\mathbf{P}}|$ is the
unit vector from receiver to source, $\mathbf{\hat{n}}_S$ and $\mathbf{\hat{n}}_P$
are the outward unit normals at source and receiver, and
$I(r_S) = \sigma T(r_S)^4/\pi$ is the specific intensity of a
Lambertian emitter at temperature $T(r_S)$.  The $1/\pi$ in
$I$ encodes the cosine-weighted hemispherical radiation pattern
$\int I\cos\theta\,{\rm d}\Omega = \pi I = \sigma T^4$.

The two cosine factors have simple physical origins,
$|\mathbf{\hat{d}}\cdot\mathbf{\hat{n}}_S|$ is the cosine of the emission angle at the
source (governing the projected area of ${\rm d}A_S$ as seen from $\vec{\mathbf{P}}$),
and $|\mathbf{\hat{d}}\cdot\mathbf{\hat{n}}_P|$ is the cosine of the incidence angle
at the receiver (governing the projected area of ${\rm d}A_P$ presented
to the incoming ray).  The absolute values account for the fact
that both the top and bottom faces of the disk can participate in
radiative exchange (and that radiated power is a strictly positive quantity).

The irradiation flux (power per unit receiver area) at each grid point is obtained by summing equation~(\ref{eq:view_factor}) over all source elements and dividing by ${\rm d}A_P$
\begin{equation}
  Q_{\rm irr}(\vec{\mathbf{P}})
  = \sum_{S \neq P} (1 - {\cal S}(S, P))
    I(r_S)\,
    \frac{|\mathbf{\hat{d}}\cdot\mathbf{\hat{n}}_S|\;|\mathbf{\hat{d}}\cdot\mathbf{\hat{n}}_P|}
         {|\mathbf{S} - \mathbf{P}|^2}\,{\rm d}A_S.
  \label{eq:Q_irr}
\end{equation}
The self-term ($S = P$) is excluded, and ${\cal S}$ is a self-shadowing term which we will describe shortly.

We note that equation~(\ref{eq:view_factor}) contains two cosine
factors, as required by the standard view-factor derivation for
Lambertian surfaces \citep[e.g.][]{Howell2021}.  A
three-cosine variant including an additional factor
$|\mathbf{\hat{n}}_S\cdot\mathbf{\hat{n}}_P|$ has appeared in the literature
\citep[e.g., equation~17 of][]{Speicher2025}, inherited from
equation~(5) of \citet{Fukue1992}.  This third factor is
spurious, and we believe it was a typographic error in \cite{Fukue1992}. To see this clearly, note that for a disk with a right-angle kink, the two surface
normals are perpendicular and the three-cosine kernel predicts
zero irradiation identically, even when the connecting ray
strikes both surfaces at $45^\circ$.  As shown in
Appendix~\ref{app:kernel}, the two-cosine result is the one
consistent with \citeauthor{Fukue1992}'s own explicit
evaluation.  We adopt the two-cosine kernel throughout this work. 

\subsection{Self Shadowing}
On the surface of a warped disk, not every source element is visible from every receiver.  The
warped disk can shadow parts of itself: a source--receiver pair separated by the warp crest may have their connecting ray
intercepted by the disk surface.  We implement a ray--surface
intersection test that checks, for each $(S, P)$ pair, whether
the ray from $S$ to $P$ passes through any intervening disk
element. To be explicit, we compute the signed height of the ray above the disk surface via 
\begin{equation}
    h(t) = \vec{\mathbf{ r}}(t) \cdot {\mathbf{\hat l}}(r) ,
\end{equation}
where $\vec{\mathbf{r}}(t)$ is the (straight line) vector joining the points $\vec{\mathbf{S}}$ and $\vec{\mathbf{P}}$, $\vec{\mathbf{r}} = \vec{\mathbf{P}}-\vec{\mathbf{S}}$ (parameterized by a photon time $t$), and ${\mathbf{\hat l}}$ is the warped disk normal evaluated at spherical radius $r \equiv |\vec{\mathbf{r}}|$. A photon which is intercepted by an intervening disk radius can be identified through a change in sign of $h$ across the trajectory. 

If the two points are blocked by an intermediate disk region, then we take ${\cal S}=1$, otherwise ${\cal S}=0$. This means that shadowed pairs are excluded from the
sum~(\ref{eq:Q_irr}).  The effect of shadowing is a
$\sim 10$--$20\%$ reduction in $Q_{\rm irr}$ at radii just beyond $r_{\rm warp}$, where the warp crest blocks the line of sight to the inner disk.  The integrated effect on the broadband SED is smaller.

\subsection{Temperature profiles}
\label{sec:temperature}

The local disk temperature is set by the balance of turbulent dissipation and irradiative heating.  In the absence of a zero-torque inner boundary condition (chosen here for simplicity, but it is also possibly more physically accurate), the turbulent
dissipation rate per unit area is
\begin{equation}
  Q_{\rm turb}(r)
  \equiv 2\sigma T_{\rm in}^4 \left(\frac{r}{R_{\rm in}}\right)^{-3},
  \label{eq:Q_visc}
\end{equation}
where $T_{\rm in}$ is the temperature at the inner disk edge $r = R_{\rm in}$, and the factor of 2 accounts for radiation from both disk faces.

Adding the irradiation heating and solving for the total
temperature\footnote{Formally one should perform this calculation iteratively, as the incoming irradiated energy will eventually be re-emitted which could then re-irradiate another disk element (etc.). We found this was always a small effect, and so drop it for computational efficiency.} leads to 
\begin{equation}
  \sigma T_{\rm tot}^4(r, \phi)
  = \frac{Q_{\rm turb}(r) + Q_{\rm irr}(r, \phi)}{2},
\end{equation}
which, expressed in terms of the dimensionless irradiation
$\widetilde Q_{\rm irr}(r, \phi) \equiv Q_{\rm irr}/(2\sigma T_{\rm in}^4)$,
gives
\begin{equation}
  T_{\rm tot}(r, \phi)
  = T_{\rm in}\left[
    \left(\frac{r}{R_{\rm in}}\right)^{-3}
    + \widetilde Q_{\rm irr}(r, \phi)
  \right]^{1/4}.
  \label{eq:T_tot}
\end{equation}
The dimensionless irradiation heating $\widetilde Q_{\rm irr}(r, \phi)$ has several notable properties:

\begin{enumerate}
  \item \textit{Radial structure.}
        Irradiation is negligible at small radii (where the warp
        has vanishingly small tilt and no surface is presented
        to the inner disk) and declines at very large radii
        (where the solid angle subtended by the inner disk
        diminishes as $r^{-2}$).  The irradiation peaks at
        the warp radius $r \sim r_{\rm warp}$, where the
        warp presents the maximum cross-section to the luminous
        inner disk.

  \item \textit{Azimuthal asymmetry.}
        The side of each annulus that faces toward the inner disk
        receives substantially more irradiation than the far side.
        When the Bardeen--Petterson twist is included, this
        hot spot rotates with radius following the twist angle
        $\gamma(r) = 2\sqrt{r_{\rm warp}/r}$.

  \item \textit{Dominance over turbulent heating.}
        At sufficiently large radii, irradiation dominates over
        turbulent dissipation: $\widetilde Q_{\rm irr} \gg (r/R_{\rm in})^{-3}$.
        The crossover radius depends on the warp amplitude
        $\theta_{\rm warp}$ and the ratio $r_{\rm warp}/R_{\rm in}$.
        Beyond this radius, the temperature profile flattens
        relative to the $T \propto r^{-3/4}$ turbulent dissipation law. Pure irradiative heating from a central point source would have $T\propto r^{-1/2}$.

  \item \textit{Shadowing.}
        Self-shadowing by the warp crest reduces irradiation in a
        band of azimuths just behind the warp peak.  This creates
        a ``shadow'' on the disk surface where $\widetilde Q_{\rm irr}$
        is suppressed.  The shadow is most pronounced for large
        $\theta_{\rm warp}$ and small $r_{\rm warp}$.
\end{enumerate}



We highlight this physics in Figures \ref{fig:irr_temp} and \ref{fig:irr_temp_1d}. In Figure \ref{fig:irr_temp} we show the dimensionless irradiation profile (in units of $\sigma T_{\rm in}^4$, left panel)  of a warped disk ($r_{\rm warp} = 5r_g$ and $\theta_{\rm warp} = 70^\circ$) and the resulting two dimensional temperature profile (right panel; we take $T_{\rm in}=2\times 10^5$ K, a canonical inner TDE disk temperature). The irradiation profile directly tracks the warp, and makes the twist clear (the twisting blue, low irradiation, region are parts of the disk hidden behind the twist). The temperature profile shows a clear two component behavior, with the temperature along the steepest part of the warp falling off much more slowly with radius than the part of the disk behind the twist (which follows $\sim r^{-3/4}$). 

In more detail we plot one dimensional (i.e., versus radius at fixed $\phi$) projections of the dimensionless irradiative heating rates (upper, plotted in units of $\sigma T_{\rm in}^4$) and temperature profiles (lower, plotted in units of $T_{\rm in}$) for two different warped disks, both with $\theta_{\rm warp}=50^\circ$ but one with $r_{\rm warp}=10r_g$ (left column) and $r_{\rm warp} = 100 r_g$ (right column). The black dashed lines in all plots are the flat disk turbulence-only heating rates and temperatures. Starting at roughly the warp radius (where the disk gains non-trivial elevation) irradiation starts to dominate over turbulent heating for some azimuthal angles, approximately following the $r^{-2}$ profile expected of an inner point source (a reasonable approximation to an accretion flow which has luminosity dominated by $\sim R_{\rm in}$). This cause some regions of the disk to follow a fall off which is better approximated by a $T \sim r^{-1/2}$  profile, which will have observational implications (to be discussed shortly). Averaging over azimuthal angles gives the black dotted (upper) and dot-dashed (lower) profiles, which deviate noticeably from a flat disk. The non-monotonic behavior at some azimuthal angles results from the inner disk twist, which leads to non-trivial blocking of inner disk radiation from reaching outer regions (e.g., the angular structure of the blue stripe in Figure \ref{fig:irr_temp}).

\newpage
\subsection{The observed spectral energy distribution}
The frequency-specific flux density $F_\nu$ of  disk radiation, as observed by a distant observer, is  by definition 
\begin{equation}
F_{\nu}(\nu) = \int I_\nu (\nu) \, \text{d}\Theta .
\end{equation} 
Here, $\nu$ is the photon frequency and $I_\nu(\nu)$ the specific intensity,  both measured at the location of the distant observer.   The differential element of solid angle subtended by the disk area element on the observer's sky is $\text{d}\Theta$. We will for the remainder of this paper neglect relativistic corrections to this expression, such as Doppler and gravitational red-shifting, and gravitational lensing. This is for computational simplicity and not necessarily because it is a good approximation in all regimes. 

As discussed above, the local effective temperature at each surface element $(r, \varphi)$ follows from energy balance between turbulent dissipation and reprocessed irradiation. We shall then assume that each disk annulus radiates as a colour-corrected blackbody with spectral hardening factor $f_{\rm c}(T)$ \citep[we use the explicit model from][]{Done2012}, so that the colour temperature is
\begin{equation}
    T_{\rm col}(r, \varphi) = f_{\rm c}\bigl(T_{\rm tot}\bigr) \, T_{\rm tot}(r, \varphi) .
\end{equation}
This choice is motivated by the fact that disk surfaces are not perfect blackbodies. Indeed, the balance between scattering opacities and absorption opacities generically leads to the observed temperatures of disk being slightly ($\sim 2$ times) higher than $T_{\rm tot}$, as photons sourced in the (much hotter as $\kappa\Sigma\gg1$) disk midplane do not fully thermalise their energy on their journey through the disk atmosphere, and so have harder energies more reflective of their midplane structure than would naively be expected from pure energetic balance and pure thermalisation grounds.

\begin{figure*}
    \centering
    \includegraphics[width=0.95\linewidth]{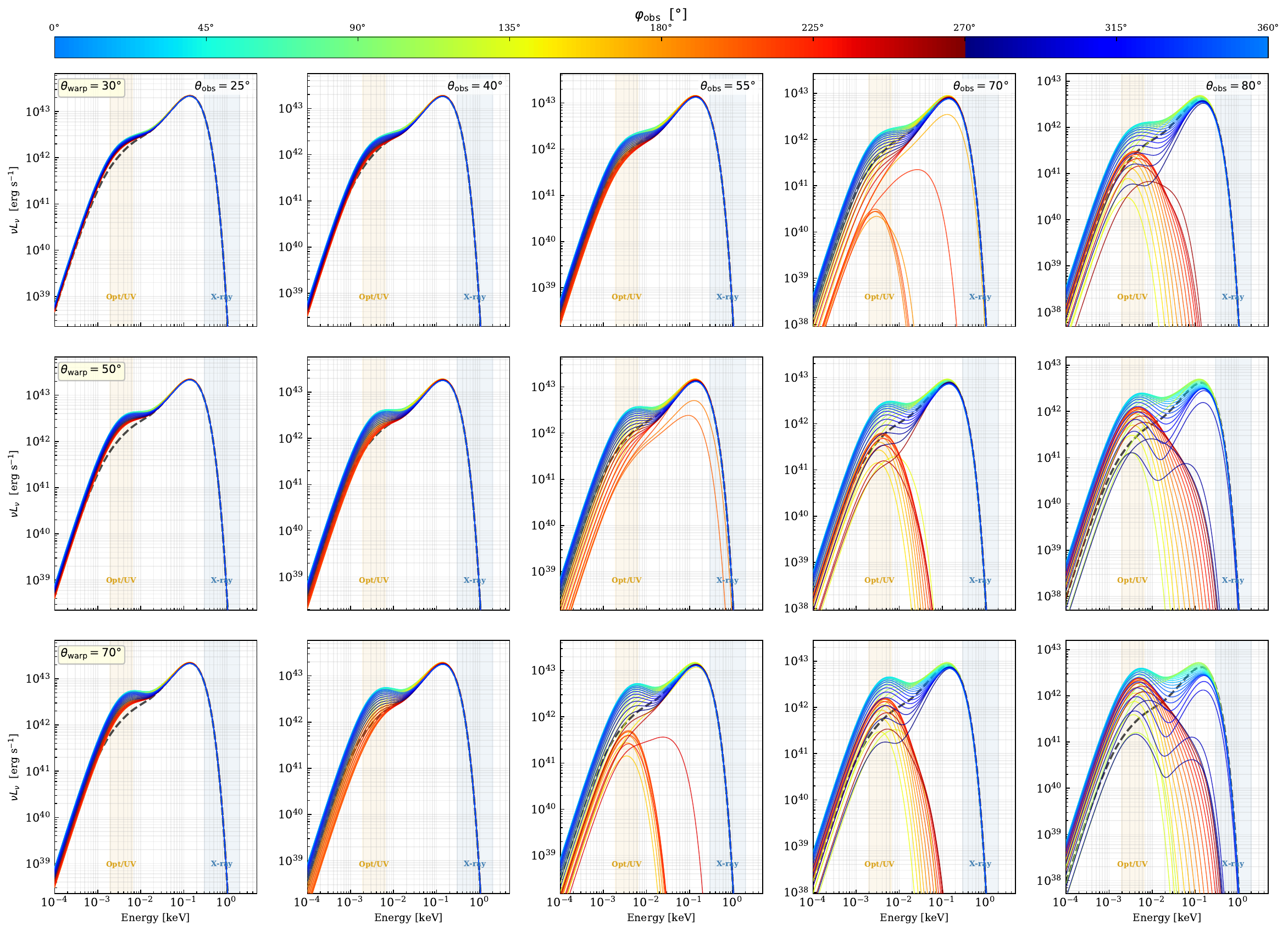}
    \caption{A sequence of warped disk spectral energy distributions for different $\theta_{\rm warp}$ (rows), observer inclinations $\theta_{\rm obs}$ (columns) as a function of observer azimuth (color bar). Note that the color bar changes from red to blue at $\phi_{\rm obs} = 270^\circ$, roughly the point at which highly inclined observers may end up behind the warp. The disks here had $r_{\rm warp} = 10\,r_g$ and $R_{\rm out} = 500\,r_g$ (with $M_\bullet = 10^7\,M_\odot$, $R_{\rm in} = 2\,r_g$ and inner temperature $T_{\rm in} = 2.5\times10^5$\,K). For roughly face on observers (left two columns) all observer azimuths see a spectrum fairly similar to a flat disk (black dashed line). For more inclined observers (e.g., right hand most panel) different observers see a much broader range of possible behaviors. This includes anomalous optical/UV colors (contrast the spectral slope in the optical/UV for some of the warped disk spectra with that of the flat disk), and X-ray suppression (outer disk blocks inner disk from sight). For a given $\phi_{\rm obs}$ there is a critical inclination $\theta_{\rm obs} = i_{\rm crit}$ at which the inner disk is blocked from view.   }
    \label{fig:theta_phi}
\end{figure*}

The specific intensity of the locally emitted radiation is then given by a modified Planck function 
\begin{equation}\label{planck}
I_\nu(\nu) = {1\over f_{\rm c}^{4}} B_\nu(\nu, f_{\rm c} T) \equiv \frac{2h\nu^3}{f_{\rm c}^4 c^2} \left[ \exp\left( \frac{h\nu}{k f_{\rm c} T} \right) - 1\right]^{-1} .
\end{equation}
so that
the spectral colour temperature is what an observer ``sees'' $T_{\rm col} = f_c T_{\rm tot}$, while the
bolometric luminosity of each annulus is preserved at
$\sigma T_{\rm tot}^4$ per face (as enforced by energy conservation).  

The observed isotropic-equivalent spectral luminosity is obtained by integrating the colour-corrected Planck function over all visible surface elements,
\begin{equation}
    \nu L_\nu = \frac{8 \pi h \nu^4}{c^2} \sum_{i,j} \frac{\mathcal{V}_{ij} \, |\cos\Upsilon_{ij}| \, \mathrm{d}A_{ij}}{f_{{\rm c},ij}^4 \left[\exp\!\left(\dfrac{h\nu}{k_{\rm B} \, f_{{\rm c},ij} \, T_{{\rm tot},ij}}\right) - 1\right]} ,
    \label{eq:nuLnu}
\end{equation}
where the sum runs over all grid cells $(r_i, \varphi_j)$ and:
\begin{itemize}
    \item $\mathbf{\hat{o}} = (\sin\theta_{\rm obs}\cos\varphi_{\rm obs},\; \sin\theta_{\rm obs}\sin\varphi_{\rm obs},\; \cos\theta_{\rm obs})$ is the unit vector toward the observer;
    \item $\cos\Upsilon_{ij} = \mathbf{\hat{l}}_{ij} \cdot \mathbf{\hat{o}}$ is the projection of the local surface normal onto the observer direction, with the absolute value  $|\cos\Upsilon|$ covering whichever disk face is visible;
    \item $\mathrm{d}A_{ij}$ is the proper area element of the warped surface, computed as discussed in the Appendix,
    \item $\mathcal{V}_{ij} \in \{0, 1\}$ is the observer-visibility mask: $\mathcal{V}_{ij} = 0$ if element $(i,j)$ is occluded from the observer's line of sight by an intervening part of the warped disk, and $\mathcal{V}_{ij} = 1$ otherwise (discussed above)
    \item $f_{{\rm c},ij} = f_{\rm c}(T_{{\rm tot},ij})$ is the local spectral hardening factor.
\end{itemize}

The leading factor of $4\pi$ converts from the directional (Poynting-flux) luminosity  \citep[e.g.,][Eq.~19]{Speicher2025}  to the observational convention $\nu L_\nu \equiv 4\pi D^2 \nu F_\nu$ used in the TDE literature \citep[e.g.][]{Mummery_et_al_2024}.

For a flat disk (i.e., every disk element has $\Upsilon_{ij} = \theta_{\rm obs}$, $\mathcal{V}_{ij} = 1$, $\widetilde {Q}_{\rm irr} = 0, {\rm d}A_{ij} = \Delta r_i \Delta \phi_j$) and Eq.~\eqref{eq:nuLnu} reduces to the standard multi-colour blackbody spectrum.  The warped geometry modifies the observed SED in three ways: (i) irradiation heating raises $T_{\rm tot}$ on the illuminated face of the warp; (ii) the varying projected area $|\cos\Upsilon|$ varies across the disk which modifies the energy-dependent flux; and (iii) occultation ($\mathcal{V} = 0$) removes the contribution of elements hidden behind the warp, preferentially suppressing the hottest (innermost) regions for observers near the maximum-occultation azimuth $\varphi_{\rm obs} \approx 180^\circ$. In Figure \ref{fig:theta_phi} we show a library of warped disk SED profiles for different observing angles, and different warp angles. There is a rich structure in this library, and we walk through some of the observational implications in the following section.

\section{Observational consequences}
In this section we discuss three plausible observational signatures of a global warp in a TDE accretion flow. 
\subsection{Anomalous (reddened) optical colors}
As was highlighted in Figure \ref{fig:irr_temp_1d}, a warped disk generally has a temperature profile which deviates from the classic $T\propto R^{-3/4}$ power-law dependence owing to irradiative heating. As the optical colors (or power law slope $b$ in $\nu L_\nu \propto \nu^b$) depend on the precise temperature profile of the disk, a disk with strong irradiative heating will show a modified spectral slope in the optical/UV.

\begin{figure*}
    \centering
    \includegraphics[width=0.95\linewidth]{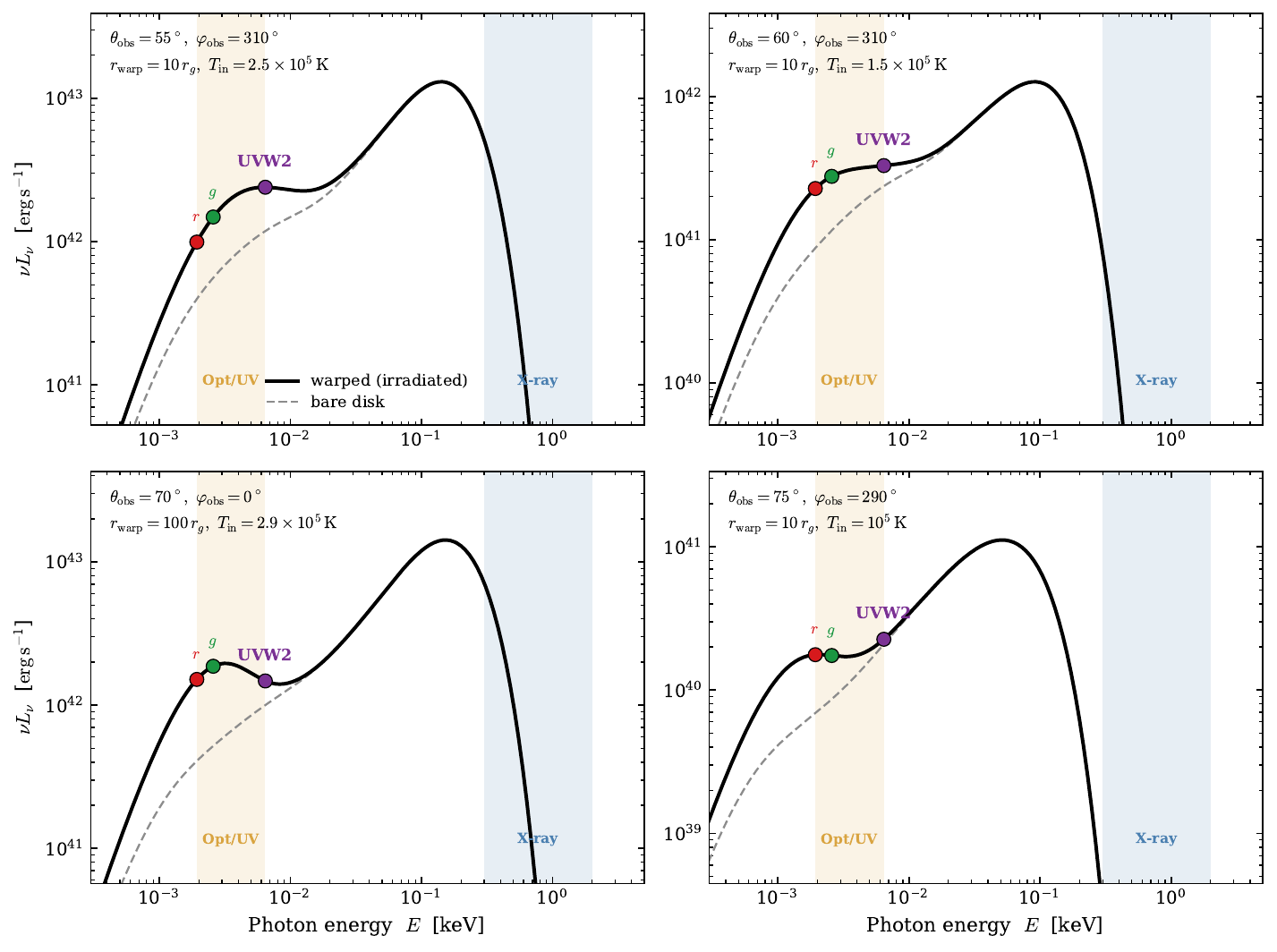}
    \caption{Some specific (cherry picked) examples of interesting optical/UV spectral slopes, for different warped disk profiles, shown to highlight the range of possibilities. The shared parameters for all of these plots are $\theta_{\rm warp} = 50^\circ$, $M_\bullet = 10^7\,M_\odot$ and $R_{\rm in} = 2\,r_g$, other parameters are shown on each plot. A warped disk can introduce a steepening of the optical emission (upper left), a flattening (upper right), an explicit bump (lower left) or more of a kink (lower right).  }
    \label{fig:cols}
\end{figure*}

Let us for analytical simplicity assume that we  can treat this as a one-dimensional problem. This is, of course, a terrible assumption. All of the interesting warped disk physics occurs because the surface of the disk is fundamentally three dimensional. However, a one dimensional approximation is very simple to handle analytically and, perhaps surprisingly, captures key physical scalings which are present in the full (numerical) calculations. 

Let us take the liberty of assuming that the observer is viewing the disk with $\varphi_{\rm obs}\approx 0^\circ$ (i.e., looking face on at the warp; numerical computations will of course have general $\varphi_{\rm obs}$, this is merely intended to highlight the physical scalings) such that for an observer at a large distance $D$ from the source, the differential solid angle into which the radiation is emitted is approximately (again, neglecting relativistic corrections)
\begin{equation}
 \text{d}\Theta \approx \frac{2\pi R \cos(i) \, {\rm d} R }{D^2} ,
\end{equation} 
where $R$ is not really the disk radius, instead it is better thought of as a rough coordinate describing the temperature fall off in the direction on the warped disk surface which captures a particular disk temperature scaling (e.g., looking at the right hand side of Figure \ref{fig:irr_temp} there are two different temperature scalings in different directions on the warped disk surface). 

We have assumed azimuthal symmetry, and $i$ is the inclination of the source with respect to the observers line of sight ($i = 0$ corresponding to face-on). Of course for a warped disk, $i$ is in reality a function of radius, and the disk is not azimuthally symmetric.

Within this approximation scheme, the observed flux from the disk surface  is therefore
\begin{equation}\label{flux}
F_\nu(\nu) \approx  \frac{2\pi \cos(i)}{D^2}\int_{R_{\rm in}}^{R_{\rm out}}  {R \over f_{\rm c}^{4}} B_\nu (\nu, f_{\rm c} T) \, {\rm d}R, 
\end{equation}
or in full
\begin{equation}\label{flux}
F_\nu(\nu) \approx  \frac{4\pi h \nu^3 \cos(i)}{c^2 D^2}\int_{R_{\rm in}}^{R_{\rm out}}  {R \over f_{\rm c}^{4}} {1 \over \exp\left( {h\nu}/{k f_{\rm c} T} \right) - 1}  \, {\rm d}R. 
\end{equation}
Assume that the colour-correction factor varies with disk temperature $T$ as a power-law (this is actually a pretty good assumption, as the balance of scattering and absorption opacities are strongly temperature dependent)
\begin{equation}
f_{\rm c} = \left({T \over T_0}\right)^p, 
\end{equation}
where the index $p$ may be either positive or negative. Further assume that the temperature in the disk varies as some power-law with radius (also a good approximation)
\begin{equation}
T = T_0 \left({R \over R_0}\right)^{-q},
\end{equation}
where $q > 0$, and $T_0$ is the same in both expressions (so that the $T_0$ in the above expression is set by the physics of colour-correction). Then 
\begin{multline}
F_\nu(\nu) \approx \frac{4\pi h \nu^3 \cos(i)}{c^2 D^2} \int_{R_{\rm in}}^{R_{\rm out}}  R \, \left({R \over R_0}\right)^{4pq}  \\  \left[ \exp\left( \frac{h\nu}{k T_0} \left({R\over R_0}\right)^{q + pq} \right) - 1\right]^{-1}  \, {\rm d}R. 
\end{multline}
Define 
\begin{equation}
y \equiv \frac{h\nu}{k T_0} \left({R\over R_0}\right)^{q + pq} ,
\end{equation}
which leaves 
\begin{align}
F_\nu(\nu) &\approx \frac{4\pi h \nu^3 \cos(i)}{c^2} \left({R_0 \over D}\right)^2 \left(\frac{h\nu}{k T_0} \right)^{-(2 + 4pq)/(q + pq)} I_{pq} ,\nonumber  \\
I_{pq} &\equiv {1\over q + pq}\int_{y_{\rm in}}^{y_{\rm out}}  {y^{(2+ 4pq)/(q + pq) - 1} \over \exp (y) - 1} \, {\rm d} y .
\end{align}
Note that this formula (in terms of $I_{pq}$) holds completely in general (within our approximation scheme), and the integral $I_{pq}$ is a function of frequency through the integration limits $y_{\rm in}, y_{\rm out}$. To specialize to the ``mid-frequency'' regime, we consider frequencies where 
\begin{equation}
k f_{\rm col} T_{\rm out} \ll h\nu \ll kf_{\rm col} T_{\rm in},
\end{equation}
where $T_{\rm in}, T_{\rm out}$ are the inner and outer disk temperatures respectively. In terms of the variable $y$ this implies 
\begin{equation}
y_{\rm in} \ll 1 \ll y_{\rm out}.
\end{equation}
Extending the integration limits so that we may approximate both as $y_{\rm} \to 0, y_{\rm out} \to \infty$ allows us to solve the integral $I_{pq}$ exactly
\begin{multline}
I_{pq} = {1\over q + pq}\int_{0}^{\infty}  {y^{(2+ 4pq)/(q + pq) - 1} \over \exp (y) - 1} \, {\rm d} y \\ =   {1\over q + pq} \Gamma\left({2+ 4pq \over q + pq }\right) \zeta_{\cal R} \left({2+ 4pq \over q + pq }\right) ,
\end{multline}
where $\Gamma$ and $\zeta_{\cal R}$ are the gamma and Riemann zeta functions respectively \citep[this result goes back to][]{LyndenBell69}. Note that in the ``mid-frequency'' limit the function $I_{pq}$ is independent of frequency, and the observed flux from the disk scales as a power-law with frequency 
\begin{equation}\label{keyR}
F_\nu \propto \nu^{3 - (2+ 4pq)/(q + pq) }. 
\end{equation}
Which means that the spectral slope is 
\begin{equation}
    \log \nu L_\nu \propto \left({4q - 2 \over q + pq}\right) \log \nu 
\end{equation}
The impact of irradiation is to make the disk temperature more shallow, with the no-irradiation limit being $q=3/4$, while the central point irradiation dominated limit being $q=1/2$. While this may appear a somewhat subtle shift in temperature profile, it can have a dramatic impact on the optical/UV colors. 

For total thermalisation (no color-correction, $p=0$), then the no-irradiation disk has very blue colors 
\begin{equation}
    \log \nu L_\nu \propto {4\over 3} \log \nu,\quad\quad {\rm (no\,\, irradiation, \,\,} p =0 ),
\end{equation}
while for a temperature profile dominated by irradiation $q\to 1/2$, one has 
\begin{equation}
    \log \nu L_\nu \approx {\rm constant},\quad {\rm (irradiation \,\, dominated,}\,\, {\rm all}\,\,p).
\end{equation}
One sees that irradiation has the effect of significantly reddening the disk spectrum.


\begin{figure*}
    \centering
    \includegraphics[width=0.7\linewidth]{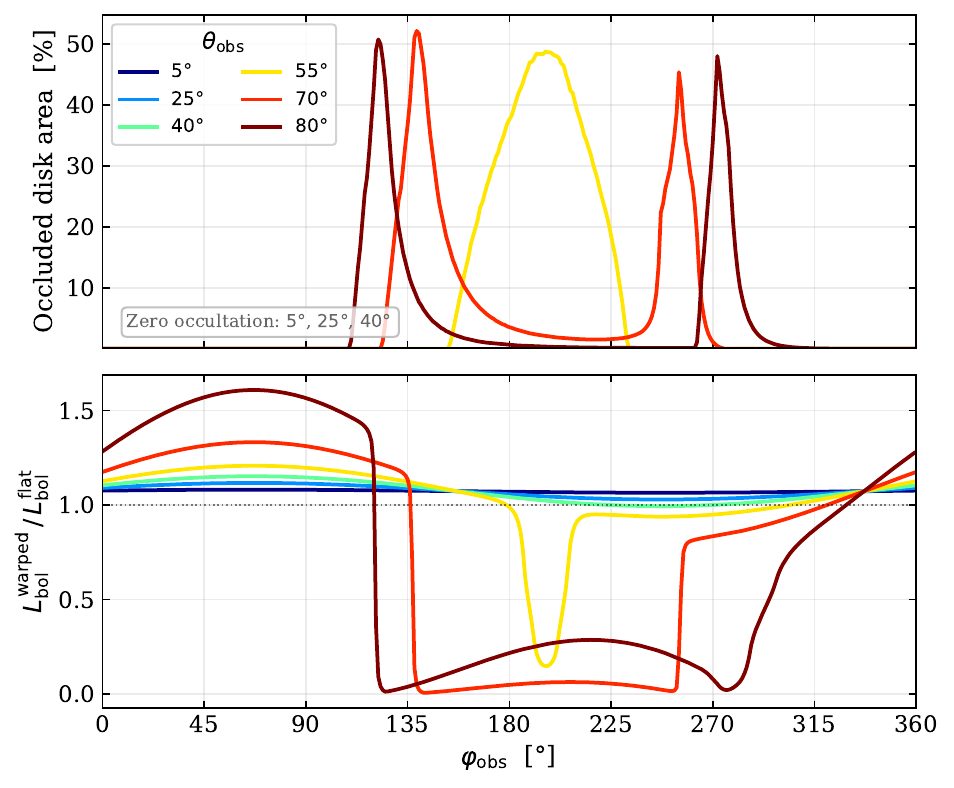}
    \caption{The fraction of the total disk area (upper panel) and the bolometric luminosity ratio (versus a flat disk, lower panel) as a function of $\phi_{\rm obs}$ for different polar observing angles $\theta_{\rm obs}$. It is interesting that occluding a large fraction of the disk area does not necessarily impact the bolometric luminosity to a great degree, as the bolometric luminosity cares primarily about the inner disk. When the inner disk is obscured the bolometric luminosity drops strongly. Generically, warped disks (which are not obscured from view) are brighter than non-warped disks, owing to irradiative heating. The warp parameters are $r_{\rm warp} = 10\,r_g$, $R_{\rm out} = 500\,r_g$ and $\theta_{\rm warp} = 50^\circ$ (with $M_\bullet = 10^7\,M_\odot$, $R_{\rm in} = 2\,r_g$ and $T_{\rm in} = 1.5\times10^5$\,K). }
    \label{fig:occ}
\end{figure*}

Despite the severe assumptions employed in computing this spectral slope, one can see that it captures the essence of what is seen in the full numerical warped disk SED calculations (Figures \ref{fig:theta_phi} and \ref{fig:cols}). 

Examining the left hand column of Figure \ref{fig:theta_phi}, which shows a sequence of warped disk spectral energy distributions for different $\theta_{\rm warp}$ (rows), at roughly face on  observer inclinations $\theta_{\rm obs}=25^\circ$ as a function of observer azimuth (color bar). Note that the color bar changes from red to blue at $\phi_{\rm obs} = 270^\circ$, roughly the point at which highly inclined observers may end up behind the warp. One sees that as the warp angle is increased (lower panels), and the irradiation begins to contribute more to the energy balance, the mid-frequency part of the spectrum increases in amplitude (as there is more energy available to radiate) and flattens (reddens) in its slope. 


Of course, the full warped disk SED is more complicated than being dominated by any one simple SED slope. The reason for this is again clear by inspection of Figures \ref{fig:irr_temp} and \ref{fig:irr_temp_1d}, there are multiple different azimuthal slices through a warped disk which have different temperature profiles. One can think therefore of the full spectral energy distribution being comprised of different sub-regions which all produce $\sim$ power law profiles (in the mid frequency regime), which then superimpose in non-trivial ways, depending on the size of the disk ($R_{\rm out}/R_{\rm in}$), the temperature of the inner disk (relative to optical/UV), the observer viewing angles, and the warp structure. Much of this range in possible behaviors can be seen by inspection of Figure \ref{fig:theta_phi}. 

However, to highlight cleanly some of the possible behaviors, we cherry pick some interesting optical/UV SED slopes in Figure \ref{fig:cols}. The shared parameters for all of these plots are denoted in the Figure caption, other parameters are shown on each plot. A warped disk can introduce a steepening of the optical emission (upper left), a flattening (upper right), an explicit bump (lower left) or more of a kink (lower right). As can be seen by the gray dashed lines on each plot, each of these SED signatures are quite distinct from what one would expect from a flat (equatorial) disk, and the observation of any one of these signatures would comprise interesting evidence for warped flows in TDEs.

\subsection{Inner-disk (X-ray) occultation}\label{sec:blocking}

\begin{figure*}
    \centering
    \includegraphics[width=0.95\linewidth]{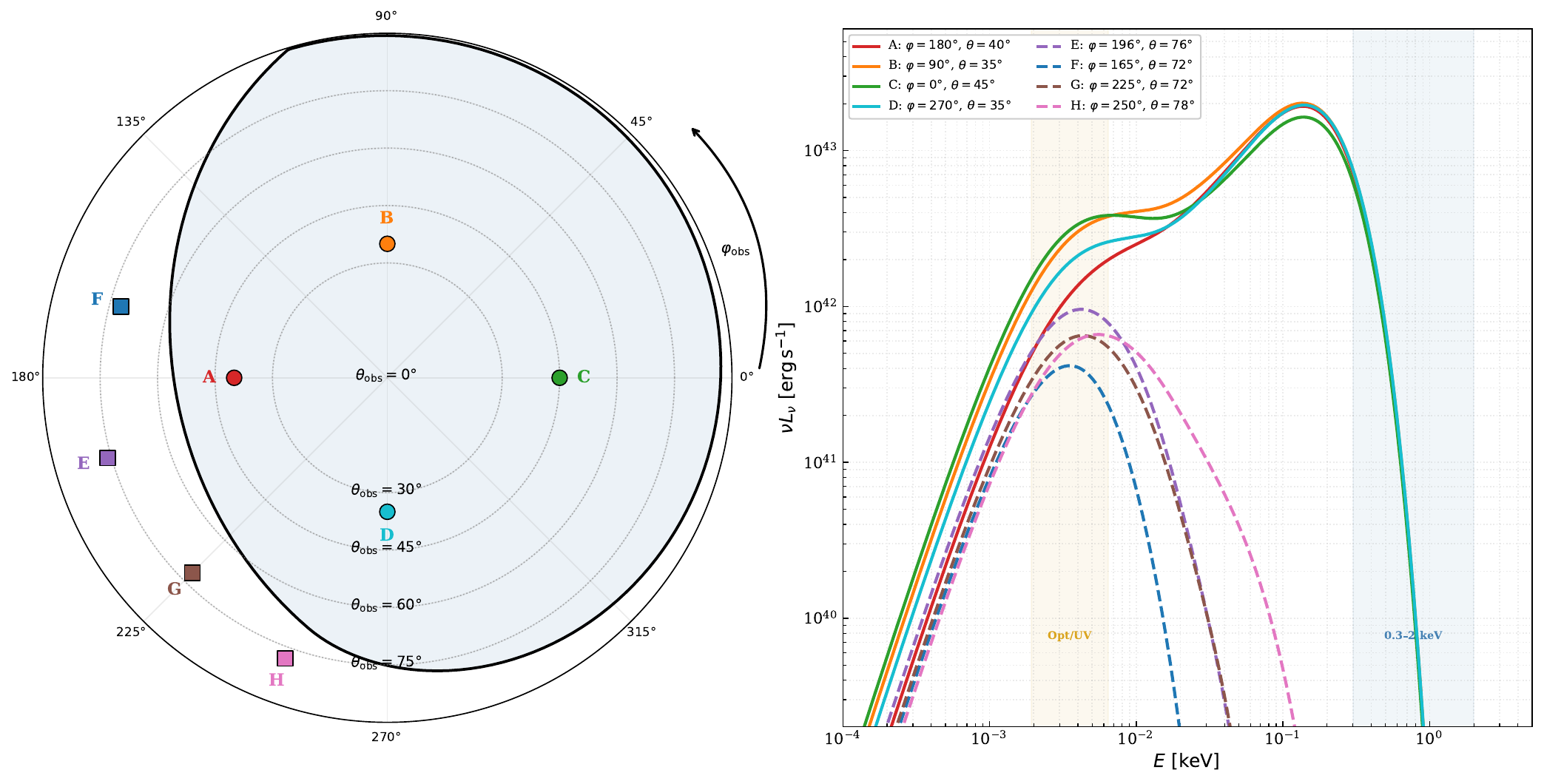}
    \caption{The critical inclination curve on the observers sky (black curve, left panel, see text for details), and the resulting difference in observed spectra. An observer within the occultation curve (i.e., observers A, B, C and D) can see the inner disk, and a relatively ``normal'' disk spectrum. On the contrary, observers (E, F, G and H) who are outside of the occultation curve cannot see the inner disk, and see strong X-ray suppression. The warp parameters are $r_{\rm warp} = 10\,r_g$, $R_{\rm out} = 500\,r_g$ and $\theta_{\rm warp} = 50^\circ$ (with $M_\bullet = 10^7\,M_\odot$, $R_{\rm in} = 2\,r_g$ and $T_{\rm in} = 2.5\times10^5$\,K).  }
    \label{fig:icrit}
\end{figure*}

\begin{figure*}
    \centering
    \includegraphics[width=0.95\linewidth]{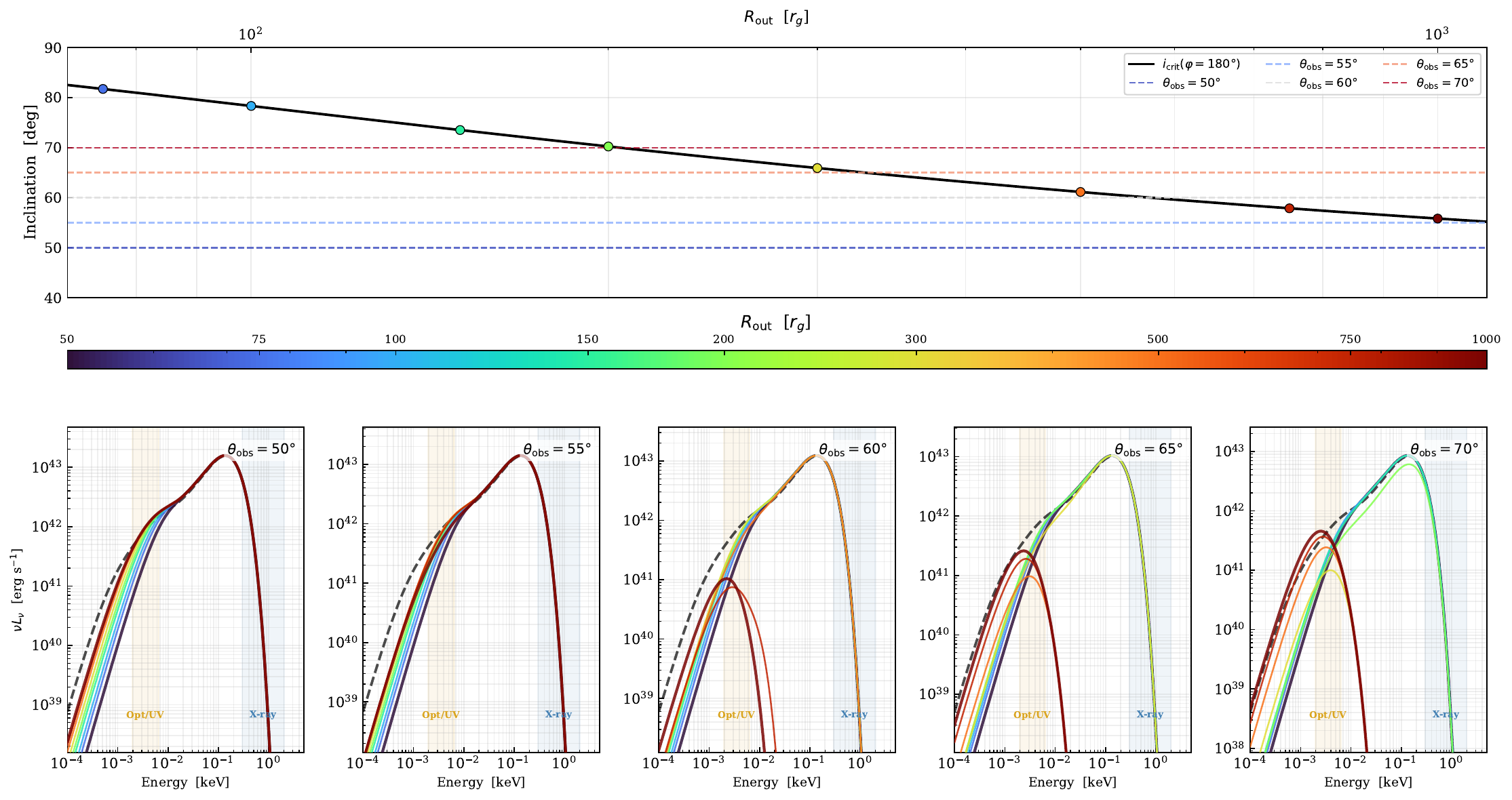}
    \caption{The evolution of the critical inclination (black curve, upper panel) as the outer disk edge grows at fixed warp parameters, and the result on the observed spectra at a range of inclinations (lower panel). The colored points on the top panel show the outer radii used to produce the SED in the lower panels (with matching color), while the horizontal dashed lines show the observer inclinations of the five panels below (furthest left panel corresponds to the lowest observer inclination line). When $i_{\rm crit}>\theta_{\rm obs}$ (black line above horizontal dashed line) the observer can see the inner disk. Increasing the outer disk size reduces the critical inclination, and so fewer observers can see the inner disk. The fixed warp parameters are $r_{\rm warp} = 20\,r_g$ and $\theta_{\rm warp} = 50^\circ$ (with $M_\bullet = 10^7\,M_\odot$, $R_{\rm in} = 2\,r_g$, $T_{\rm in} = 2.5\times10^5$\,K), with the outer radius swept over $R_{\rm out} = 50$--$1000\,r_g$. These curves are made for $\phi_{\rm obs}=180^\circ$, i.e., asymptotically behind the warp, the worst-case scenario for seeing the inner disk. This sequence of SEDs cannot be read purely as time evolution (even though the outer disk does grow with time), as one must also include the evolution of the warp parameters and disk temperature. }
    \label{fig:rout}
\end{figure*}
\begin{figure*}
    \centering
    \includegraphics[width=0.95\linewidth]{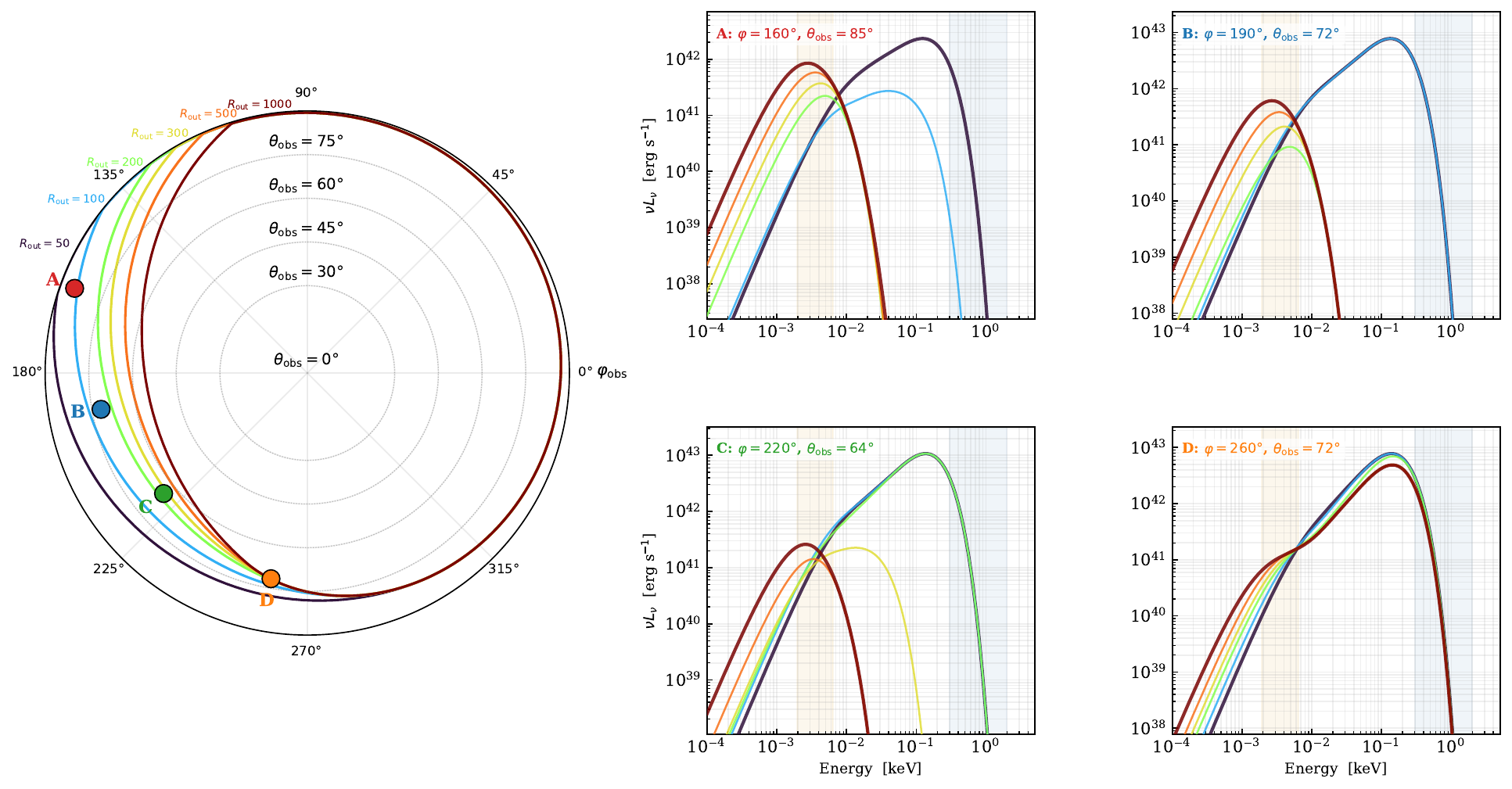}
    \caption{A similar plot to Figure \ref{fig:rout}, except now expressed in observer sky space. As $R_{\rm out}$ grows, the occultation curve shrinks in the observer sky, meaning that more observers can no longer see the inner disk (being outside of the occultation curve means the observer cannot see the inner disk). We show SEDs for three observers for whom the occultation curve sweeps past their location on the observer sky (A, B, C), and one who remains (just) inside of the occultation curve (D), highlighting the sensitivity of the observed SED to observer location close to the occultation curve. This sequence of SEDs cannot be read purely as time evolution (even though the outer disk does grow with time), as one must also include the evolution of the warp parameters and disk temperature. The fixed warp parameters are $r_{\rm warp} = 20\,r_g$ and $\theta_{\rm warp} = 50^\circ$ (with $M_\bullet = 10^7\,M_\odot$, $R_{\rm in} = 2\,r_g$, $T_{\rm in} = 2.5\times10^5$\,K); the occultation curves shown correspond to $R_{\rm out} = 50, 100, 200, 300, 500$ and $1000\,r_g$. }
    \label{fig:rout2}
\end{figure*}

Perhaps the most striking SED change highlighted in Figure \ref{fig:theta_phi} is not the optical/UV colors however, it is the behavior shown in the right hand two (high observer inclination) columns. One  observes that for certain observer azimuths the high energy emission component completely disappears.  The physical origin of this effect is simple enough to understand, and can be seen explicitly in the bottom right panel of Figure \ref{fig:examples} --- when a disk is highly warped, and an observer is highly inclined, it is possible to put the observer behind the warp, completely obscuring the inner disk from view. This angular dependence can be seen clearly in Figure \ref{fig:occ}, where we show both the fraction of the total disk area (upper panel) and the bolometric luminosity ratio (versus a flat disk, lower panel) as a function of $\phi_{\rm obs}$ for different polar observing angles $\theta_{\rm obs}$. It is interesting that occluding a large fraction of the disk area does not necessarily impact the bolometric luminosity to a great degree, as the bolometric luminosity cares primarily about the inner disk (the turbulent heating rate falls as $r^{-3}$). When the inner disk is obscured the bolometric luminosity drops strongly. Generically, warped disks (which are not obscured from view) are brighter than non-warped disks, owing to irradiative heating (which can also be seen in Figures \ref{fig:theta_phi}, \ref{fig:cols}). 

In this section we derive an analytic formula for the critical observer inclination $i_{\rm crit}$ at which the warped outer disk begins to occlude the inner disk from view.  Observers at inclinations $\theta_{\rm obs} > i_{\rm crit}$ (more edge-on) see the inner disk blocked by the warp while observers at $\theta_{\rm obs} < i_{\rm crit}$ see the inner disk unobstructed.

The condition for a disk element at position $\vec{\mathbf{r_0}}$ to be occluded from view is the following
\begin{equation}
    h(t) = (\vec{\mathbf{r_0}} + \mathbf{\hat o} t)\cdot \mathbf{\hat l}(|\vec{\mathbf{r_0}} + \mathbf{\hat o} t|)= 0, \quad t \in (0, \infty),
\end{equation}
where $\mathbf{\hat o}$ points towards the observer and $t$ parametrizes how long along the light rays path to the observer the intersection is (for rays which are not blocked  there will be no $t\in (0, \infty)$ that satisfies this equation).  

The critical ray which is only just occluded by the warped surface is the tangent ray, or 
\begin{equation}
    h'(t) = \mathbf{\hat o} \cdot \mathbf{\hat l} + {\vec{\mathbf{r}}(t)\cdot \mathbf{\hat o} \over |\vec{\mathbf{r}}(t)| } (\vec{\mathbf{r}}(t) \cdot \mathbf{\hat l}')= 0, \quad t\in(0, \infty), 
\end{equation}
where $\mathbf{\hat l}' = {\partial \mathbf{\hat l}}/\partial r$ and we have used the short hand notation $\vec{\mathbf{r}}(t) = \vec{\mathbf{r_0}} + \mathbf{\hat o} t$. Simultaneously solving the twin conditions $h(t_\star, i_{\rm crit}) = h'(t_\star, i_{\rm crit}) = 0$ is painful, and not particularly illuminating (though can be done, leaving a transcendental equation for $t_\star$ which ultimately gives $i_{\rm crit}$). 

It is easier to simply solve $h(t, i_{\rm crit}) = 0$ for every radius in the disk between the element and the observer (which forces $t$), and then simply select the lowest $i_{\rm crit}$ of all solutions. This then has to be the critical ray. 

For simplicity let us assume that the critical inclination at which X-rays (or generally emission from the inner disk) starts to get blocked by the warp  for a given observer is when the observer can no longer see the origin (i.e., when roughly half the inner disk is blocked). This is then $\vec{\mathbf{r_0}}=0$, and $h(t)=0$ becomes 
\begin{multline}
    t(\sin i_{\rm crit} \cos \phi_{\rm obs} \sin \beta(r) \cos \gamma(r) \\ + \sin i_{\rm crit} \sin \phi_{\rm obs} \sin \beta(r) \sin \gamma(r) + \cos i_{\rm crit} \cos \beta(r) ) = 0,
\end{multline}
which, as $t> 0$, implies that 
\begin{equation}\label{eq:icrit}
    i_{\rm crit} = \min_{r \in (R_{\rm in}, R_{\rm out})} \left\{ \tan^{-1}\left[ {\cos \beta(r) \over - \sin\beta(r) \cos (\phi_{\rm obs} - \gamma(r))} \right] \right\},
\end{equation}
which as $i_{\rm crit} \leq 90^\circ$ (by convention; we set any solutions which want $\tan i_{\rm crit}<0$ to be $i_{\rm crit} = 90^\circ$) requires that the phase $\phi_{\rm obs} - \gamma(r) \in (\pi/2, 3\pi/2)$, or in other words some part of the disk warp must be raised up between the observer and the origin. It is interesting to note that as this critical inclination depends only on the angles $(\beta, \gamma)$, it is in reality only a function of the dimensionless ratio $R_{\rm out}/r_{\rm warp}$ and $\theta_{\rm warp}$ (which is what determines these two angles). 


In Figure \ref{fig:icrit} we show that this analytical occultation curve indeed splits X-ray bright from X-ray dim warped disk SEDs. On the left hand panel we show with a black solid curve we show the critical inclination curve on the observers sky, and on the right hand panel we show the resulting difference in observed spectra. An observer within the occultation curve (i.e., observers A, B, C and D) can see the inner disk, and a relatively ``normal'' disk spectrum. On the contrary, observers (E, F, G and H) who are outside of the occultation curve cannot see the inner disk, and see strong X-ray suppression.

This inner disk occultation is of obvious observational importance, particularly if the warped disk profile is time dependent (as the occultation can come and go). It is worth examining on which parameters (and how strongly) this effect depends.

We start with the dependence of the occultation on the outer disk radius $R_{\rm out}$. The sign of the dependence is obvious --- a larger disk has a bigger surface area to work with when it comes to blocking lines of sight, and at fixed everything else is a more effective occluder (lower $i_{\rm crit}$). 

We show this in two (equivalent) ways in Figures \ref{fig:rout} and \ref{fig:rout2}. Figure \ref{fig:rout} shows the evolution of the critical inclination (black curve, upper panel) as the outer disk edge grows at fixed warp parameters, and the result on the observed spectra at a range of inclinations (lower panel). The colored points on the top panel show the outer radii used to produce the SED in the lower panels (with matching color), while the horizontal dashed lines show the observer inclinations of the five panels below (furthest left panel corresponds to the lowest observer inclination line). When $i_{\rm crit}>\theta_{\rm obs}$ (black line above horizontal dashed line) the observer can see the inner disk. Increasing the outer disk size reduces the critical inclination, and so fewer observers can see the inner disk.  These curves are made for $\phi_{\rm obs}=180^\circ$, i.e., asymptotically behind the warp, the worst-case scenario for seeing the inner disk. As $R_{\rm out}$ is increased (at fixed warp), more observers end up with lines of sight blocked by the disk, and fewer observers see bright X-rays. 

In Figure \ref{fig:rout2} we show a similar plot to Figure \ref{fig:rout}, except now expressed in observer sky space. As $R_{\rm out}$ grows, the occultation curve shrinks in the observer sky, meaning that more observers can no longer see the inner disk (being outside of the occultation curve means the observer cannot see the inner disk). We show SEDs for three observers for whom the occultation curve sweeps past their location on the observer sky (A, B, C), and one who remains (just) inside of the occultation curve (D), highlighting the sensitivity of the observed SED to observer location close to the occultation curve. 

In both plots we stress that these sequences of SEDs cannot be read purely as time evolution (even though the outer disk does grow with time), as one must also include the evolution of the warp parameters and disk temperature. These plots are intended to isolate the impact $R_{\rm out}$ has on the disk blocking. 

\begin{figure*}
    \centering
    \includegraphics[width=.95\linewidth]{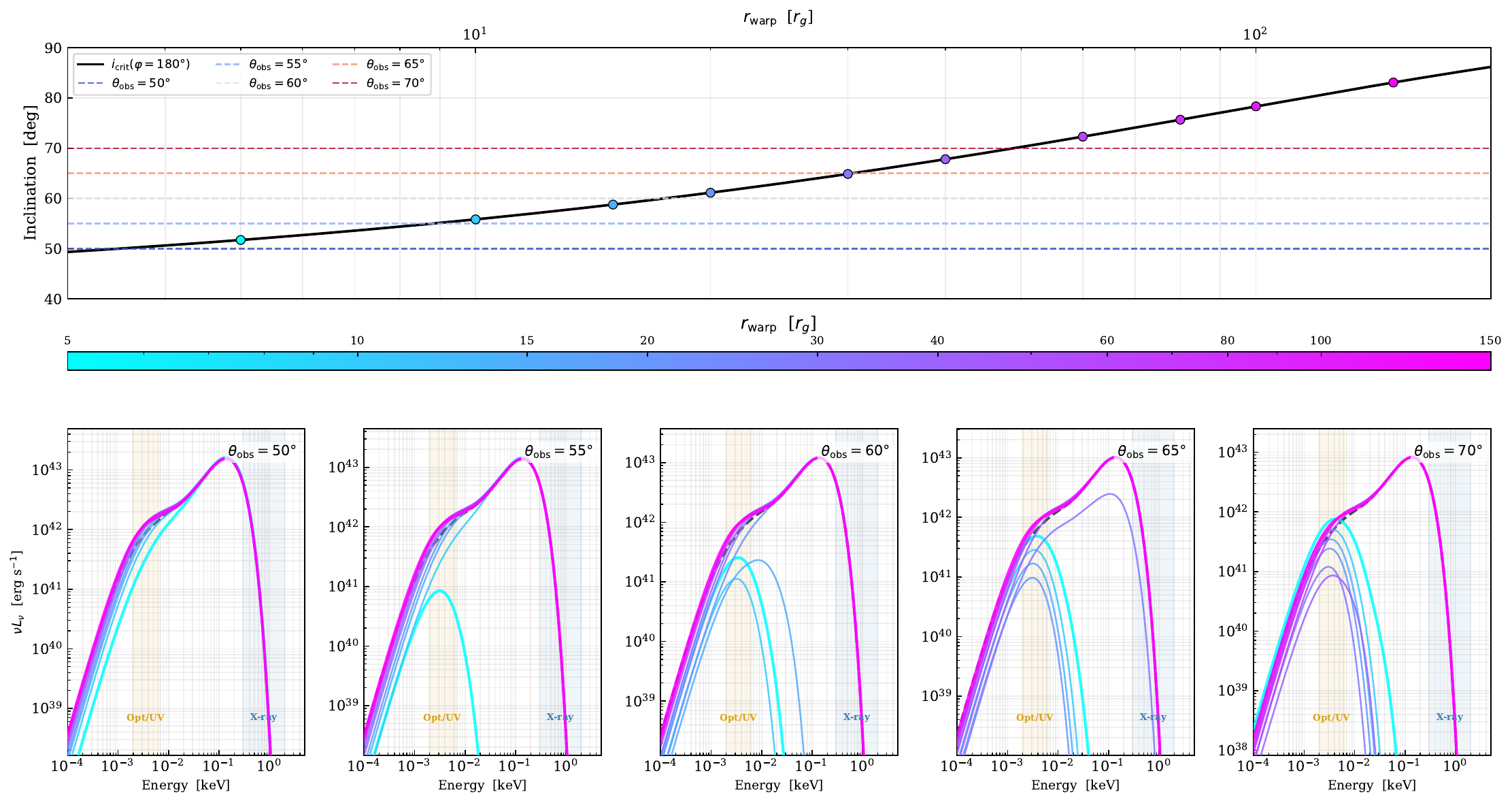}
    \caption{An analogous plot to Figure \ref{fig:rout}, except now for a growing warp radius at fixed $R_{\rm out}$. As $r_{\rm warp}$ grows, the inner disk is progressively revealed to distant observers (equivalently, the critical inclination grows). This is the opposite behavior (in a time dependent sense)  from the $R_{\rm out}$ evolution, as both are expected to grow with time. As before, the colored points on the top panel show the warp radii used to produce the SED in the lower panels (with matching color), while the horizontal dashed lines show the observer inclinations of the five panels below (furthest left panel corresponds to the lowest observer inclination line). When $i_{\rm crit}>\theta_{\rm obs}$ (black line above horizontal dashed line) the observer can see the inner disk. The fixed parameters are $R_{\rm out} = 500\,r_g$ and $\theta_{\rm warp} = 50^\circ$ (with $M_\bullet = 10^7\,M_\odot$, $R_{\rm in} = 2\,r_g$, $T_{\rm in} = 2.5\times10^5$\,K), with the warp radius swept over $r_{\rm warp} = 5$--$150\,r_g$. These curves are made for $\phi_{\rm obs}=180^\circ$, i.e., asymptotically behind the warp, the worst-case scenario for seeing the inner disk. }
    \label{fig:rwarp}
\end{figure*}
\begin{figure*}
    \centering
    \includegraphics[width=0.95\linewidth]{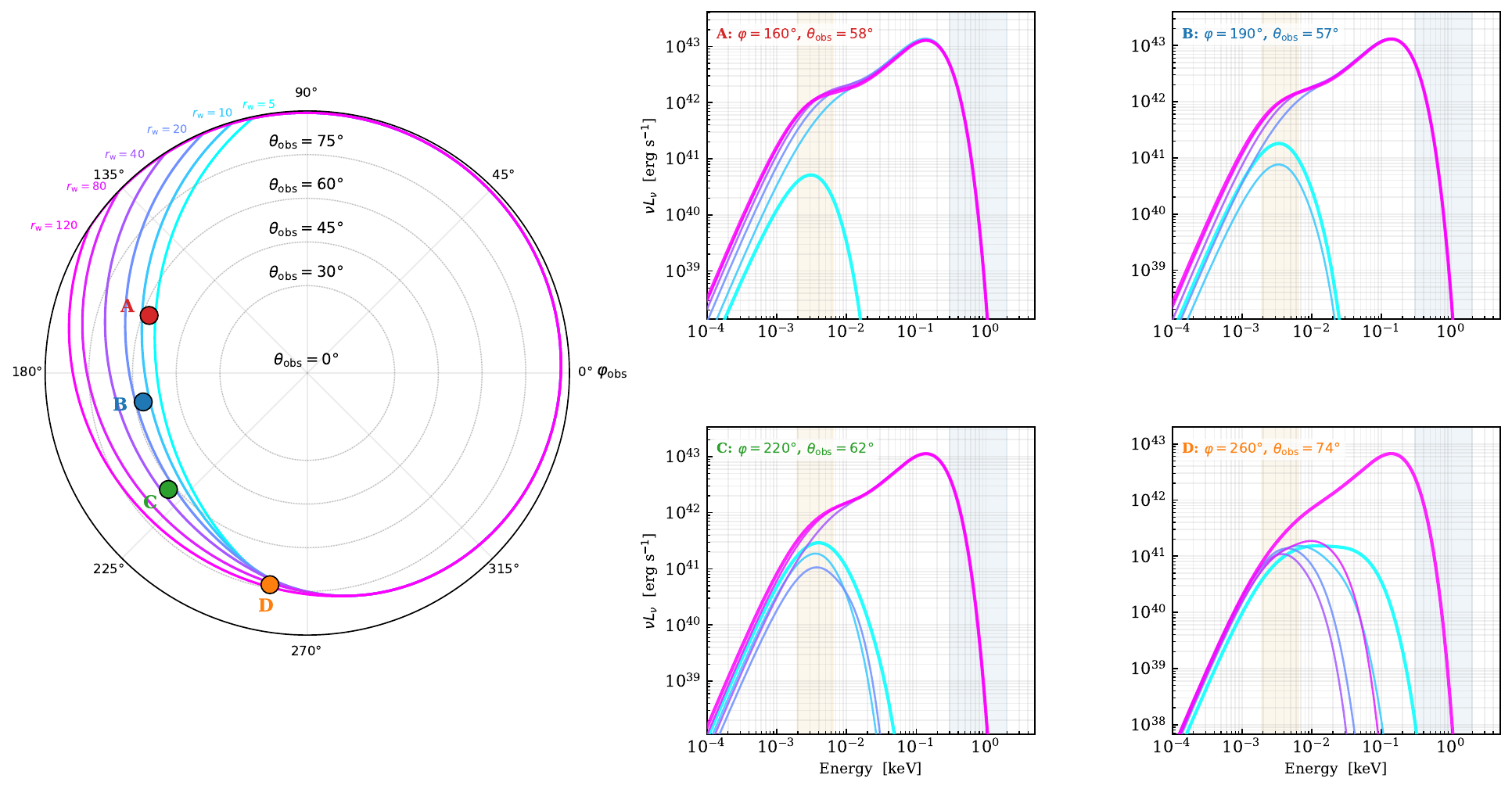}
    \caption{A similar plot to Figure \ref{fig:rwarp}, except now expressed in observer sky space. As $r_{\rm warp}$ grows, the occultation curve grows in the observer sky, meaning that more observers can see the inner disk (being outside of the occultation curve means the observer cannot see the inner disk). This is the opposite dependence as a growing $R_{\rm out}$. We show SEDs for four observers for whom the occultation curve sweeps past their location on the observer sky (A, B, C, D),  This sequence of SEDs cannot be read purely as time evolution (even though the warp radius does grow with time), as one must also include the evolution of the other disk parameters. The fixed parameters are $R_{\rm out} = 500\,r_g$ and $\theta_{\rm warp} = 50^\circ$ (with $M_\bullet = 10^7\,M_\odot$, $R_{\rm in} = 2\,r_g$, $T_{\rm in} = 2.5\times10^5$\,K); the occultation curves shown correspond to $r_{\rm warp} = 5, 10, 20, 40, 80$ and $120\,r_g$. }
    \label{fig:rwarp2}
\end{figure*}
The warp radius $r_{\rm warp}$ has an {\it opposite} effect on disk occultation --- as $r_{\rm warp}$ grows more and more of the disk is flat, and more and more observers can see the innermost (X-ray bright) disk. 

In Figure \ref{fig:rwarp} we produce an analogous plot to Figure \ref{fig:rout}, except now for a growing warp radius at fixed $R_{\rm out}$. As $r_{\rm warp}$ grows, the inner disk is progressively revealed to distant observers (equivalently, the critical inclination grows). This is the opposite behavior (in a time dependent sense)  from the $R_{\rm out}$ evolution, as both are expected to grow with time. As before, the colored points on the top panel show the warp radii used to produce the SED in the lower panels (with matching color), while the horizontal dashed lines show the observer inclinations of the five panels below (furthest left panel corresponds to the lowest observer inclination line). When $i_{\rm crit}>\theta_{\rm obs}$ (black line above horizontal dashed line) the observer can see the inner disk.  These curves are made for $\phi_{\rm obs}=180^\circ$, i.e., asymptotically behind the warp, the worst-case scenario for seeing the inner disk. 

Figure \ref{fig:rwarp2} shows this same behavior in observer sky space. As $r_{\rm warp}$ grows, the occultation curve grows in the observer sky, meaning that more observers can see the inner disk (being outside of the occultation curve means the observer cannot see the inner disk).  We show SEDs for four observers for whom the occultation curve sweeps past their location on the observer sky (A, B, C, D). Again, we stress that this sequence of SEDs cannot be read purely as time evolution (even though the warp radius does grow with time), as one must also include the evolution of the other disk parameters. 

We postpone a discussion of explicit time evolution until a later section, but note here that generically the growth of $r_{\rm warp}$ ``beats'' the growth of $R_{\rm out}$ (the disk generally aligns faster than it spreads), meaning that generically more observers can see the inner disk as time progresses. 

\subsection{Reverberation}
As a warped disk is not flat, a fluctuation in the temperature profile of the disk at some radius and time $T_{\rm turb}(r) \to T_{\rm turb}(r) + \delta T(r, t)$ will be propagated via irradiative heating to all other disk elements that have a line of sight view to the perturbed region. In general, therefore, a perturbation in the inner disk temperature (perhaps via a turbulent temperature fluctuation) which produces a potentially observable X-ray flare will produce a (potentially) observable optical/UV perturbation at a later time. 

In this sub-section we derive a simplified model for the response of the entire disk to a perturbation in the intrinsic disk temperature profile in the innermost (brightest) regions. We make the following simplifying assumptions 
\begin{itemize}
    \item A perturbation of the disk luminosity profile by fractional amount $\delta L(t)/L_0$ occurs due to a fluctuation confined only to the very inner disk 
    \item The light travel time from the inner disk to an outer ring of the flow at spherical radius $r$ is well approximated by $\Delta t = |\vec{\mathbf{r}}-\vec{\mathbf{r}}_{\rm in}|/c\approx r/c$. 
\end{itemize}
Neither of these approximations are particularly restrictive (or particularly poor), and could easily be generalized if data ever required it. 

We will normalise the luminosity perturbation by the quiescent bolometric (turbulent) luminosity of the disk 
\begin{equation}
    L_0 = \int_{R_{\rm in}}^{R_{\rm out}} 4\pi r \, \sigma T_{\rm in}^4 \left(\frac{r}{R_{\rm in}}\right)^{\!-3} \mathrm{d}r
        \approx 4\pi R_{\rm in}^2 \sigma T_{\rm in}^4 \,,
    \label{eq:L0}
\end{equation}
where the factor $4\pi r$ accounts for the annular area $2\pi r\,\mathrm{d}r$ on both disk faces, and we have assumed the disk is large $R_{\rm in}/R_{\rm out}\ll 1$.

Suppose the inner disk undergoes a luminosity fluctuation 
\begin{equation}
    L(t) = L_0 + \delta L \, \times G(t) \,,
    \label{eq:Lpulse}
\end{equation}
where $G(t)$ is some profile of the flare which peaks at unity. Our assumption is that this perturbation is localised to the inner regions (so does not modify the outer turbulent temperature profile), meaning that we shall perturb the turbulent temperature via a functional form 
\begin{equation}
    T_{\rm turb}^4(r) \to T_{\rm turb}^4(r) \times \left(1 + {\delta L \over L_0} G(t) F(r)\right),
\end{equation}
where $F(r)$ is some peaked function of radius that again has maximum of unity (e.g., a Gaussian localized to the inner disk). 

Each outer-disk element $(r, \varphi)$ receives the luminosity pulse after a light-travel delay from the inner disk.
Since $R_{\rm in} \ll r$ for the reprocessing region, we approximate the source as a point at the origin
\begin{equation}
    \tau_{\rm src}(r, \varphi) = \frac{|\vec{\mathbf{r}}(r, \varphi)|}{c} \,.
    \label{eq:tau_src}
\end{equation}
Since each ring is a great circle on a sphere of radius $r$, every point on ring $r$ has $|\vec{\mathbf{r}}| = r$ exactly, regardless of the tilt $\beta$ or twist $\gamma$.  Therefore
\begin{equation}
    \tau_{\rm src}(r) = \frac{r}{c} \,,
\end{equation}
which is azimuthally symmetric as for a flat disk.

Photons reprocessed at element $\vec{\mathbf{r}}(r, \varphi)$ travel to a distant observer in direction $\mathbf{\hat o}$.  Elements closer to the observer along $\mathbf{\hat o}$ arrive earlier.  The relative delay (with respect to photons from the origin) is
\begin{equation}
    \tau_{\rm obs}(r, \varphi) = -\frac{\vec{\mathbf{r}}(r, \varphi) \cdot \mathbf{\hat o}}{c} \,,
    \label{eq:tau_obs}
\end{equation}
where the sign convention is such that $\tau_{\rm obs} > 0$ for elements on the far side of the disk (delayed).

At observer time $t_{\rm obs}$, element $(r, \varphi)$ was emitting at coordinate time
\begin{equation}
    t_{\rm emit}(r, \varphi) = t_{\rm obs} + \frac{\vec{\mathbf{r}}(r, \varphi) \cdot \mathbf{\hat o}}{c} \,,
\end{equation}
and saw the source luminosity at retarded time $t_{\rm ret} = t_{\rm emit} - r/c$, meaning
\begin{equation}
    \frac{L(t_{\rm ret})}{L_0} = 1 + \frac{\delta L}{L_0} \, G\!\left(t_{\rm obs} + \frac{\vec{\mathbf{r}} \cdot \mathbf{\hat o}}{c} - \frac{r}{c}\right) .
\end{equation}
As the irradiation is absorbed and re-emitted at the $\tau = 1$ photosphere of the irradiated face, the perturbation $\delta L$ only modifies the temperature of the face on which it is absorbed.  In steady state ($L = L_0$), both faces are in thermal equilibrium at the same temperature (the irradiation has had time to diffuse through the disk column).  For the perturbation, however, only the face visible to the observer responds on the light-crossing timescale.

\begin{figure*}
    \centering
    \includegraphics[width=0.95\linewidth]{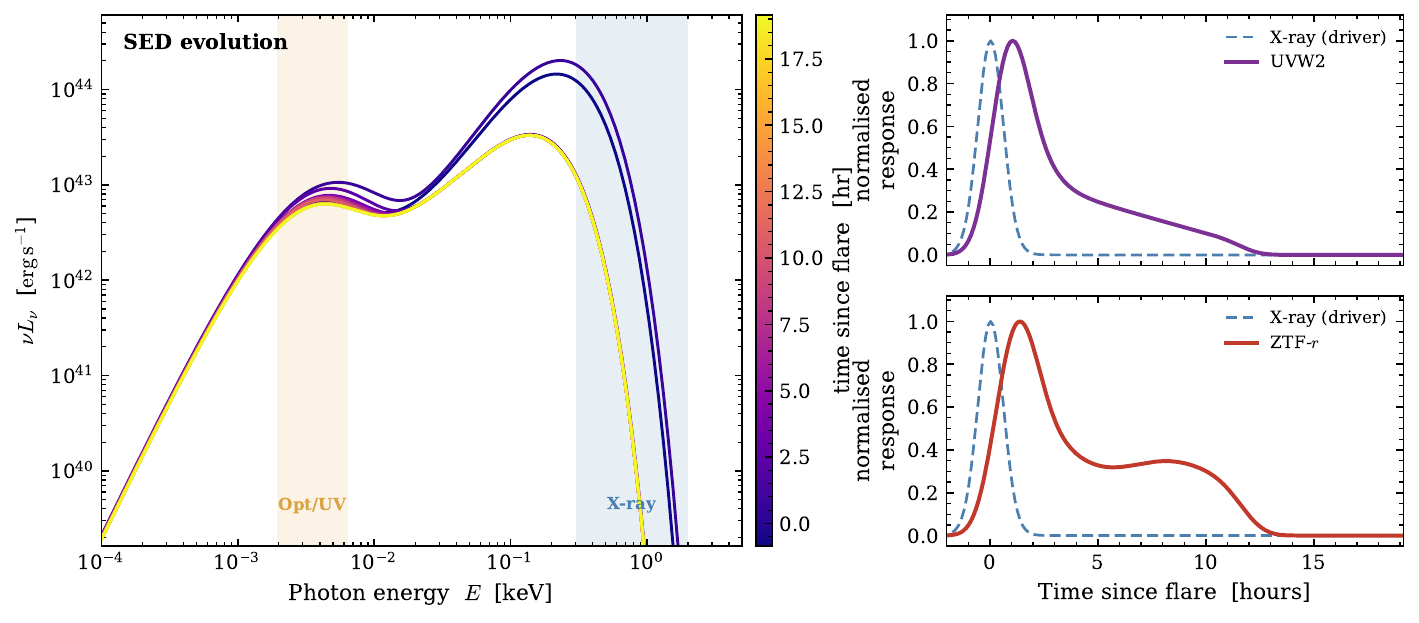}
    \caption{Evolution of the disk spectral energy distribution (left) and light curves in UV and optical bands (right) under the assumption of a localized inner disk temperature perturbation. The (large amplitude, but certainly not unprecedented in the TDE literature) X-ray flare is driven by the initial perturbation to the turbulent disk temperature, while the delayed optical/UV response is reverberation, with the disk responding on the $\sim$ light travel time to irradiative heating. The disk has $r_{\rm warp} = 40\,r_g$, $R_{\rm out} = 500\,r_g$ and $\theta_{\rm warp} = 50^\circ$ (with $M_\bullet = 10^7\,M_\odot$, $R_{\rm in} = 2\,r_g$ and baseline $T_{\rm in} = 2.5\times10^5$\,K), viewed at $\theta_{\rm obs} = 70^\circ$, $\phi_{\rm obs} = 0^\circ$; the perturbation is a localized Gaussian pulse of amplitude $\Delta L = 3\times10^{44}$\,erg\,s$^{-1}$, radial width $\sigma_R = 6\,r_g$ and temporal width $\sigma_{\rm flare} = 50\,r_g/c$. The double peaked $r$-band light curve represents the response of the near (first) and far (second) sides of the disk. The optical/UV response to a factor $\sim 10$ X-ray flare is a factor $\sim 2$ for this set of parameters.   }
    \label{fig:reverbA}
\end{figure*}

We decompose the irradiation kernel by face.  For each receiver element at radius $r$ with disk normal $\mathbf{\hat{l}}(r)$, define
\begin{align}
    \widetilde Q_{\rm irr}^{\,{\rm flare}, +}(r, \varphi) &= \!\!\!\sum_{\substack{{\rm sources~with} \\ \vec{\mathbf{r}}' \cdot \mathbf{\hat{l}}(r) > 0}} \!\!\! K(r, \varphi|r', \varphi') \, I_{\rm flare}(r') \, {\rm d}A' \,, \qquad \\
    \widetilde Q_{\rm irr}^{\,{\rm flare}, -}(r, \varphi) &= \!\!\!\sum_{\substack{{\rm sources~with} \\ \vec{\mathbf{r}}' \cdot \mathbf{\hat{l}}(r) < 0}} \!\!\! K(r, \varphi|r', \varphi') \, I_{\rm flare}(r') \, {\rm d}A' \,,
    \label{eq:Qface}
\end{align}
where 
\begin{equation}
    K(r, \varphi|r'\varphi') = {|\mathbf{\hat d}\cdot \mathbf{\hat l}(r)||\mathbf{\hat d}\cdot \mathbf{\hat l}(r')| \over |\vec{ \mathbf{r}} - \vec{\mathbf{r}}'|^2 } \, (1-{\cal S}(\vec{\mathbf{r}},\vec{\mathbf{r}}')),
\end{equation}
is the standard radiation transfer Kernel (introduced and discussed above), and the vector $\vec{\mathbf{d}} = \vec{ \mathbf{r}} - \vec{\mathbf{r}}'$.
The functions $\widetilde Q_{\rm irr}^{\,{\rm flare}, \pm}$  are the irradiation received on the top ($\mathbf{\hat{l}}$) and bottom ($-\mathbf{\hat{l}}$) faces respectively from the flare.   The $180^\circ$ point symmetry of the warped disk model ($\vec{\mathbf{r}} \to -\vec{\mathbf{r}}$, $\mathbf{\hat{l}} \to \mathbf{\hat{l}}$) ensures $\widetilde Q_{\rm irr}^{\,{\rm flare}, +}(r, \varphi) = \widetilde Q_{\rm irr}^{\,{\rm flare}, -}(r, \varphi + \pi)$, so the global split is exactly $50/50$.  At individual elements, however, the split ranges by a large fraction as the face pointing toward the inner disk receives substantially more irradiation than the face pointing away from it. The intensity function $I_{\rm flare}(r')$ is itself simply a function of the choice of $F(r)$.

\begin{figure*}
    \centering
    \includegraphics[width=0.95\linewidth]{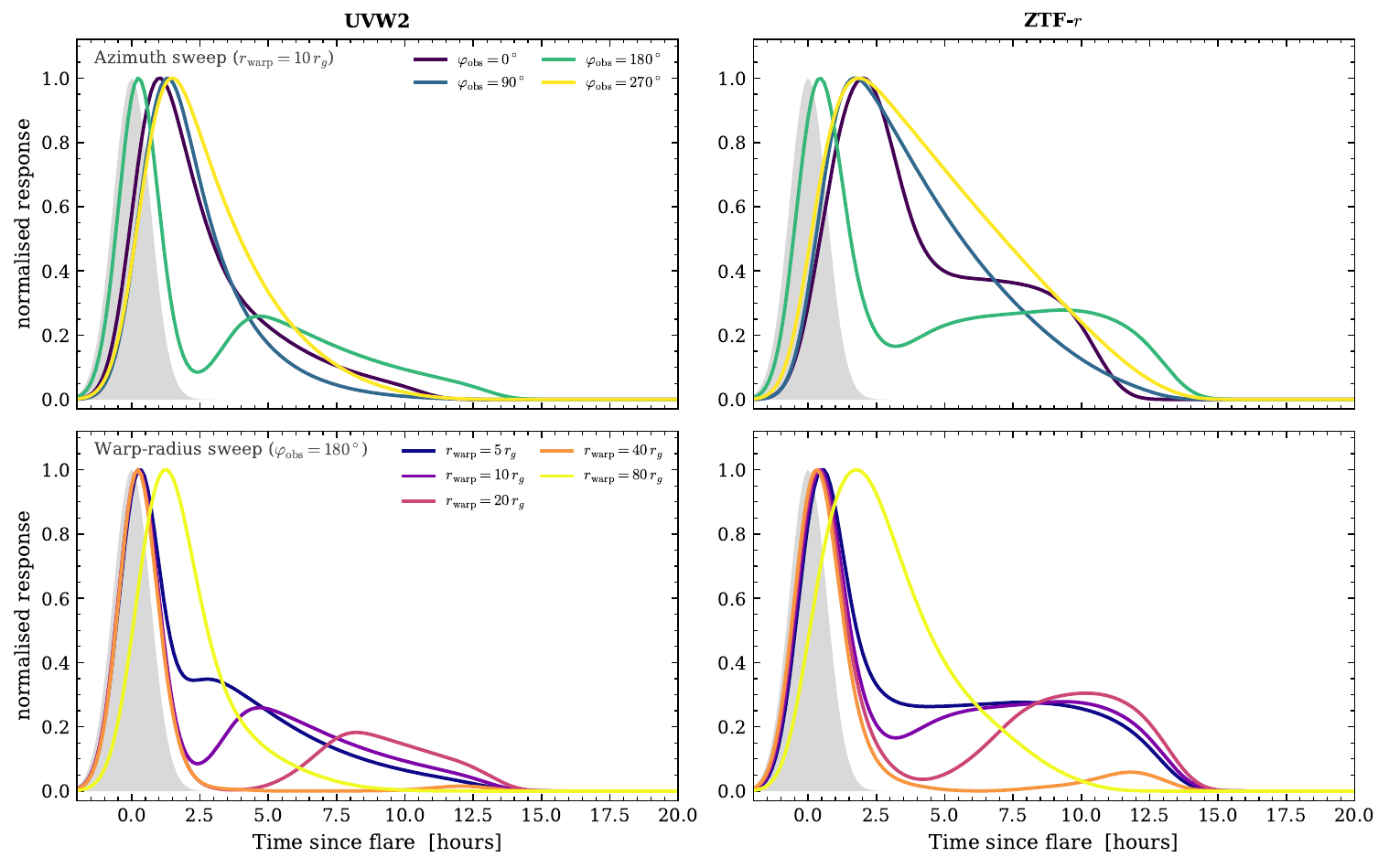}
    \caption{Some reverberation light curves for different observer viewing angles (upper) and different warp radii (lower) for a similar X-ray flare to that discussed above (Fig. \ref{fig:reverbA}) for a UV band (left column) and the ZTF $r$-band (right column). The disk has $R_{\rm out} = 500\,r_g$, $\theta_{\rm warp} = 50^\circ$, $M_\bullet = 10^7\,M_\odot$, $R_{\rm in} = 2\,r_g$ and $T_{\rm in} = 2.5\times10^5$\,K, viewed at $\theta_{\rm obs} = 70^\circ$; the upper row sweeps the observer azimuth $\phi_{\rm obs} = 0, 90, 180, 270^\circ$ (at fixed $r_{\rm warp} = 10\,r_g$) and the lower row sweeps the warp radius $r_{\rm warp} = 5, 10, 20, 40, 80\,r_g$ (at fixed $\phi_{\rm obs} = 180^\circ$). The X-ray perturbation is a Gaussian pulse with $\Delta L = 10^{44}$\,erg\,s$^{-1}$, $\sigma_R = 6\,r_g$ and $\sigma_{\rm flare} = 50\,r_g/c$. The phenomenology of warped-disk reverberation is complex, with single and double peaked flares possible, with different amplitudes and offsets from the primary response, all controlled by multiple parameters. For compact TDE disks, the light crossing time is generically short $\sim {\cal O}$(hours), making observability challenging, but likely not impossible. }
    \label{fig:reverbB}
\end{figure*}

The face visible to the observer is determined by the sign of $\mathbf{\hat{l}} \cdot \hat{\mathbf{o}}$.  The time-dependent temperature at each element is then
\begin{multline}
    T_{\rm tot}^4(r, \varphi, t_{\rm obs}) = T_{\rm in}^4 \Bigg[ \left(\frac{r}{R_{\rm in}}\right)^{\!-3} \left(1 + {\delta L\over L_0} G(t_{\rm ret}) F(r)\right)
    \\ + \widetilde Q_{\rm irr} 
    + \widetilde Q_{\rm irr}^{\,{\rm flare}, \pm} \times \frac{\delta L}{L_0} G\bigl(t_{\rm ret}\bigr) \Bigg] ,
    \label{eq:Ttot_t}
\end{multline}
where
\begin{equation}
    \widetilde Q_{\rm irr}^{\,{\rm flare},\pm}(r, \varphi) =
    \begin{cases}
        \widetilde Q_{\rm irr}^{\,{\rm flare},+}(r, \varphi) \quad\quad\quad \text{if } \mathbf{\hat{l}} \cdot \hat{\mathbf{o}} > 0 \\\quad\quad\quad  \text{(observer sees $+\mathbf{\hat{l}}$ face)} \\ \\[4pt]
        \widetilde Q_{\rm irr}^{\,{\rm flare},-}(r, \varphi) \quad\quad\quad \text{if } \mathbf{\hat{l}} \cdot \hat{\mathbf{o}} < 0 \\ \quad\quad\quad \text{(observer sees $-\mathbf{\hat{l}}$ face)}
    \end{cases}
    \label{eq:Qpm}
\end{equation}
and $t_{\rm ret} = t_{\rm obs} + \vec{\mathbf{r}} \cdot \hat{\mathbf{o}}/c - r/c$ as before.  In steady state ($\delta L = 0$) Eq.~\eqref{eq:Ttot_t} reduces to the equilibrium temperature $T_{\rm in}^4[(r/R_{\rm in})^{-3} + \widetilde Q_{\rm irr}]$, which is face-independent.

The observed isotropic-equivalent spectral luminosity at observer time $t_{\rm obs}$ is
\begin{equation}
    \nu L_\nu(t_{\rm obs}) = \frac{8\pi h\nu^4}{c^2}
    \sum_{i,j} \frac{\mathcal{V}_{ij}\, |\cos\Upsilon_{ij}|\, {\rm d}A_{ij}}
    {f_{\rm col}^4\!\left[\exp\!\left(\dfrac{h\nu}{k_{\rm B} f_{\rm col} T_{\rm tot}^\pm(t_{\rm obs})}\right) - 1\right]} \,,
    \label{eq:nuLnu}
\end{equation}
where as before $\cos\Upsilon_{ij} = \mathbf{\hat l}_{ij} \cdot \mathbf{\hat o}$ is the projected area factor, $\mathcal{V}_{ij} \in \{0, 1\}$ is the observer-visibility mask (accounting for occultation by the warp), $f_{\rm col}$ is the spectral hardening factor, and ${\rm d}A_{ij}$ is the proper area element of the warped surface. Note that when $\cos\Upsilon_{ij} >0$ the upper face $T_{\rm tot}^+$ is taken, while $T_{\rm tot}^-$ is taken when  $\cos\Upsilon_{ij} <0$.  

At each timestep, $T_{\rm tot}(r, \varphi, t_{\rm obs})$ is evaluated from Eq.~\eqref{eq:Ttot_t} and the sum is performed over the full grid.
The visibility mask $\mathcal{V}_{ij}$ and area elements ${\rm d}A_{ij}$ are time-independent (i.e., we assume that the disk geometry is static on the light-crossing timescale; this is an excellent approximation).

For elements on the near side ($\vec{\mathbf{r}} \cdot \mathbf{\hat o} > 0$), the two terms partially cancel, compressing the lag.  For elements behind the warp ($\vec{\mathbf{r}} \cdot \mathbf{\hat o} < 0$), they add, extending the lag. For certain observing angles this can lead to a double peaked reprocessing profile. 


In Figure \ref{fig:reverbA} we show the evolving SED produced by taking Gaussian profiles for both the time pulse
\begin{equation}
    G(t) \propto \exp(-t^2/2\sigma_{\rm flare}^2)
\end{equation}
and the turbulent fluctuation radial profile 
\begin{equation}
    F(r) \propto \exp(-(r-R_{\rm in})^2 / 2\sigma_R^2),
\end{equation}
where we took $\sigma_{\rm flare} = 50 r_g/c$ and $\sigma_R = 6r_g$. This second radial Gaussian ensures that the optical/UV response is entirely driven by reprocessing, not just raising the turbulent temperature profile at all radii.

In Figure \ref{fig:reverbA} we see the evolution of the disk spectral energy distribution (left) and light curves in selected UV and optical bands (right). The (large amplitude, but certainly not unprecedented in the TDE literature) X-ray flare is driven by the initial perturbation to the turbulent disk temperature, while the delayed optical/UV response is reverberation, with the disk responding on the $\sim$ light travel time to irradiative heating.  The double peaked $r$-band light curve represents the response of the near (first) and far (second) sides of the disk. The optical/UV response to a factor $\sim 10$ X-ray flare is a factor $\sim 2$ for this set of parameters. 

The precise reverberation response to an X-ray flare is a function of effectively ever parameter in the warped TDE disk problem, meaning that the phenomenology of warped-disk reverberation is complex, with single and double peaked flares possible, with different amplitudes and offsets from the primary response, all controlled by multiple parameters (see Figure \ref{fig:reverbB} for the impacts of changing $\phi_{\rm obs}$ and $r_{\rm warp}$). For compact TDE disks, the light crossing time is generically short $\sim {\cal O}$(hours), making observability challenging, but likely not impossible.

\newpage
\section{Temporal evolution}
The results of the proceeding sections have examined the behavior of the disk SED for a fixed warp geometry, appropriate for comparison to a given snapshot of a warped TDE disks emission. 

Examining the classical estimate of the characteristic warp radius
\begin{equation}
  \frac{r_{\rm warp}}{r_g}
  \approx (4a\alpha)^{2/3}\left(\frac{R}{H}\right)^{4/3},
\end{equation}
one sees that a TDE disk should slowly sweep through a sequence of quasi-steady warp profiles, as the disk cools and thins on the ``viscous'' time. 

This is because the disk aspect ratio $H/R$ is likely to drop asymptotically with time in a disk with a fixed mass budget like that of a TDE. 

To see this, examine the solutions of the equation for vertical hydrostatic equilibrium,
for a steady 
flow with no vertical velocity.
Again, we are pushing simple flat disk physics into a somewhat uncomfortable regime here, so that we can access the physical scalings of parameters, and fully non-linear warps do not really sit in a simple hydrostatic equilibrium set up \citep[e.g.,][]{Ogilvie1999}. Nonetheless these nonlinearities will perturb the system around some equilibrium value, and that equilibrium value will be reasonably well constrained by a linear hydrostatic equilibrium argument. We are interested only in broad brush time scalings, and not at this stage the detailed physics of the precise solutions of the warped disk equations. 


The conventional approach in flat disk theory  is to take $z \sim (\partial/\partial z)^{-1} \sim H$, leading to 
\begin{equation}
    H = \sqrt{P_g + P_r + P_B\over \rho\Omega^2}, 
\end{equation}
note that 
\begin{equation}
    \rho = {\Sigma \over H}, 
\end{equation}
such that 
\begin{equation}
    H = {1\over \Sigma \Omega^2} (P_g + P_r + P_B).
\end{equation}
The magnetic pressure term $B^2/8\pi$ is the most poorly constrained here, while simultaneously almost certainly being the most important. 


\begin{figure*}
    \centering
    \includegraphics[width=0.95\linewidth]{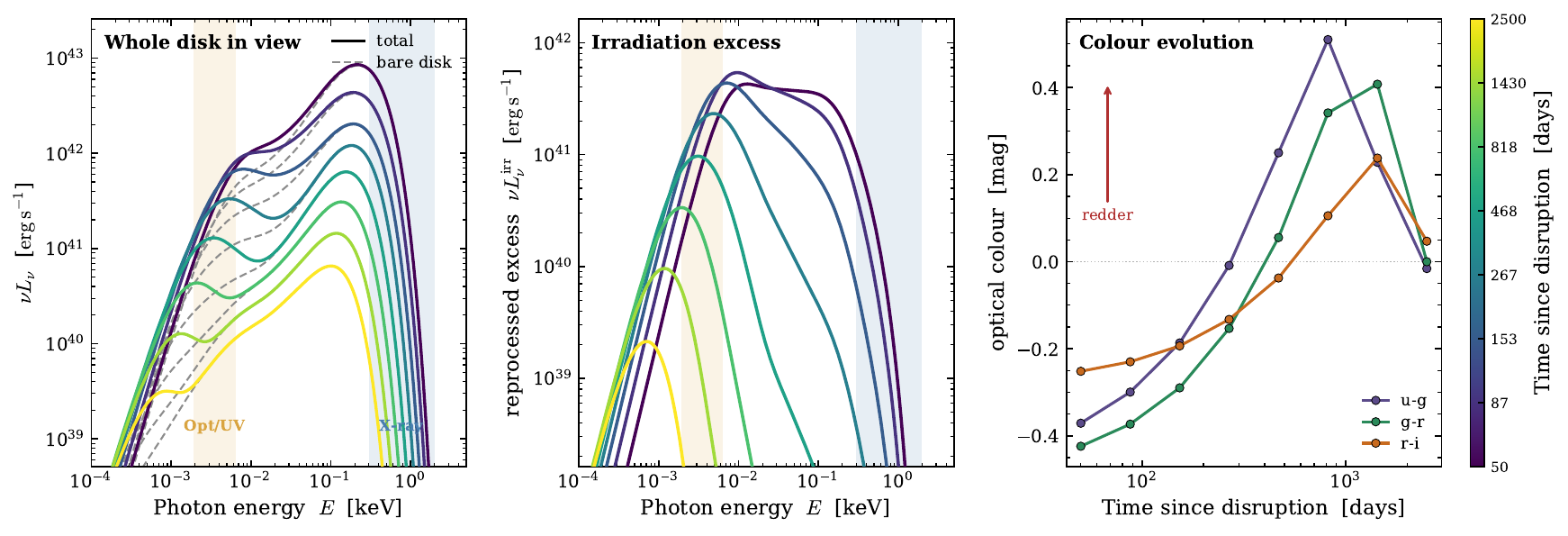}
    \caption{An example of an interesting transient observational signature which can be induced by an evolving warp profile: a transient irradiation ``bump'' in the optical/UV which throws off the naively expected optical colors of the flow. This behavior is not especially generic, and has been chosen to highlight a particular observable. The left panel shows the observing SED for a particular observer, with the dashed gray lines showing the bare (no irradiation) disk evolution, while the solid lines show the  full warped disk evolution. One can see a bump driven by irradiation which sweeps through the optical/UV band with time (denoted by color, see color bar on the right). The explicit excess is shown in the central panel, while the color evolution (in magnitude differences) is shown in the right hand panel. There is a clear transient reddening, for this set of parameters: an evolving disk with $\theta_{\rm warp} = 50^\circ$, $M_\bullet = 10^6\,M_\odot$, $R_{\rm in} = 2\,r_g$, initial warp radius $r_{\rm warp}^0 = 1\,r_g$ and initial outer radius $R_{\rm out}^0 = 250\,r_g$, peak temperature $T_{\rm max} = 5\times10^5$\,K, viscous time $t_0 = 50$\,d and decay index $n = 4/3$, viewed at $\theta_{\rm obs} = 70^\circ$, $\phi_{\rm obs} = 0^\circ$.   }
    \label{fig:bumps}
\end{figure*}

\begin{figure*}
    \centering
    \includegraphics[width=0.8\linewidth]{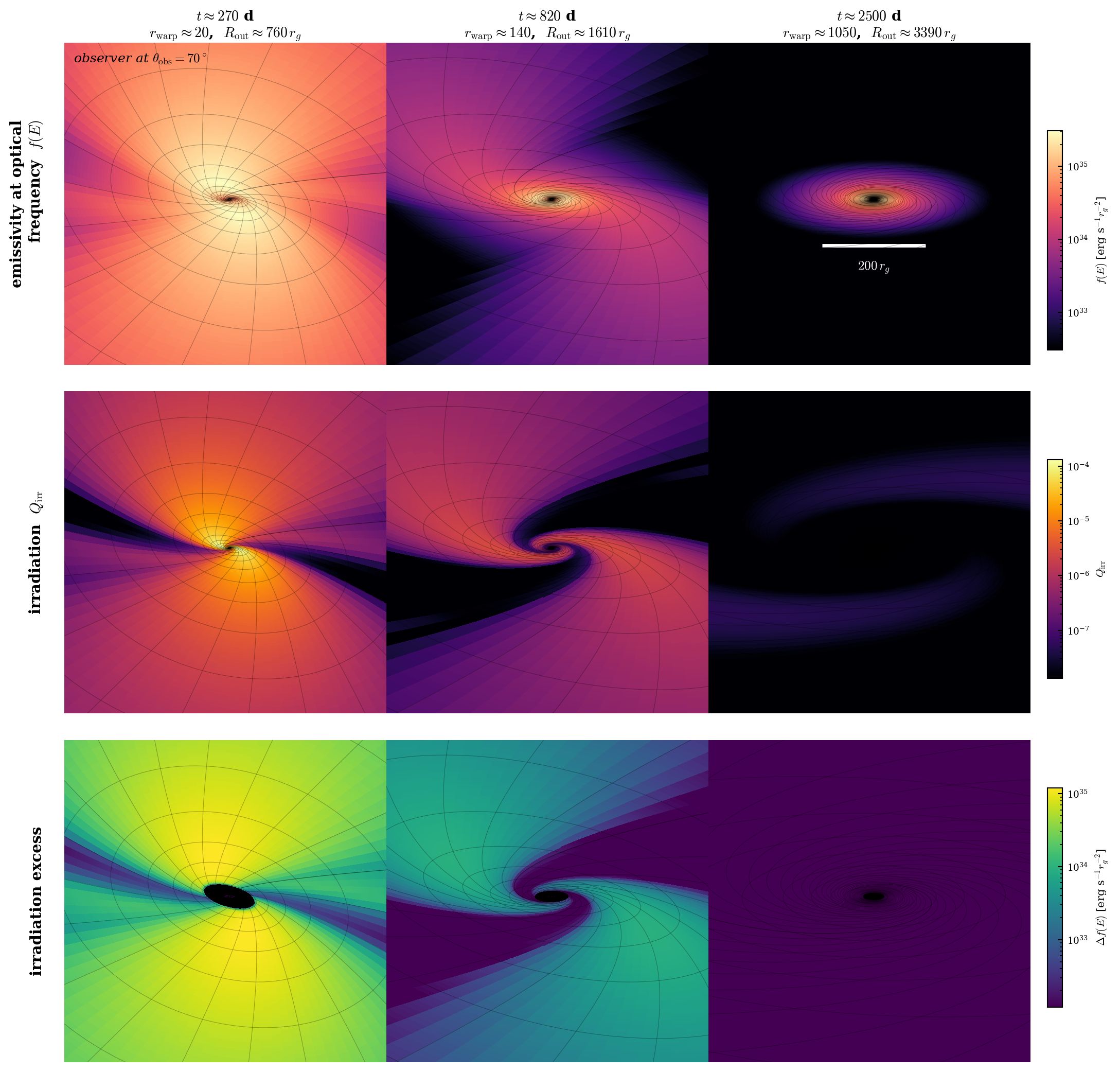}
    \caption{The warp geometry that drives the observational signature seen in Fig. \ref{fig:bumps}. For this disk, at early times (left column) there are two main regions which contribute non-trivially to the emission in the optical band. These are the usual $\sim$ ring from the flat-disk region which has characteristic temperature in the optical (the bright ring in the top panel), and a long irradiatively heated streak up and down the warp (the $\sim$ pink region in the top panel). This second emission region tracks the warped and twisted profile (as can be seen in the lowest, flux excess plot), and is driven entirely by irradiation (middle). As the disk evolves (middle and right columns) the inner disk starts to align, and the irradiative excess disappears and the optical emissivity reverts to the standard  flat disk profile. This figure is made from the observers perspective, and the white bar on the top right panel shows a extent of $\sim 200 r_g$.   }
    \label{fig:bump}
\end{figure*}


It is very simple to show that for the typical observed disk temperatures of TDEs ($T_R \sim 100$ eV), and with the surface density of a TDE disk well constrained by mass conservation $\Sigma \pi R_{\rm out}^2 \lesssim M_\star$, then a TDE disk inevitably has $P_r\gg P_g$ until $H/R \sim 10^{-3}$. 
Given that canonical thin disk models suffer from violent \cite{LightmanEardley74} thermal and viscous instabilities in the $P_r\gg P_g$ limit (in the absence of magnetic support $P_B\gg P_r$, \citealt{Begelman07}), while the inner regions of TDE disks are observed to be stable over many thermal times \citep[see][for analysis of the short timescale variability of X-ray bright TDEs]{Chakraborty26} one infers that magnetic support likely plays a key role in TDE disk evolution (or the $\alpha$ model is badly wrong). 

An exciting possibility arising from warped disk evolution in TDEs is therefore that we may have an observational probe of the evolution of $r_{\rm warp}$, and therefore $H/R$, as the accretion rate drops over many decades in a highly misaligned disk. It seems almost certain that the disk aspect ratio will care about the accretion rate in the disk, i.e., $H/R = f(\dot M)$, and so we parameterize the disk scale height in this manner.

To be precise, in this paper we consider two models of $r_{\rm warp}$ evolution, and show that they make distinct predictions for the evolution of observables.  The exact details of these two models are almost certainly not right but they capture the two qualitative regimes in which the warp radius shrinks or grows relative to the outer radius of the disk.   The first model we take is 
\begin{equation}
    {H\over R} \sim \dot M \sim t^{-n}, 
\end{equation}
which leads to a moderately strongly growing warp radius with time 
\begin{equation}
    r_{\rm warp}(t) \sim (H/R)^{-4/3} \sim t^{4n/3}.
\end{equation}
Here and throughout $n>1$ is the index at which the bolometric luminosity of the disk decays, and depends on ones choice of (radial) stress-tensor closure \citep[e.g.,][]{Cannizzo90}. The second model is a self-similar constant scale height model 
\begin{equation}
    {H\over R} \sim {\rm constant}.
\end{equation}
in which the location of the warp radius is independent of time.   

The evolution of the outer edge of the disk meanwhile follows from simple global angular momentum conservation, and is much more robustly understood. Namely, while the mass drops with time
\begin{equation}
    M_{\rm disk} = M_{\rm disk, 0} - \int \dot M\, {\rm d}t \sim t^{1-n},
\end{equation}
the angular momentum stays approximately constant 
\begin{equation}
    J_{\rm disk} \propto M_{\rm disk}\sqrt{GM_\bullet R_{\rm out}} \sim {\rm constant}
\end{equation}
implying 
\begin{equation}
    R_{\rm out} \sim t^{2n-2}. 
\end{equation}
We see that in models where $H/R$ is constant, the outer radius increases in time while the warp radius does not.   By contrast, when $H/R$ depends linearly on the accretion rate, while both the warp radius and outer radius are expected to grow with time, 
we expect $r_{\rm warp}/R_{\rm out}$ to grow with time (i.e., eventually the entire disk aligns with the spin axis out to its outer edge). This is because 
\begin{equation}
    r_{\rm warp}/R_{\rm out} \sim t^{2-2n/3}, \quad (H/R \sim \dot M),
\end{equation}
which grows with time for $n < 3$, easily satisfied for all canonical models of radial angular momentum transport in disks, which typically have $1 < n < 3/2$.    Indeed, the ratio $r_{\rm warp}/R_{\rm out}$ grows with time provided that $H/R$ depends on accretion rate $\dot M$ to a power $H/R \sim \dot M^\beta$ with index $\beta$ larger than $\beta > (3/2)(1-1/n)$. 

This is an important result, as it implies that the features induced by the warp (irradiation, X-ray obscuration, reverberation, etc.) are expected to be transient. To turn this into a zeroth order model for examining time evolution of warped TDE disks we use the following expressions 
\begin{align}
    T_{\rm in}(t) &= T_{\rm in, 0} \times \left[1 + \left({t\over t_{\rm visc}}\right)^{-n/4}\right], \\
    R_{\rm out}(t) &= R_{\rm out, 0}\times \left[1 + \left({t\over t_{\rm visc}}\right)^{2n-2}\right], \\
    r_{\rm warp}(t) &= r_{\rm warp, 0} \times \left[1 + \left({t\over t_{\rm visc}}\right)^{4n/3}\right] ,
\end{align}
and specify $n = 4/3$ (the canonical self-similar value of this index). One could of course change the index on the $r_{\rm warp}$ evolution to $n'$ (any other choice), and recompute evolutionary tracks under different disk scale height scalings.

The behavior of the disk is self similar for different choices of $t_{\rm visc}$, which really just change the time of the observer (i.e., changing $t_{\rm visc}$ by a factor 10 would just change the length of time one needs to observe the system to see the transient behavior by a factor 10). One can then pick a set of parameters $\{T_{\rm in, 0}, R_{\rm out, 0}, r_{\rm warp, 0}, \theta_{\rm warp}\}$ and compute time dependent spectra and light curves. 

The plots which we show in the next few sections are all computed under the $H/R \sim \dot M$ assumption, while $H/R \sim {\rm constant}$ results are discussed in the text to highlight the qualitatively different evolution that occurs depending on whether $r_{\rm warp}/R_{\rm out}$ increases or decreases with time. 

\subsection{Transient irradiation signals in the optical/UV}
As we discussed above, a strong irradiative flux on the outer disk generally reddens the optical/UV spectra, and can for some regions of parameter space lead to bumps/kinks throughout the observed region (e.g., Figures \ref{fig:theta_phi}, \ref{fig:cols}). 

This is a transient feature of the evolution -- eventually ($r_{\rm warp}/R_{\rm out}\to 1$) these behaviors go away, and one is left with a simple mid-frequency slope $\sim \nu^{4/3}$ (with possible color-correction corrections, which are mild), and relatively constant optical colors. 

This means that for some TDEs one might expect to observe a transient period of ``anomalous'' optical colors, as the warp relaxes. At early times (the precise meaning of ``early'' depends on the viscous time and other parameter values) there is a large wall of disk density which can be irradiated by the hot inner disk, causing a prominent irradiation signature. At later times (again the precise meaning of ``late'' is source dependent) the disk has largely flattened into the equatorial plane, there is no irradiation signature,  and the colors revert to more standard values. 

There is obviously a huge parameter space which can be numerically explored here, and so we just show one such disk evolution scenario which highlights the broad point, while noting that this is not some unique prediction of the model, merely illustrative. 

In Figure \ref{fig:bumps} we show  a transient irradiation ``bump'' in the optical/UV which throws off the naively expected optical colors of the flow.  The left panel shows the observing SED for a particular observer, with the dashed gray lines showing the bare (no irradiation) disk evolution, while the solid lines show the  full warped disk evolution. One can see a bump driven by irradiation which sweeps through the optical/UV band with time (denoted by color, see color bar on the right). The explicit excess is shown in the central panel, while the color evolution (in magnitude differences) is shown in the right hand panel. There is a clear transient reddening, for this set of parameters (which are displayed in the Figure caption).  

If one could spatially resolve such a disk, one would see something like that plotted in Figure \ref{fig:bump}. We show three snapshots from the evolution (from the observers $\theta, \phi$ perspective) showing the innermost $\sim \pm 330r_g$ of the disk, with the top panel showing the emissivity at optical frequencies ($\nu = 7\times10^{14}$ Hz), the middle panel showing the dimensionless irradiation, and the lowest panel showing the difference in the emissivity from a disk with irradiation switched off.  The disk's warp and outer radii are shown in the each columns title. 

The structure and physical location of region producing the optical/UV emission changes dramatically over the course of this disk evolution history. At early times (left hand column) there is a large and extended region which non-trivially contributes to the optical emission. This extended region is entirely driven by inner disk irradiation, and dominates over the usual turbulently heated region. 

As the flow evolves, and both the warp and outer radius change with time, the disk regions producing the optical emission reverts to its flat disk state. The middle column ($t\sim 800$ days) shows emission dominated (at optical frequencies) by a flat-ish classical disk region, but with still non-trivial contribution from a large-scale warped structure. By $t\sim 2500$ days the disk is effectively flat, and one reverts to the standard flat disk $T\sim r^{-3/4}$ profile with a canonical emissivity to radius mapping (i.e., the flux peaks at a radius where $h\nu_{\rm obs} \sim kT(r)$ and falls either side). 

{One can generate transient optical colour changes in a disk which has a static $r_{\rm warp}$ as the outer radius grows (our constant $H/R$ model), but these are driven on much longer timescales. The reason for this is that a static irradiation signature (i.e., a single snapshot from Figure \ref{fig:bumps}) with an evolving inner disk temperature will sweep through different optical bands as the disk regions with large irradiation signatures contribute to different bands. However, the evolution of the inner disk temperature is slow, owing to its $t^{-n/4}$ dependence, meaning that very many viscous times must pass before a distant observer perceives a dramatic change in optical colours. The robust detection of transient colour changes (particularly a redder to blue evolution) favours a time dependent $r_{\rm warp}$, and a growing $r_{\rm warp}/R_{\rm out}$.    }

\newpage
\subsection{Delayed X-ray rises and inner disk revealing}

\begin{figure}
    \centering
    \includegraphics[width=0.99\linewidth]{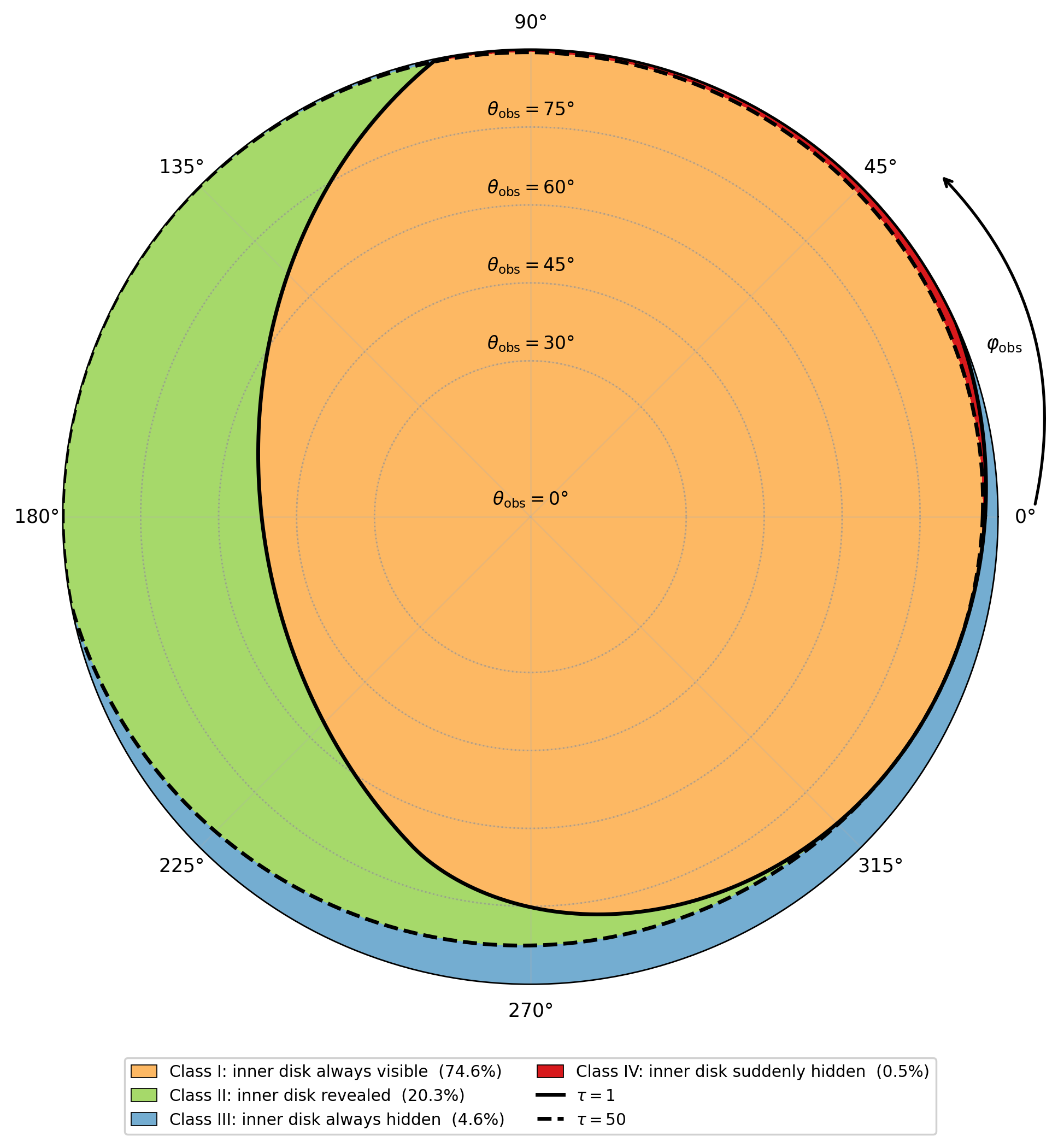}
    \caption{The evolution of the occultation curve in the observers sky for a disk with a growing warp radius and disk size over 50 viscous timescales (see text). The initial occultation curve is shown by the solid black line, while the final occultation curve is shown by the black dashed curve. The occultation curve expands, showing that the growing warp radius ``wins'' over the growing disk outer edge, and progressively more observers can see the inner disk with time. This evolution splits the observer sky into four classes, (i) those who always can see the inner disk, orange, $\sim 75\%$, (ii) those who have the inner disk revealed to them, green, $\sim 20\%$, (iii) those who can never see the inner disk, blue, $\sim 5\%$, and (iv) those who have the inner disk suddenly hidden, red, $<1\%$. All percentages are the fractions of the entire sky, i.e.,  the relevant fractions for isotropic observers. The precise values depend on the warp profile chosen, but it is generic that more observers see the inner disk with time.  }
    \label{fig:icrit_xray}
\end{figure}

\begin{figure*}
    \centering
    \includegraphics[width=0.9\linewidth]{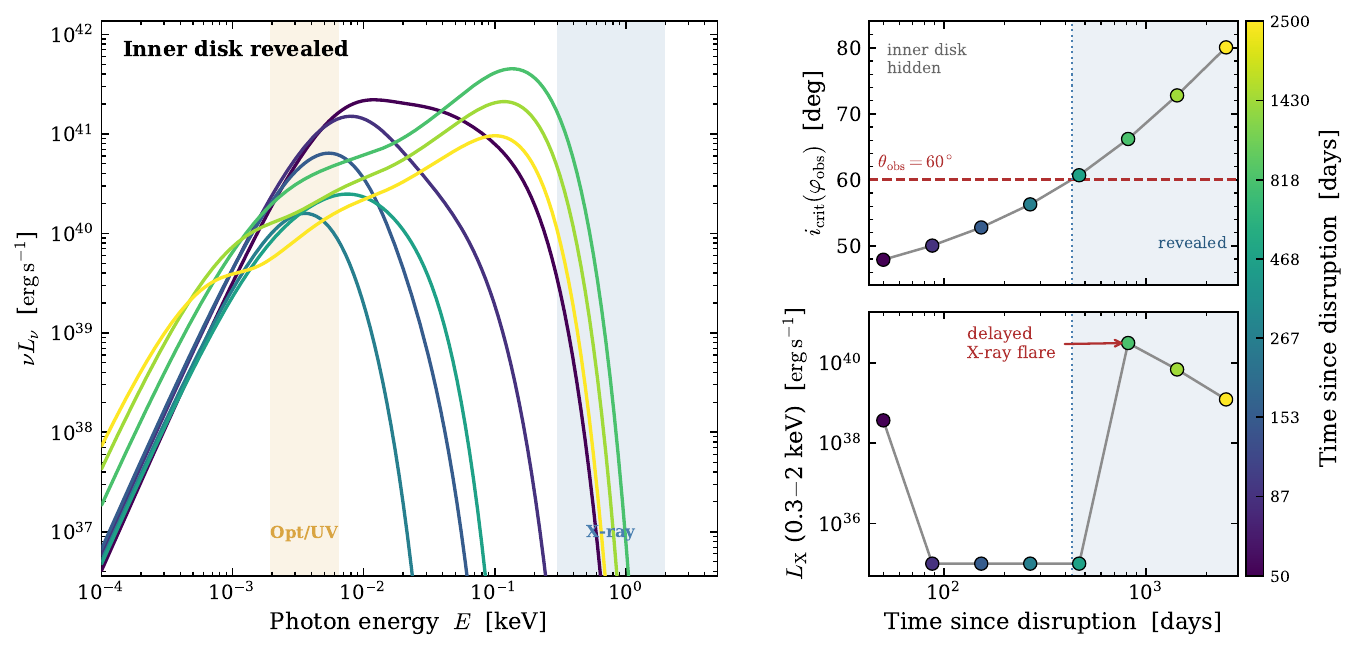}
    \caption{A second observational signature of an evolving warped disk profile, the revealing of the inner disk to a distant observer as the warp radius grows and the disk relaxes to the equatorial plane. The left hand panel shows evolving SED of the disk as observed at $\theta_{\rm obs} = 60^\circ$, $\phi_{\rm obs} = 196^\circ$, for an evolving disk with $\theta_{\rm warp} = 50^\circ$, $M_\bullet = 10^6\,M_\odot$, $R_{\rm in} = 2\,r_g$, initial warp radius $r_{\rm warp}^0 = 1\,r_g$ and initial outer radius $R_{\rm out}^0 = 250\,r_g$, peak temperature $T_{\rm max} = 5\times10^5$\,K, viscous time $t_0 = 50$\,d and decay index $n = 4/3$. This highlights how as the warp radius grows (and the critical inclination grows beyond the observer inclination, upper right panel) the inner disk is revealed, and an X-ray flare occurs. This results in a delayed X-ray flare (lower right panel), as is routinely observed from TDEs.  }
    \label{fig:irr_time}
\end{figure*}

\begin{figure}
    \centering
    \includegraphics[width=0.9\linewidth]{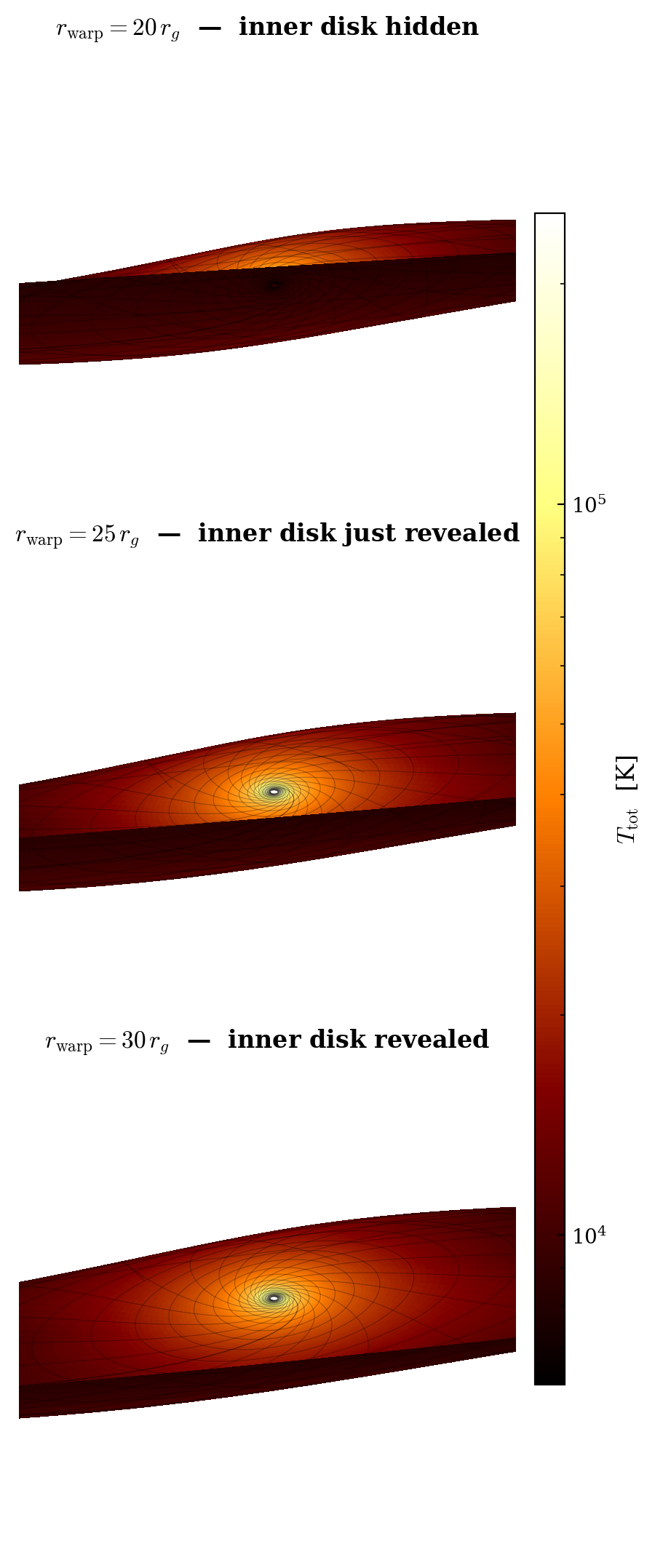}
    \caption{The warp geometry that drives the observational signature seen in Fig. \ref{fig:irr_time}. For this disk, the inner disk is originally hidden from view by the warp itself (upper panel), but as the  warp grows through a critical radius $\sim 25r_g$ (middle panel), the inner disk is revealed. The inner disk then gets progressively easier to observe as the warp radius continues to grow. }
    \label{fig:viewX}
\end{figure}


There are two components which go into determining how much X-ray emission a given observer sees from a TDE accretion flow as a function of time. The first is how much the disk cools with time (which exponentially decreases the observed Wien-tail X-ray luminosity), and whether or not the inner disk is obscured from view from the outer disk. The first (disk cooling) is a well understood property of all disks \citep[e.g., in the TDE literature][]{MumBalb20a, Guolo25b}, and so in this work we focus on the second. 

We begin with the evolution of the occultation curve (defined in equation \ref{eq:icrit}), which splits the observers sky into regions which can and cannot see the inner disk. In Figure \ref{fig:icrit_xray} we display the evolution of the occultation curve in the observers sky for a disk with a growing warp radius and disk size over 50 viscous timescales. We take $r_{\rm warp, 0} = 1r_g$ and $R_{\rm out, 0}=100r_g$. The initial occultation curve is shown by the solid black line, while the final occultation curve is shown by the black dashed curve. The occultation curve expands, showing that the growing warp radius ``wins'' over the growing disk outer edge, and progressively more observers can see the inner disk with time. This evolution splits the observer sky into four classes: (i) those who always can see the inner disk, which we denote in orange, and which contains about $\sim 75\%$ of the sky by area; (ii) those who have the inner disk revealed to them, denoted in green, comprising $\sim 20\%$ of the sky; (iii) those who can never see the inner disk, denoted blue, comprising $\sim 5\%$ of the sky; and (iv) those who have the inner disk suddenly hidden, denoted in red, comprising  $<1\%$ of the sky. All percentages are the fractions of the entire sky, i.e.,  the relevant fractions for isotropic observers. The precise values depend on the warp profile chosen, but it is generic that more observers see the inner disk with time.  

This has a clear observational implication, as we show in Figure \ref{fig:irr_time}. An observer located in the ``green'' (class ii) part of the sky, will generally start the evolution of the flow behind the warp, and will then at later times have the inner disk revealed. The left hand panel shows evolving SED of the disk (the warp and observer parameters are listed in the caption), which highlights how as the warp radius grows (and the critical inclination grows beyond the observer inclination, upper right panel) the inner disk is revealed, and an X-ray flare occurs. This results in a delayed X-ray flare (lower right panel), as is routinely observed from TDEs \citep[e.g.,][]{Gezari17}. 

The specific warp geometric effect which drives this observational signature is shown in Fig. \ref{fig:viewX}. For this disk, the inner disk is originally hidden from view by the warp itself (upper panel), but as the  warp grows through a critical radius $\sim 25r_g$ (middle panel), the inner disk is revealed. The inner disk then gets progressively easier to observe as the warp radius continues to grow. 

{If the warp radius does not grow with time (i.e., the self similar $H/R=$ constant model), then no inner disks are revealed and the only dramatic warp-induced transients are  systems where the inner disk is suddenly hidden from view (see for example Fig. \ref{fig:rout2}). As far as the authors are aware, such an observational transition has not been observed in a TDE system, while delayed X-ray flares (i.e., the inner disk is revealed) are relatively common (as will be discussed more in section 6). This is perhaps the best evidence, we believe, for an evolution (specifically an increase) of the ratio $r_{\rm warp}/R_{\rm out}$ in the TDE context.   }

We wish to stress that while we have shown this X-ray reveal for one particular observer and one particular disk in the evolving warp case, this behavior is actually surprisingly common for isotropically sourced TDEs, as we now discuss. 

\subsection{How many observers are blocked?}
We can make this inner disk revealing quantitative on the population level, under the following assumptions. For an isotropic population of TDEs:
\begin{itemize}
    \item The spin axis $\mathbf{\hat s}$ is random on the sphere,
    \item The stellar orbit $\mathbf{\hat l_\infty}$ is random and independent of $\mathbf{\hat s}$,
    \item The observer $\mathbf{\hat o}$ is fixed, and therefore equivalently also random relative to the source frame.
\end{itemize}
We note that the second assumption, that $\mathbf{\hat l_\infty}$ is random and independent of $\mathbf{\hat s}$, is not true in the details. Even if stars are distributed isotropically in the galactic center (perhaps a good assumption within the sphere of influence where stars involved in TDEs originate), the fact that tidal forces in general relativity depend on the inclination between the stellar orbit and the black hole's spin axis \citep[e.g.,][for applications to TDEs]{Kesden12, Mummery24, Xin26} means that at high black hole masses (close to the \citealt{Hills75} mass where tidal forces are relativistic), there will be a coupling between $\mathbf{\hat l_\infty}$ and $\mathbf{\hat s}$, breaking the independence of the two. We shall neglect this complication (though it could be computed exactly). 

As before we are always free to define two of our axes, namely $\mathbf{\hat s} = \mathbf{\hat z}$, and  $\mathbf{\hat x}$ so that $\mathbf{\hat l_\infty}$ lies in the $xz$ plane. This means that in this coordinate system, all three angles are independently distributed:
\begin{align}
    P(\theta_{\rm warp}) &= \tfrac{1}{2}\sin\theta_{\rm warp} \,, & \theta_{\rm warp} &\in [0, \pi] \,, \label{eq:Ptmax} \\
    P(\theta_{\rm obs}) &= \tfrac{1}{2}\sin\theta_{\rm obs} \,, & \theta_{\rm obs} &\in [0, \pi] \,, \\
    P(\phi_{\rm obs}) &= \tfrac{1}{2\pi} \,, & \phi_{\rm obs} &\in [0, 2\pi) \,.
\end{align}
The warp amplitude $\theta_{\rm warp}$ is drawn from the isotropic distribution because the stellar orbit is random relative to the spin, implying $\cos \theta_{\rm warp} = \mathbf{\hat l_\infty} \cdot \mathbf{\hat s} \sim {\cal U}(0, 1)$.

\begin{figure}
    \centering
    \includegraphics[width=0.95\linewidth]{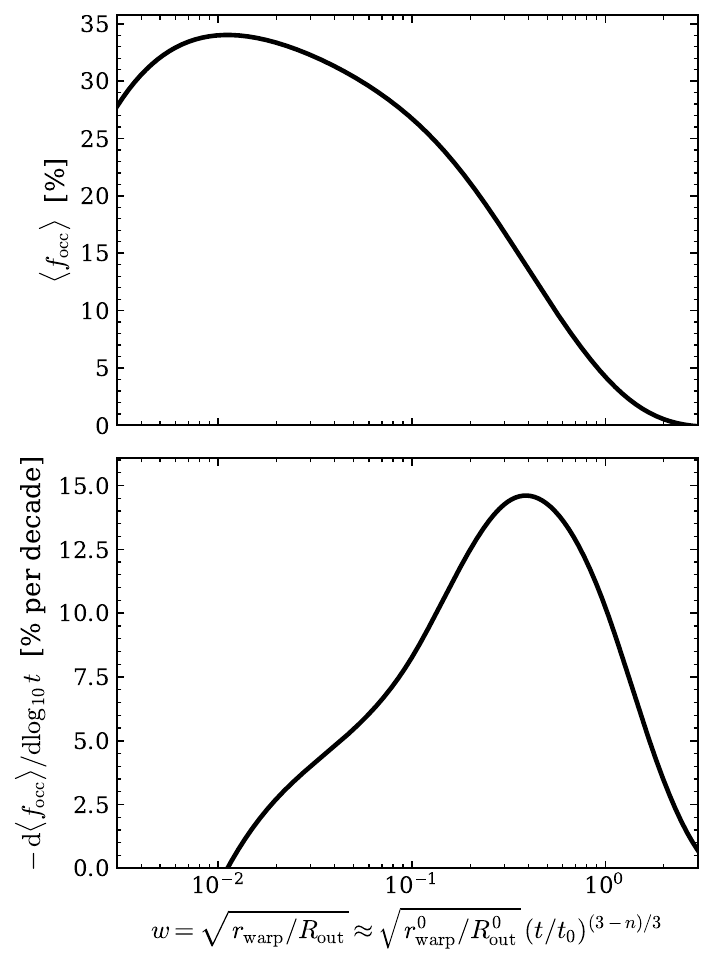}
    \caption{The self-similar observer- and warp-angle averaged inner disk obscuration fraction (upper), and the logarithmic inner-disk reveal rate (lower) as a function of the self-similar variable $w$. If the warp radius is a small fraction of the disk size $\lesssim 1\%$ then a large fraction of random observers for random disks will not be able to see the inner disk. The logarithmic reveal rate peaks at $\sim 15\%$ per decade of time, at $r_{\rm warp}\approx R_{\rm out}/4$.  }
    \label{fig:focc_self}
\end{figure}

For a single source with warp profile $\{\theta_{\rm warp}, r_{\rm warp}, R_{\rm out}\}$, the observer is occluded when
\begin{equation}
    \theta_{\rm obs} > i_{\rm crit}(\phi_{\rm obs}, \theta_{\rm warp}, r_{\rm warp}, R_{\rm out}) \,.
\end{equation}
The disk possesses a $180^\circ$ point symmetry, an observer at $(\theta_{\rm obs}, \phi_{\rm obs})$ sees the same occultation as one at $(\pi - \theta_{\rm obs}, \phi_{\rm obs} + \pi)$.  This means the occultation pattern in the lower hemisphere ($\theta_{\rm obs} > \pi/2$) is a copy of the upper hemisphere shifted by $\pi$ in azimuth.  We may therefore restrict to $\theta_{\rm obs} \in [0, \pi/2]$ without loss of generality, provided we normalize over the upper hemisphere only.

The fraction of (isotropic) observers that are occluded, for a fixed warp configuration (i.e., as a function of ${\rm warp} = \{\theta_{\rm warp}, r_{\rm warp}, R_{\rm out}\}$), is 
\begin{equation}
    f_{\rm occ}({\rm warp}) = \frac{1}{2\pi} \int_0^{2\pi} \int_{i_{\rm crit}}^{\pi/2} \sin\theta_{\rm obs} \, \mathrm{d}\theta_{\rm obs} \, \mathrm{d}\phi_{\rm obs} \,.
    \label{eq:focc_single}
\end{equation}
The $1/(2\pi)$ normalizes over the upper hemisphere.  Performing the $\theta_{\rm obs}$ integral
\begin{equation}
    f_{\rm occ}({\rm warp}) = \bigl\langle \cos\, i_{\rm crit}\bigr\rangle_{\phi_{\rm obs}}
    \equiv \frac{1}{2\pi} \int_0^{2\pi} \cos\bigl(i_{\rm crit}\bigr) \, \mathrm{d}\phi_{\rm obs} \,.
    \label{eq:focc_cos}
\end{equation}
This is the azimuthal average of $\cos i_{\rm crit}$ --- directly readable from the observer sky $i_{\rm crit}$ plots as the fractional area outside of the contour (e.g., Figures \ref{fig:icrit}, \ref{fig:rout2}, \ref{fig:rwarp2}, \ref{fig:icrit_xray}).  It ranges from $0$ (no occultation; $i_{\rm crit} = \pi/2$ everywhere, a flat disk) to $1$ (full occultation; $i_{\rm crit} = 0$ everywhere, although this second limit never actually occurs).

For a population of TDEs with isotropically distributed stellar orbits, the warp amplitude $\theta_{\rm warp}$ is drawn from Eq.~\eqref{eq:Ptmax}.  The population-averaged occultation fraction is
\begin{align}
    \bigl\langle f_{\rm occ} \bigr\rangle &= \int_0^{\pi} \frac{\sin\theta_{\rm warp}}{2} \, f_{\rm occ}(\theta_{\rm warp}) \, \mathrm{d}\theta_{\rm warp} \\
    &= \int_0^{\pi} \frac{\sin\theta_{\rm warp}}{2} \, \bigl\langle \cos\, i_{\rm crit} \bigr\rangle_{\phi_{\rm obs}} \, \mathrm{d}\theta_{\rm warp} \,.
    \label{eq:focc_pop}
\end{align}
Written in full as a triple integral:
\begin{multline}
    \bigl\langle f_{\rm occ} \bigr\rangle = \frac{1}{4\pi} \int_0^{\pi} {\sin\theta_{\rm warp}} \\ \int_0^{2\pi} \int_{i_{\rm crit}}^{\pi/2} \sin\theta_{\rm obs} \, \mathrm{d}\theta_{\rm obs} \, \mathrm{d}\phi_{\rm obs} \, \mathrm{d}\theta_{\rm warp}\,.
    \label{eq:focc_triple}
\end{multline}
The three integrals correspond to the three independent angles of the three-vector problem: $\theta_{\rm warp}$ (warp amplitude, from the stellar orbit), $\phi_{\rm obs}$ (observer azimuth around the spin), and $\theta_{\rm obs}$ (observer inclination from the spin).

Then one needs to specify the warp and outer radii. Interestingly, by inspection of equation \ref{eq:icrit} one can see that the critical inclination is only a function of the ratio $r_{\rm warp}/R_{\rm out}$ (through $\beta(r)$ and $\gamma(r)$), and the observer angles, at a given instance of time.  As we then integrate over both observing angles in forming the average occultation fraction $\bigl\langle f_{\rm occ} \bigr\rangle$, then this observer averaged quantity is only a function of  $w(t) = \sqrt{r_{\rm warp}(t)/R_{\rm out}(t)}$. 

\begin{figure*}
    \centering
    \includegraphics[width=0.69\linewidth]{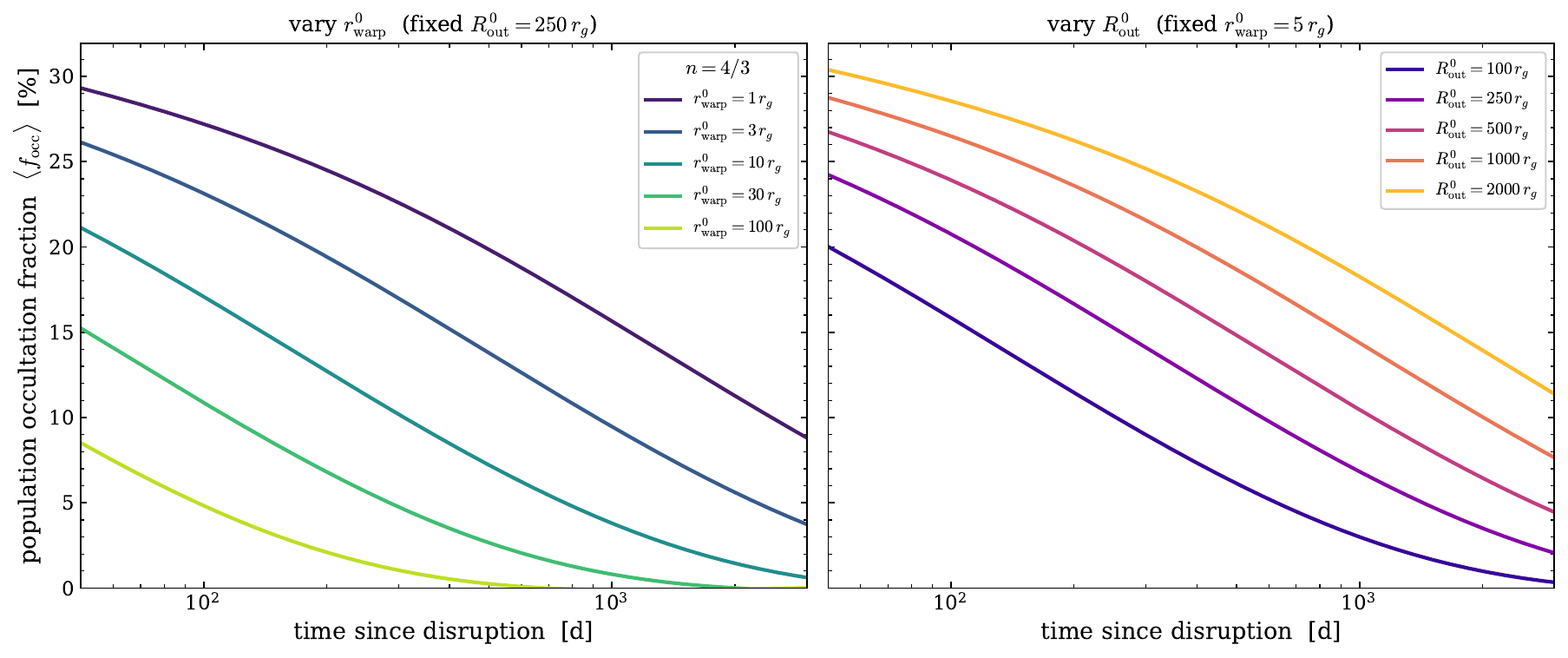}
    \includegraphics[width=0.69\linewidth]{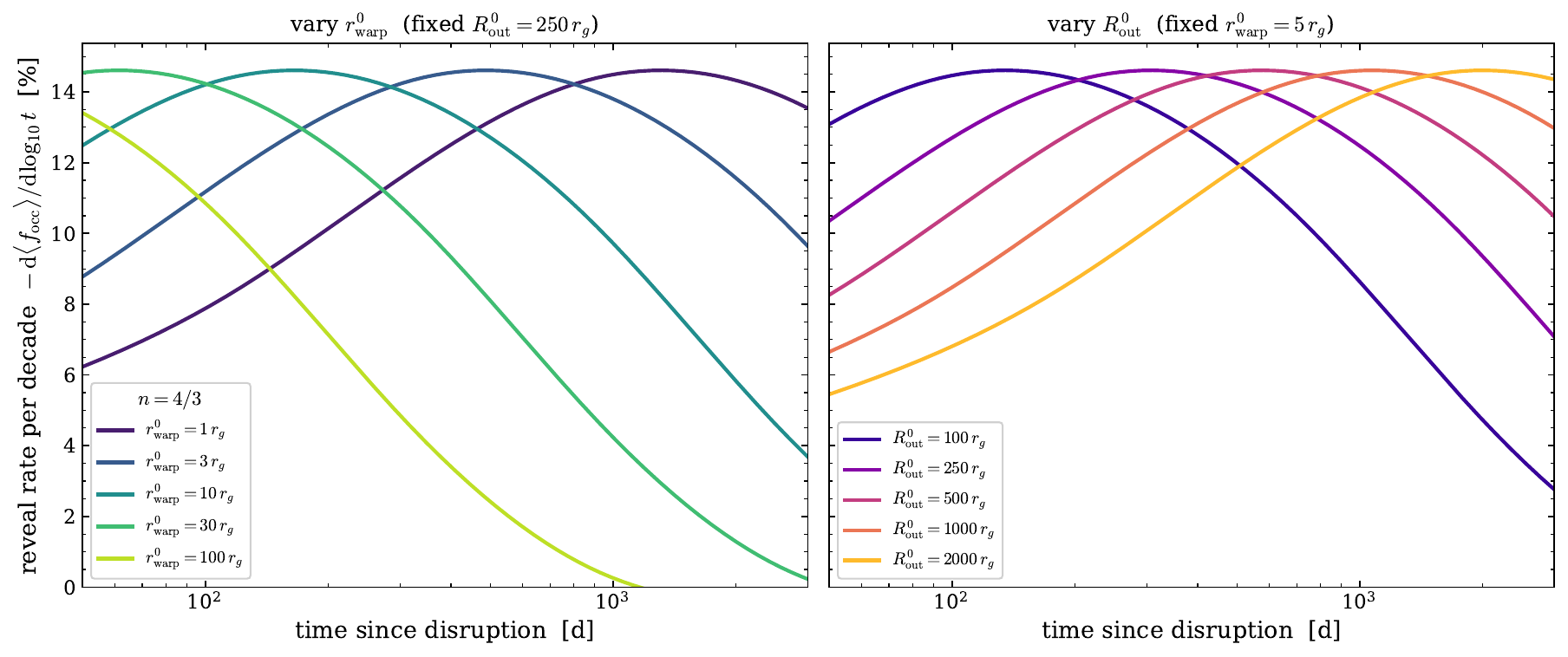}
    \caption{The absolute value of the average disk obscuration fraction (upper) 
    and the logarithmic reveal rate (lower) for different assumptions about the initial warp and outer disk radii. The viscous time was chosen to be $50$ days, but the time dependence is self-similar and so this merely scales the horizontal axis of each plot (and the vertical axis of the linear reveal plot). If canonical estimates of both the disk formation scale and warp radius (discussed in this paper) are broadly accurate, then a sizable fraction of TDEs may show delayed X-ray flares via this warped disk mechanism. 
  }
    \label{fig:focc_t}
\end{figure*}

We plot the function $\bigl\langle f_{\rm occ} \bigr\rangle(w)$ in Figure \ref{fig:focc_self}, showing that the average fraction of observers who cannot see the inner disk edge in an isotropic-TDE universe peaks at $\sim 1/3$rd, which occurs when the warp radius is $\sim 0.01\% - 1\%$ of the outer disk edge. While the lower end of this range is unlikely to be satisfied, the typical circularisation radius (a proxy for the disk formation radius in a TDE) is 
\begin{equation}
    {R_{\rm circ} \over r_g} \approx {94 } \left(\frac{M_\bullet}{10^6M_{\odot}}\right)^{-2/3} \left(\frac{R_\star}{R_\odot}\right) \left(\frac{M_\star}{M_\odot}\right)^{-1/3} ,
\end{equation}
meaning that an initially compact warp radius $\sim {\cal O}(r_g)$ would lead to a substantial fraction of observers obscured from the inner disk in a typical TDE. A back of the envelope estimate of the warp radius puts it squarely in the $\sim 1-10 r_g$ scale for $H/R \sim \alpha \sim 0.1$, so this may actually be very relevant for TDE disks. 

Perhaps the most relevant quantity for the discovery of delayed X-ray flares caused by this mechanism is the change in the fraction of obscured TDEs as a function of logarithmic time, where we choose logarithmic time because TDEs are not  observed $\sim$ daily by X-ray instruments (especially if they are X-ray faint), but are observed more stochastically on $\sim$ week-month-year timescales (with the anticipation that they may rebrighten).  

We show this ``logarithmic reveal rate'' in the lower panel of Figure \ref{fig:focc_self}, which peaks at $\sim 15\%$ per decade at $w(t)\approx 1/2$, or $r_{\rm warp}\approx R_{\rm out}/4$. 

The absolute scales of the reveal rate and occultation fractions are also of interest, and in Figure \ref{fig:focc_t} we show a sequence of profiles for different choices of $r_{\rm warp, 0}$ and $R_{\rm out, 0}$, assuming a canonical viscous time of $t_0 = 50$ days (though this just scales the horizontal axis). The upper panel shows $\bigl\langle f_{\rm occ} \bigr\rangle(t)$, while reveal rates per unit logarithmic time (lower) are also shown. If canonical estimates of both the disk formation scale and warp radius (discussed in this paper) are broadly accurate, then a sizable fraction of TDEs may show delayed X-ray flares via this warped disk mechanism. 

{Again, we stress that the above results were computed under the assumption that the disk aspect ratio tracks the accretion rate. The precise time evolution of $H/R$ (or really $r_{\rm warp}$) in a magnetically supported transient flow is not well understood on a first principles level. Future studies examining the population level distribution of the ratios of X-ray and (e.g.,) UV luminosities of late time TDE disks may well be able to place constraints on the fraction of sources that remain obscured by warp profiles, which may (ambitiously) be used in the future to constrain the evolutionary physics of disk warp profiles.  }

\subsection{Accelerated evolution of the optical/UV plateau}
{Perhaps the most simple evolutionary property of a warped TDE disk occurs if the observer is roughly edge on (with respect to the black holes equatorial plane), and the disk has a moderate $\theta_{\rm warp}$. In this limit TDE disks will (if the warp radius grows with time) show an accelerated evolution of their optical/UV light curves during the plateau phase. The physics behind this is simple, the optical/UV emission is sourced from the main body of the disk for which the flux (from a flat disk surface) is proportional to $\sim \cos i_{\rm eff}$, where $i_{\rm eff}$ is the effective inclination between the observer and the {\it disk} axis that produces the bulk of the optical/UV signal. For small warp radii (i.e., early times) the outer disk will be aligned with the initial stellar angular momentum vector, i.e., }
\begin{multline}
    \cos i_{\rm eff} = \mathbf{\hat o} \cdot \mathbf{\hat l} \\ \approx \sin\theta_{\rm warp}\sin \theta_{\rm obs} \cos \phi_{\rm obs} + \cos\theta_{\rm obs} \cos\theta_{\rm warp} .
\end{multline}
{As the warp radius grows (i.e., late times) the outer disk aligns with the black hole spin axis, and then $i_{\rm eff} \approx \theta_{\rm obs}$.  }

{If these two projected area elements differ by a significant factor, then a distant observer will infer an accelerated evolution of the optical/UV flux (note that this can be either  a decay or a rise, depending on the precise values of the three angles in the problem). We do not plot this explicitly, as it is already visible in the evolution seen in Figure \ref{fig:bumps} where (even just in the evolution of the dashed ``bare disk'' spectra ignoring the transient irradiation signature) a factor $\sim 10$ evolution in the plateau is observed, which is very much at the high end of observations.  Attempts have been made in the literature to compute and characterize the time evolution of the plateau phase of TDE disks \citep{Yael25b}, and this dataset may well contain interesting information regarding warp evolution.    }

\begin{figure*}
    \centering
    \includegraphics[width=0.95\linewidth]{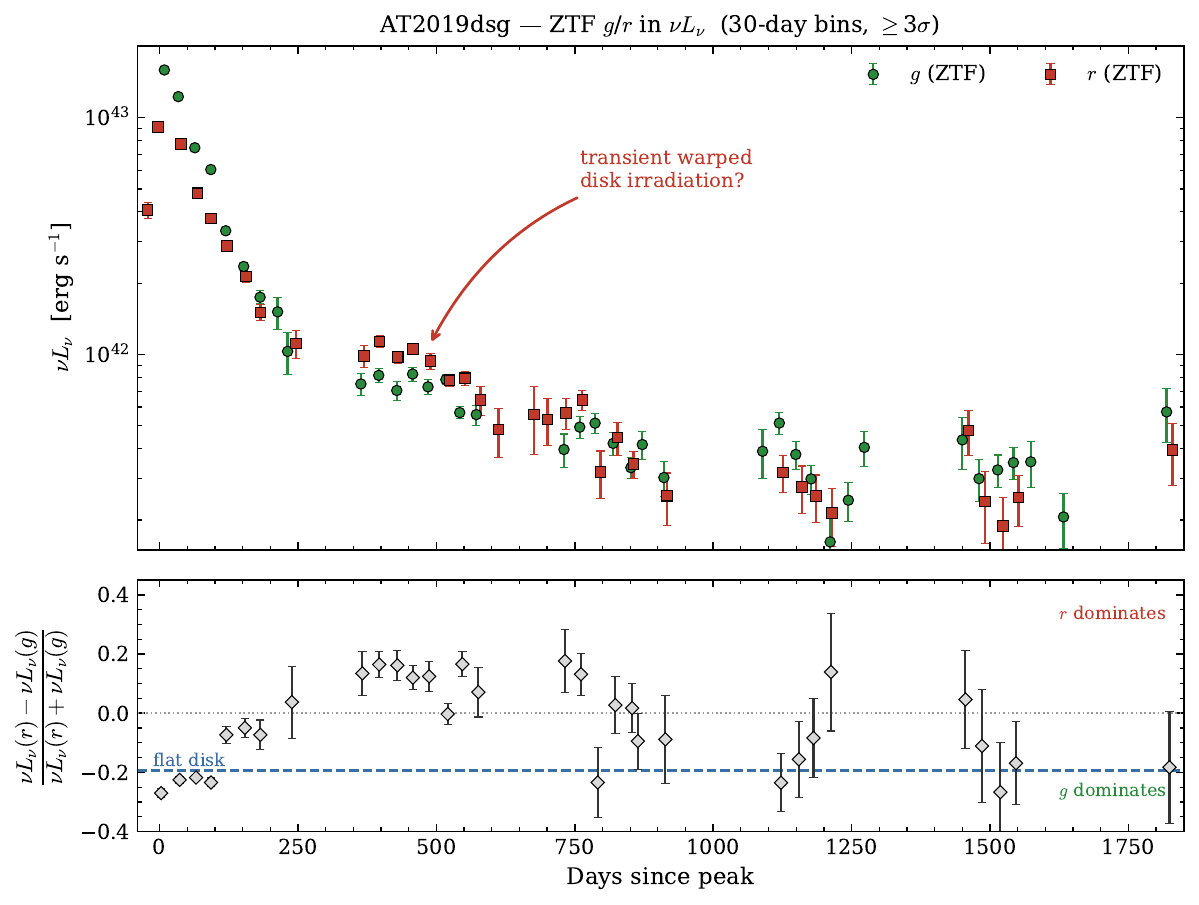}
    \caption{A possible signature of a warped TDE disk, in the optical light curve of AT2019dsg. In the top panel we show the evolving ZTF $r$- and $g$-band light curves, binned on 30 day timescales, with only bins with a signal to noise $\geq3$ retained. The initial optical flare is blue, with $g>r$ robustly. As the emission transitions into the disk phase (the start of the plateau at $\sim 250$ days) the flow transitions into an $r>g$ state, which then inverts at yet later times ($\gtrsim 1000$ days) to $g>r$. This is consistent with a transient phase of irradiation heating (see Fig. \ref{fig:bumps}) when the disk is in its initially warped state, before relaxing. The lower panel shows the normalised color ($(r-g)/(r+g)$) in physics units ($\nu L_
    \nu)$, showing that beyond  $\sim 1000$ days the color reverts to the expected $(r-g)/(r+g) \approx -0.19$ expected for a classical (flat) multi-temperature disk. }
    \label{fig:at2019dsg}
\end{figure*}
\begin{figure*}
    \centering
    \includegraphics[width=0.95\linewidth]{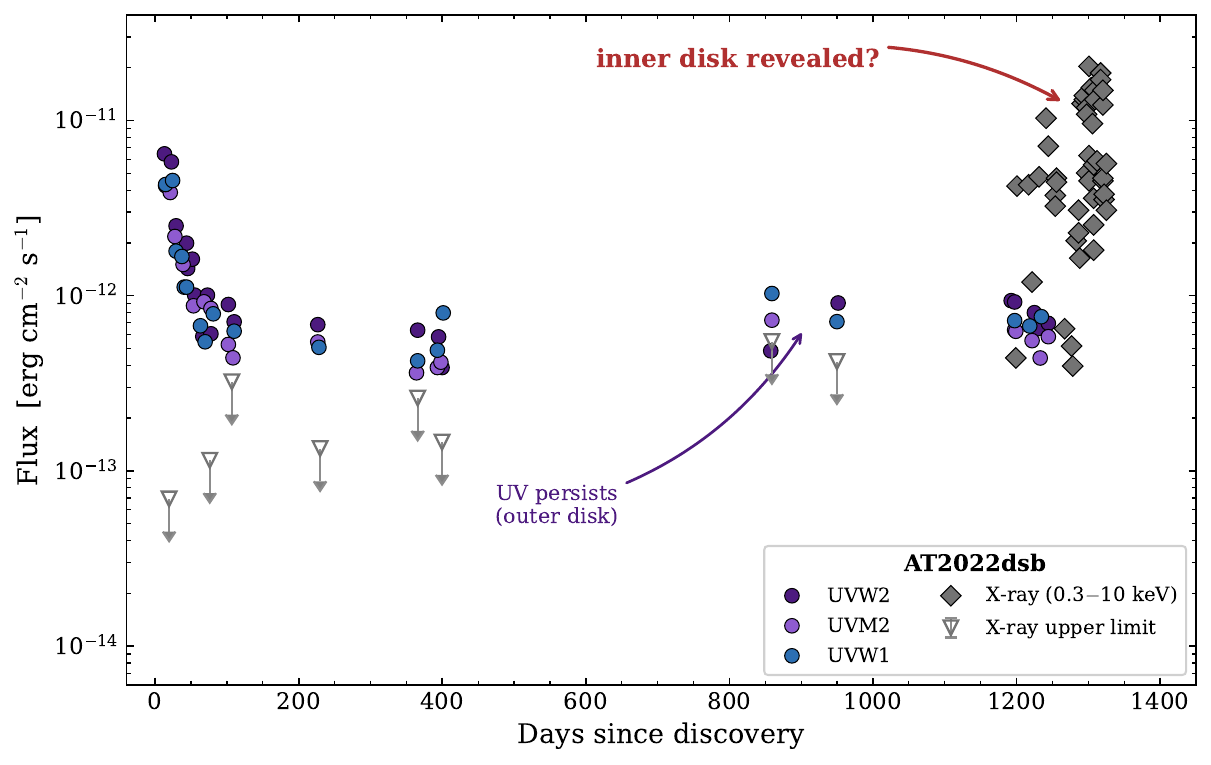}
    \caption{A possible signature of a warped disk in the TDE AT2022dsb, which shows a pronounced delayed X-ray flare at $\sim 1000$ days. What is particularly interesting about AT2022dsb is that it has relatively deep X-ray upper limits covering a reasonable baseline before the delayed flare, and also shows a clear UV plateau which is robustly detected from day $\sim 200$ onwards. This means that the disk had clearly formed at least $\sim 800$ days prior to the X-ray flare, meaning that the outer disk was very much in view of the observer. A relaxing disk warp may explain this behavior.  }
    \label{fig:at2022dsb}
\end{figure*}
\begin{figure*}
    \centering
    \includegraphics[width=0.95\linewidth]{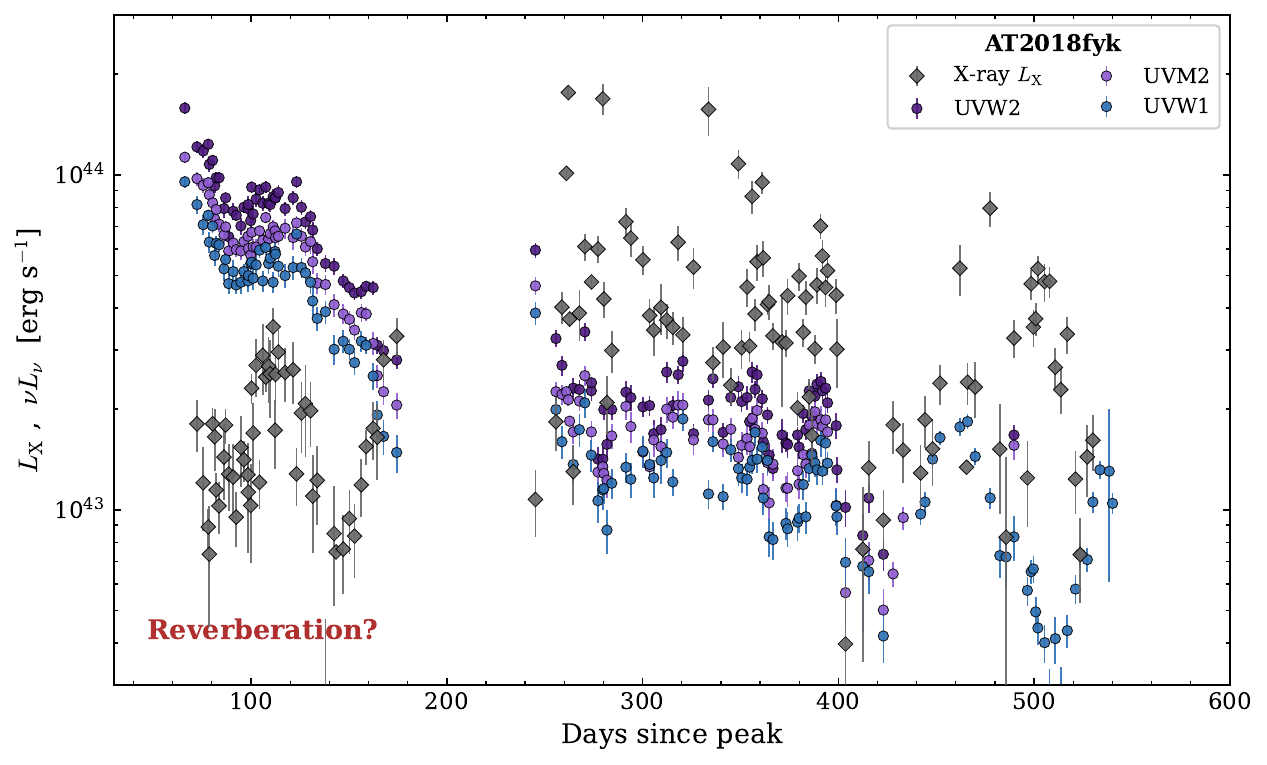}
    \caption{A possible signature of a warped disk in the TDE AT2018fyk, which shows  pronounced UV/X-ray variability. The X-ray variability is at the factor $\sim 20$ level, while the UV variability is closer to a factor $\sim 2-3$, consistent with our simulations of warped disk reverberation (Fig. \ref{fig:reverbA}). The light crossing time of a TDE disk (even around a massive black hole like AT2018fyk) is only $\sim 1$ day, and so no actual correlation is expected (or strongly detected) in the low-cadence light curves seen here. We believe this behavior is suggestive, however.  }
    \label{fig:at2018fyk}
\end{figure*}

\section{Observational discussion}
Having discussed theoretical possibilities of the impacts of warped accretion disks on observed TDE emission, here we discuss possible signatures already found within the TDE literature. We do not, at this stage, fit any individual TDE. Although this is obvious interest, a more complete model of warped disk evolution (rather than the simple self similar profiles assumed here) is perhaps warranted before detailed comparisons to data are performed.

\subsection{AT2019dsg -- transient irradiation signatures?}

AT2019dsg is a canonical multi-wavelength tidal disruption event \citep{Stein21}. It is bright across optical/UV wavelengths, showing a bright flare which transitions to a plateau (disk dominated) state after $\sim 1$ year \citep{Mummery_et_al_2024}. It peaks at $\sim$ near Eddington accretion rates at early times \citep{Stein21}, with bright $L_X\sim 10^{43}$ erg/s X-ray emission with the canonical super soft X-ray spectrum \citep{Guolo25b, Guolo26}. In the radio it has shown two distinct flares \citep{Cendes23}, likely associated with a super-Eddington wind (at early times) and an inner disk state transition (at late times, \citealt{GoodwinMummery26}). 

One very interesting but perhaps underappreciated element of its evolution is shown in Figure \ref{fig:at2019dsg}. Here we show only the ZTF $r$- and $g$-band light curves, which we have binned on 30 day timescales and only kept those bins with a signal to noise ratio $\geq 3$. We follow \citep{Mummery_et_al_2024} in subtracting the host component in this light curve, and plot in rest-frame $\nu L_\nu$. At early times (which, we stress, are non-disk dominated in the optical)  AT2019dsg shows the classical blue TDE spectrum, with $g>r$ at peak. This is completely standard but nonetheless important as it rules out (for example) a high dust obscuration fraction which uniformly reddens the spectrum for what follows. 

Starting at day $\sim 300$, when the optical emission is transitioning to the disk dominated state, there is a phase at which the $r$-band flux exceeds the $g$ band flux. This, as we discussed earlier, is not something which is possible for a flat disk. Indeed a flat disk (in  the mid-frequency) spectral region, has a normalised luminosity ratio of 
\begin{equation}
    {\nu L_\nu(r) - \nu L_\nu(g) \over \nu L_\nu(r) + \nu L_\nu(g)} = {1 - (\nu_g/\nu_r)^{4/3}\over 1 + (\nu_g/\nu_r)^{4/3}} \approx -0.19 ,
\end{equation}
shown by a dashed horizontal line in the lower panel of Figure \ref{fig:at2019dsg}. Interestingly, this is roughly the value to which the very late time behavior of the optical emission asymptotes. 

In the warped disk case however (see Figures \ref{fig:theta_phi}, \ref{fig:cols}, \ref{fig:bumps}, \ref{fig:bump}) a much wider range of behaviors are possible at early times. Our interpretation of what we are seeing in the optical light curve of AT2019dsg is analogous to Figure \ref{fig:bumps} --- as the disk transitions into its disk-dominated optical phase ($\sim 300$ days; note that the disk was dominating the X-ray from $\sim$ day 0), it is highly warped, and there is an irradiation heating signature in its SED. As time progresses the warp radius grows (faster than the outer disk radius), and the flow relaxes to the equatorial plane. In this later state there is no irradiative heating, and the spectrum reverts to its flat disk behavior. 

It is perhaps worth noting that AT2019dsg has the shortest ``viscous timescale'' of all modeled TDE accretion flows (a result driven by its observed X-ray evolution, \citealt{mummery2024fitted, Guolo25b}), so it is perhaps not surprising that it shows rapid warp evolution (if this interpretation is correct).

\subsection{AT2022dsb -- delayed inner disk reveal?}

In Figure \ref{fig:at2022dsb} we show what may be the observation of another  signature of warped disk evolution which we have discussed in detail, the delayed revealing of the inner disk from a relaxing warp.  This behavior is clear in the UV/X-ray light curve of AT2022dsb \citep{Malyali24}. 

While AT2022dsb is by no means the only TDE to show a delayed X-ray flare, it is (as far as the authors are aware) the TDE which has shown the most dramatic late time brightening (in times both of delay time, and of amplitude). What is particularly interesting about AT2022dsb is the well constrained, and apparently normal, UV evolution accompanying this X-ray delay. 

This is important because it implies that the outer disk was formed and observable from (at least) $\sim 200$ days post detection, while at the same time there are a series of deep X-ray upper limits ($\sim 100$ times deeper than the subsequent peak detection flux). We believe this rules out all models which rely on delayed disk formation being the origin of delayed X-ray flares in TDEs. There was also a single detection of soft X-ray emission {\it before} the optical flare \citep{Malyali24}, which could be interpreted as further evidence for prompt disk formation. \cite{Malyali24} also found some very low level hard X-ray emission during the $\sim 200-1000$ day window, but this was interpreted by the authors as pre-existing host emission, not associated with the transient. 

We also note that once it is detected in the X-rays it again shows the canonical super-soft X-ray spectrum characteristic of TDEs, and is well modeled by a flat accretion flow \citep{Guolo26}. 

The apparently normal outer disk behavior coupled to the deep X-ray upper limits during the disk-dominated phase, followed by a rapid brightening at late times, makes us favor the warp-relaxation mechanism for the delayed X-ray flare in AT2022dsb. We cannot of course rule out other mechanisms for this delayed flare, and perhaps population statistics (as in Figures \ref{fig:focc_self}, \ref{fig:focc_t}) will help differentiate the dominant delayed flare mechanism at play in reality. 

\subsection{AT2018fyk -- possible reverberation signatures?}

Finally, we turn to AT2018fyk \citep{Wevers21, Wevers23fyk, Pasham24fyk} and possible signatures of reverberation in a TDE. As we discussed in an earlier section, to drive noticeable UV variability from a warped disk irradiation mechanism, one requires a very high amplitude of X-ray variability. 

Fortunately, there are numerous candidate TDEs with high amplitude X-ray variability \citep{Mummery25Variability, Chakraborty26}. The best source (in terms of continual X-ray and UV coverage and cadence) is AT2018fyk. AT2018fyk is a fascinating source, and is a strong repeating partial TDE candidate \citep{Wevers23fyk}. We focus here on the first $\sim 700$ days of its evolution, which is prior to any repeats. 

In Figure \ref{fig:at2018fyk} we show the X-ray (gray) and multi-band UV (blue, pink, purple) light curves. These light curves are corrected for extinction and are shown in $\nu L_\nu$ (UV) or integrated $L_X$ (X-ray, across 0.3 to 10 keV) units. One immediately sees that the X-ray light curve of AT2018fyk shows high amplitude variability, of order a factor $\sim 20$ (the observed range is $\sim 10^{43}$ erg/s up to $\sim 2\times 10^{44}$ erg/s). The initial UV flare (first $\sim 200$ days) is not of particular interest (as its physics is more poorly understood), but beyond day $\sim 300$ the disk is dominating the UV flux, and one can look for reverberation signatures.

By eye it is clear to see that the UV emission from AT2018fyk is also highly variable on short timescales, at the level of a factor $\sim 2-3$. This is interesting because this is the level of UV variability which a warped disk reprocessing an order $\sim 10$ X-ray flare produced (Figure \ref{fig:reverbA}). The light crossing time for an $\sim {\cal O}(100r_g)$ disk around a $\sim {\rm few} \times 10^7 M_\bullet$ black hole is still $<1$ day, and the X-ray variability timescale is clearly below the observing cadence. This means that we do not expect to observe UV-X-ray correlations in this data set (as the expected lag time is $<$ the observing cadence), and indeed we do not. As far as we can tell the only way in which the warped disk reverberation signature could be resolved is if simultaneous XMM and HST/{\it Swift} observations were taken with a sufficiently long baseline to capture the nature X-ray variability timescale. 

We do, however, think it is suggestive that the best observationally constrained  highly X-ray-variable TDE shows such pronounced UV variability on a similar timescale. Recall that the natural timescale for disk turbulence at the UV scale is significantly longer than at the X-ray scale (owing to the larger radii at which UV emission is sourced in a flat disk, and the fact that the natural timescale of a disk region scales as $r^{+3/2}$). The fact that we are unable to resolve the UV variability timescale in AT2018fyk makes a reprocessing interpretation a promising avenue to explore further. 



\section{Conclusions}
We have developed a model framework for computing the observational
appearance of a globally warped TDE accretion disk, and used it to identify
and quantify three signatures of warping. We then compute their expected evolution as a
TDE ages, and estimate how often they may be seen across the
population. Given the \cite{Scheuer1996} tilt-and-twist profile, we compute
the self-irradiation of the tilted outer disk by the hot inner flow, the
resulting two-dimensional temperature structure, the occultation of the
inner disk by the warp, and the reprocessed response to a time-dependent
flare, and assemble these into predicted spectral energy distributions,
light curves, and population statistics. We
summarize the main results here in turn.

\emph{Irradiation reddens the optical/UV.} The tilted outer disk intercepts
and reprocesses radiation from the hot inner flow, so its temperature falls
off more slowly with radius than in a flat disk --- tending from the
flat-disk $T\propto r^{-3/4}$ toward the $T\propto r^{-1/2}$ of a centrally
irradiated disk. This flattens and reddens the mid-frequency continuum and,
for some geometries and viewing angles, prints distinctive bumps or kinks
onto the spectrum. These are features a flat disk simply cannot produce,
and are therefore clean diagnostics of a warp.

\emph{The warp can hide and later reveal the inner disk.} A sufficiently
tilted outer disk blocks an inclined observer's view of the hot inner
regions. We derived a simple analytic criterion --- a critical inclination
$i_{\rm crit}$ --- separating sightlines that do and do not see the inner
disk, and showed that whether it is hidden is a competition between the
outer radius $R_{\rm out}$ (a larger disk hides more) and the warp radius
$r_{\rm warp}$ (a growing warp scale reveals more). The question then becomes whether $r_{\rm warp}$
generically grows faster than $R_{\rm out}$ as the disk cools. If it does, then the typical
outcome is that the inner disk is progressively unveiled with time
--- a natural, purely geometric explanation for the delayed X-ray
brightenings now seen in a growing number of TDEs, one that requires no
delay in the formation of the disk itself. Turning this around, it seems plausible that population studies of TDEs in the X-ray could reveal interesting properties of the evolution of global warped disk structures. 

\emph{The warp reprocesses X-ray variability into the optical/UV.} Since
the outer disk is not aligned with the inner disk, a flare in the inner disk illuminates and heats
it, and the energy is re-radiated after a light-travel delay. We find that
a factor $\sim\!10$ X-ray flare can drive a factor $\sim\!2$ optical/UV echo on the disk's
light-crossing time, with either a 
single- or double-peaked profile depending on the viewing geometry and precise warp profile. For a compact TDE disk the light travel time is only of order hours,
making the lag challenging --- but not impossible --- to resolve.

\emph{The warp can accelerate the evolution of the plateau.} The flux observed in the mid-frequency and Rayeligh-Jeans regions of disk spectra is proportional to the projected area of the disk as observed by an observer $\mathbf{\hat o}$, $\cos i_{\rm eff} = \mathbf{\hat o} \cdot \mathbf{\hat l}$, where $\mathbf{\hat l}$ is the disk normal. An evolving warp profile makes the disk normal a function of time $\mathbf{\hat l}(t)$, leading to a potentially accelerated evolution of the late time disk-dominated plateau phase of optical/UV emission. 

All warp-driven effects are in principle transient themselves, as the warp relaxes into the equatorial
plane the reddening fades, the inner disk is revealed, and the reprocessing
weakens, so each is expected to be relevant during the  earlier phases of
disk evolution. The temporal evolution of the warp radius itself is not robustly understood in a fundamental theoretical sense, and tidal disruption events may well offer the best astronomical probes of this frontier of accretion disk theory. 

We further quantified how common inner-disk occultation
should be across an isotropic population of TDEs, and found the
observer-averaged obscured fraction to be a single universal function of
the ratio $w=\sqrt{r_{\rm warp}/R_{\rm out}}$, and surprisingly common. Up to roughly $\sim\!1/3$ of viewing
angles have the inner disk  obscured for compact warps, and the rate at which inner disks
are revealed peaks at $\sim\!15\%$ per decade in time (near
$r_{\rm warp}\approx R_{\rm out}/4$). For plausible disk-formation and warp
radii, a sizable fraction of all TDEs should therefore exhibit a
warp-driven delayed X-ray flare. 

In the final section we pointed to candidate signatures of each effect already in the
literature, while stressing that these are suggestive rather than
definitive, as we fit no individual event here. In AT2019dsg there is a
transient episode in which the $r$-band flux rises above the $g$-band, as
expected from irradiative reddening and impossible for a flat disk. In
AT2022dsb a dramatic delayed X-ray flare is accompanied by a normal UV
plateau and deep pre-brightening X-ray upper limits, showing that the disk was
present but hidden --- disfavoring delayed disk formation as the
cause. Finally AT2018fyk exhibits high-amplitude X-ray and
optical/UV variability of just the relative amplitude our reprocessing
model predicts. We believe that the rapid evolution of (e.g.,) the optical colours of AT2019dsg, and the common occurrence of delayed X-ray flares in TDEs, support the hypothesis that the warp radius evolves relatively rapidly as the disk thins. 

Several simplifications should be borne in mind. We have neglected
relativistic corrections --- Doppler boosting, gravitational redshift, and
light bending --- which are important in the innermost regions; we have
adopted a single, smoothly-warped surface, deferring the possibility of
disk tearing; we have used simple self-similar scalings for the time
evolution of the warp rather than a solved warp-evolution equation; and the
warp microphysics remains encoded in somewhat uncertain geometric quantities $r_{\rm warp},  \theta_{\rm warp}$. Each of these is a natural target for future refinement. The
most valuable next steps would be a self-consistent model of the warp's
evolution (in place of our self-similar profiles), source-by-source fitting
of candidate events, and population tests of the predicted delayed-flare
rate against the samples now being assembled by X-ray surveys. Targeted, simultaneous X-ray and UV monitoring
at a cadence shorter than the light-crossing time could isolate the
reverberation signature directly.

The connection to black hole spin deserves particular emphasis. Every
signature discussed here vanishes as $a\to0$. A Schwarzschild black hole has no frame dragging, so there
is no Lense--Thirring torque, no warp, and hence no warp-induced reddening,
occultation, or reverberation. The detection of these signatures is thus
qualitative evidence for a spinning, misaligned system, while their
amplitude and timing encode the warp radius $r_{\rm warp}\propto a^{2/3}$.
Converting this into a spin measurement for an individual source is perhaps over ambitious, but at the population level some nuisances
partly average out, and the detection at all of warped disk signatures in a TDE sample offer a
genuinely independent probe of the spin distribution of
otherwise-quiescent supermassive black holes --- a quantity that is very
hard to access by any other means.

More broadly, TDEs may become a practical laboratory for the physics of
warped disks. The very features that make warped disks theoretically
awkward --- their non-linearity, and their sensitivity to poorly
constrained internal stresses --- are imprinted on quantities that TDE
observations can measure: colors, X-ray delays, reverberation lags, and the
statistics of inner-disk obscuration. A robust mapping between warp models
and TDE data, run in reverse, would allow these observations to inform the
theory. In this sense the warped disks of tidal disruption events are a
natural meeting point for accretion theory and time-domain observation
and, because the warp is a direct consequence of frame dragging, perhaps
our cleanest astrophysical view of strong-gravity disk warping.

\section*{Acknowledgments}
A.M. is grateful to various audience members of talks over the years who have pointed out that tidal disruption event accretion flows will not be in the equatorial plane and should show warped disk signatures. 

A.M. is grateful to Callum Fairbairn for various stimulating conversations throughout the completion of this work. A.M. acknowledges support from the Ambrose Monell Foundation, the W.M. Keck Foundation and the John N. Bahcall Fellowship Fund at the Institute for Advanced Study.   E.Q. thanks the Aspen Center for Physics for support during the completion of this work, which is supported by National Science Foundation grant PHY-2210452.

\bibliography{andy}

\appendix
\section{Geometry surfaces and areas}\label{app:geo}
This appendix describes the surface geometry of the warped disk---its local
orthonormal frame, tangent vectors, area elements, and surface normal---and
identifies which of these quantities enter the emitted spectral energy
distribution. 

\subsection{Disk Surface and Angular-Momentum Profile}
\label{ageo:profile}

At radius $r$ the angular-momentum unit vector is
\begin{equation}
\label{ageo:lhat}
\mathbf{\hat{l}}(r) = \left( l_\perp\cos\gamma,\;
                                l_\perp\sin\gamma,\;
                                l_z \right),
\end{equation}
where $\gamma(r) = 2\sqrt{r_{\rm warp}/r}$ is the \cite{Scheuer1996}
twist, $l_\perp(r) = \sin\theta_{\rm warp}\,e^{-\gamma}$, and
$l_z = \sqrt{1-l_\perp^2}$. The local tilt is
$\beta(r) = \arccos l_z$, so $\sin\beta = l_\perp$ and $\cos\beta = l_z$.
The disk surface is the union of tilted great circles, at radius $r$, the
ring lies in the plane $\mathbf{r}\cdot\mathbf{\hat{l}}(r) = 0$ with
$|\mathbf{r}| = r$.

\subsection{Orthonormal Frame on Each Ring}
\label{ageo:frame}

Construct a right-handed frame
$(\mathbf{\hat{e}}_1,\,\mathbf{\hat{e}}_2,\,\mathbf{\hat{l}})$ at each $r$:
\begin{align}
\mathbf{\hat{e}}_1 &= \frac{\mathbf{\hat{z}}\times\mathbf{\hat{l}}}
                            {|\mathbf{\hat{z}}\times\mathbf{\hat{l}}|}
        = \left(-\sin\gamma,\;\cos\gamma,\;0\right), \\[4pt]
\mathbf{\hat{e}}_2 &= \mathbf{\hat{l}}\times\mathbf{\hat{e}}_1
        = \left(-\cos\beta\cos\gamma,\;-\cos\beta\sin\gamma,\;\sin\beta\right).
\end{align}
The ring is parameterised by the intrinsic angle $\psi$,
\begin{equation}
\label{ageo:P}
\mathbf{P}(r,\psi) = r\,\mathbf{\hat{e}}(\psi),
\qquad
\mathbf{\hat{e}} \equiv \cos\psi\,\mathbf{\hat{e}}_1
                       + \sin\psi\,\mathbf{\hat{e}}_2,
\end{equation}
for which $|\mathbf{P}| = r$ and $\mathbf{P}\cdot\mathbf{\hat{l}} = 0$.
The tangent to the ring is
\begin{equation}
\mathbf{\hat{e}}_\psi \equiv \frac{\partial\mathbf{\hat{e}}}{\partial\psi}
= -\sin\psi\,\mathbf{\hat{e}}_1 + \cos\psi\,\mathbf{\hat{e}}_2.
\end{equation}

\subsection{Frame Rotation Rates}
\label{ageo:rot}

As $r$ varies, the frame
$(\mathbf{\hat{e}}_1,\mathbf{\hat{e}}_2,\mathbf{\hat{l}})$ rotates. Since
$\mathbf{\hat{e}}_1\cdot\mathbf{\hat{e}}_1
= \mathbf{\hat{e}}_2\cdot\mathbf{\hat{e}}_2 = |\mathbf{\hat{l}}|^2 = 1$,
all derivatives are perpendicular to the  vector being differentiated, and
\begin{equation}
\label{ageo:frame_rot}
\frac{\mathrm{d}\mathbf{\hat{e}}_1}{\mathrm{d}r}
   = \omega_1\,\mathbf{\hat{e}}_2 + \omega_2\,\mathbf{\hat{l}},
\qquad
\frac{\mathrm{d}\mathbf{\hat{e}}_2}{\mathrm{d}r}
   = -\omega_1\,\mathbf{\hat{e}}_1 + \omega_3\,\mathbf{\hat{l}},
\qquad
\frac{\mathrm{d}\mathbf{\hat{l}}}{\mathrm{d}r}
   = -\omega_2\,\mathbf{\hat{e}}_1 - \omega_3\,\mathbf{\hat{e}}_2,
\end{equation}
where
\begin{equation}
\label{ageo:omegas}
\omega_1 = \gamma'\cos\beta, \qquad
\omega_2 = -\gamma'\sin\beta , \qquad
\omega_3 = \beta',
\end{equation}
with a prime denoting a derivative with respect to $r$, and 
\begin{equation}
\gamma'(r) = -\sqrt{\frac{r_{\rm warp}}{r^3}},
\qquad
\beta'(r) = \tan\beta\,\sqrt{\frac{r_{\rm warp}}{r^3}}.
\end{equation}

\subsection{Tangent Vectors}
\label{ageo:tangents}

At fixed $r$ the frame is frozen, so the local azimuthal tangent is
\begin{equation}
\label{ageo:vpsi}
\mathbf{t}_\psi = \frac{\partial\mathbf{P}}{\partial\psi}
= r\,\mathbf{\hat{e}}_\psi.
\end{equation}
At fixed $\psi$ the radial tangent is
\begin{align}
\mathbf{t}_r
  &= \frac{\partial\mathbf{P}}{\partial r}
   = \mathbf{\hat{e}} + r\left(\cos\psi\,
        \frac{\mathrm{d}\mathbf{\hat{e}}_1}{\mathrm{d}r}
      + \sin\psi\,\frac{\mathrm{d}\mathbf{\hat{e}}_2}{\mathrm{d}r}\right)
  \nonumber\\[4pt]
  &= \mathbf{\hat{e}} + r\omega_1\,\mathbf{\hat{e}}_\psi
     + r\,\Omega\,\mathbf{\hat{l}},
  \label{ageo:vr}
\end{align}
where
\begin{equation}
\label{ageo:Omega}
\Omega(r,\psi)
   \equiv \omega_2\cos\psi + \omega_3\sin\psi
   = -\gamma'\sin\beta\cos\psi + \beta'\sin\psi
   = -\,\mathbf{\hat{e}}\cdot\frac{\mathrm{d}\mathbf{\hat{l}}}{\mathrm{d}r}.
\end{equation}

\subsection{Surface Area Element and Surface Normal}
\label{ageo:surface}

The cross product of the tangent vectors is
\begin{equation}
\mathbf{t}_r\times\mathbf{t}_\psi
  = r\left(\mathbf{\hat{l}} - r\Omega\,\mathbf{\hat{e}}\right),
\end{equation}
with magnitude $r\sqrt{1 + r^2\Omega^2}$ (using
$\mathbf{\hat{l}}\cdot\mathbf{\hat{e}} = 0$). The geometric surface area
element in $(r,\psi)$ coordinates is therefore
\begin{equation}
\label{ageo:dA_surf}
\mathrm{d}A_{\rm surf} = r\sqrt{1 + r^2\Omega^2}\;\mathrm{d}r\,\mathrm{d}\psi,
\end{equation}
and the top-surface unit normal is
\begin{equation}
\label{ageo:ntop}
\mathbf{\hat{n}}_{\rm top}
  = \frac{\mathbf{\hat{l}} - r\Omega\,\mathbf{\hat{e}}}
         {\sqrt{1 + r^2\Omega^2}}.
\end{equation}
The normal $\mathbf{\hat{n}}_{\rm top}$ differs from the ring normal
$\mathbf{\hat{l}}$ by a tilt of order
$r\Omega \sim r\,|\mathrm{d}\mathbf{\hat{l}}/\mathrm{d}r|$ toward
$\mathbf{\hat{e}}$: where the warp changes rapidly, adjacent rings are tilted
differently and the surface corrugates.

\subsection{Ring Area Element}
\label{ageo:ring_area}

The ring area element is simpler. The area of a thin annulus at radius $r$
with width $\mathrm{d}r$, measured in the plane of the ring (perpendicular to
$\mathbf{\hat{l}}$) and independent of the warp, is
\begin{equation}
\mathrm{d}A_{\rm ring} = 2\pi r\,\mathrm{d}r,
\end{equation}
or, in terms of the intrinsic angle $\psi$ that uniformly parametrizes the ring
\begin{equation}
\mathrm{d}A_{\rm ring} = r\,\mathrm{d}r\,\mathrm{d}\psi.
\end{equation}
The key difference from $\mathrm{d}A_{\rm surf}$ is the absence of the
stretching factor $\sqrt{1+r^2\Omega^2}$, which accounts for the surface
corrugation between adjacent rings. For fiducial warp parameters,
$\sqrt{1+r^2\Omega^2} - 1$ peaks at $\sim 1\%$ near the warp radius.

\subsubsection{Intrinsic versus Projected Azimuth}
\label{ageo:azimuth}

The intrinsic angle $\psi$ parameterises position on a single tilted ring
uniformly: $\psi = 0$ lies along $\mathbf{\hat{e}}_1$ (the line of nodes,
where the ring crosses the equatorial plane) and $\psi = \pi/2$ along
$\mathbf{\hat{e}}_2$ (the direction of maximum elevation). For a flat disk,
$\psi$ reduces to the ordinary azimuthal angle. The projected azimuth
$\varphi$ is the angle measured in the equatorial plane when the ring
position is projected along the spin axis $\mathbf{\hat{z}}$. Because a
tilted circle projects to an ellipse, the mapping $\psi\to\varphi$ is
nonlinear,
\begin{equation}
\label{ageo:psi_to_phi}
\varphi = \arctan\!\left(\tan\psi\,\cos\beta\right) + \varphi_{\rm nodes},
\qquad
\varphi_{\rm nodes} \equiv \gamma + \tfrac{\pi}{2},
\end{equation}
where $\varphi_{\rm nodes}$ is the equatorial
azimuth of the line of nodes $\mathbf{\hat{e}}_1$ and the $\cos\beta$ inside the
arctangent performs the tilt projection. The offset $\pi/2$ is purely
geometric: $\mathbf{\hat{e}}_1 = \mathbf{\hat{z}}\times\mathbf{\hat{l}}/
|\mathbf{\hat{z}}\times\mathbf{\hat{l}}|$ is perpendicular to the
projection of $\mathbf{\hat{l}}$ (at azimuth $\gamma$), so the line of
nodes sits at $\gamma + \pi/2$ in the equatorial plane. Points near the
ascending and descending nodes are compressed in $\varphi$ while points near
maximum elevation are stretched. The Jacobian of the transformation is
\begin{equation}
\label{ageo:jacobian}
\frac{\partial\psi}{\partial\varphi} = \frac{\cos\beta}{D^2},
\qquad
D^2 \equiv \sin^2\!\varphi + \cos^2\!\varphi\,\cos^2\!\beta.
\end{equation}
The projected azimuth $\varphi$ is the natural coordinate for an observer
looking along $\mathbf{\hat{z}}$, since all rings then share a common
azimuthal grid, whereas the intrinsic angle $\psi$ acquires a ring-dependent
origin once the twist is applied. In $(r,\varphi)$ coordinates the ring area
element is therefore
\begin{equation}
\label{ageo:dA_ring_phi}
\mathrm{d}A_{\rm ring}
   = r\,\mathrm{d}r\,\mathrm{d}\psi
   = \frac{r\cos\beta}{D^2}\;\mathrm{d}r\,\mathrm{d}\varphi.
\end{equation}
For a flat disk ($\beta = 0$), $D = 1$ and $\psi = \varphi$ everywhere.

\subsection{Emission: Ring Area and $\mathbf{\hat{l}}$, Not Surface Area
and $\mathbf{\hat{n}}_{\rm top}$}
\label{ageo:emission}

\subsubsection{The Ring Area Element}
Each ring at radius $r$ dissipates energy at a rate set by the turbulent stress tensor. The temperature is defined per unit ring area (equivalently, per unit
mass for a surface density $\Sigma$),
\begin{equation}
\mathrm{d}L_{\rm bol}(r) = 2\,\sigma T^4(r)\cdot 2\pi r\,\mathrm{d}r,
\end{equation}
the factor of $2$ accounting for emission from both faces. This is a
fundamental property of the accretion flow: $T^4$ is fixed by the
angular-momentum transport equation and is independent of how the rings are
assembled into a surface. Integrating $\sigma T^4$ over the surface area
instead would give
\begin{equation}
\mathrm{d}L_{\rm surf} = 2\sigma T^4\cdot 2\pi r\sqrt{1+r^2\Omega^2}\,\mathrm{d}r
   = \sqrt{1+r^2\Omega^2}\;\mathrm{d}L_{\rm bol},
\end{equation}
over-predicting the bolometric luminosity by the stretching factor (a
$\sim 1\%$ effect for fiducial parameters).

\subsubsection{The Ring Normal }
Each ring emits as an independent blackbody, with radiation directed along
its own ring normal $\pm\mathbf{\hat{l}}(r)$ rather than along the
geometric surface normal $\mathbf{\hat{n}}_{\rm top}$. The surface normal
incorporates the tilt due to adjacent rings being inclined differently, but a
ring's emission is determined only by its own angular-momentum direction. The
specific intensity toward an observer at direction $\mathbf{\hat{o}}$ is
\begin{equation}
I_\nu(r,\mathbf{\hat{o}})
  = \frac{B_\nu\!\left(f_c\,T(r)\right)}{f_c^4}\,
    |\mathbf{\hat{l}}(r)\cdot\mathbf{\hat{o}}|,
\end{equation}
and the observed SED is
\begin{equation}
\label{ageo:sed}
\nu L_\nu = 4\pi\,\frac{2h\nu^4}{c^2}
   \int \frac{|\mathbf{\hat{l}}\cdot\mathbf{\hat{o}}|\;\mathrm{d}A_{\rm ring}}
             {f_c^4\left[\exp\!\left(h\nu/k f_c T\right)-1\right]}.
\end{equation}
Using $\mathbf{\hat{n}}_{\rm top}\cdot\mathbf{\hat{o}}$ with
$\mathrm{d}A_{\rm surf}$ instead introduces errors of up to $\sim 30\%$ in the
optical band at high observer inclinations, because $\mathbf{\hat{n}}_{\rm top}$
deviates from $\mathbf{\hat{l}}$ by up to $\sim 12^\circ$ at the warp, and
the outer, cool rings that dominate the optical are precisely where the warp
correction is largest. The X-ray band is nearly unaffected ($<0.1\%$), being
dominated by the inner, $\sim$ flat disk.

\subsubsection{Energy Conservation}
Integrating the intensity over all observer directions and both faces, the
total power emitted by the ring at $r$ is
\begin{equation}
\mathrm{d}L(r)
  = 2\int_{+} I_\nu\,\mathrm{d}\Omega\cdot\mathrm{d}A_{\rm ring}
  = 2\sigma T^4\cdot 2\pi r\,\mathrm{d}r,
\end{equation}
using $\int_0^{2\pi}\!\!\int_0^{\pi/2}|\cos\vartheta|\sin\vartheta
\,\mathrm{d}\vartheta\,\mathrm{d}\alpha = \pi$ and
$\int B_\nu\,\mathrm{d}\nu = \sigma T^4/\pi$. This recovers the correct
bolometric luminosity; replacing $\mathrm{d}A_{\rm ring}$ with
$\mathrm{d}A_{\rm surf}$ would give $\sqrt{1+r^2\Omega^2}$ times the correct
answer, violating energy conservation.

\section{The irradiation kernel}
\label{app:kernel}
In this appendix we discuss a typographical error in \citet{Fukue1992} for computing the irradiation kernel for a Lambertian disk, which has propagated into later literature \citep[e.g.,][]{Speicher2025}.

We begin by recaping the derivation in the main body of the text. Consider two infinitesimal surface elements ${\rm d}A_S$ (source) and ${\rm d}A_P$ (receiver) separated by displacement $\vec{\mathbf{d}} = \vec{\mathbf{P}} - \vec{\mathbf{S}}$, with $d \equiv |\vec{\mathbf{d}}|$ and $\mathbf{\hat{d}} \equiv \vec{\mathbf{d}}/d$.  Let $\mathbf{\hat{n}}_S$ and $\mathbf{\hat{n}}_P$ be their outward unit normals.

A Lambertian emitter radiates power $I\cos\theta_S\,{\rm d}\Omega\,{\rm d}A_S$ into solid angle ${\rm d}\Omega$ at emission angle $\theta_S$, where $I$ is the specific intensity.  The solid angle subtended by ${\rm d}A_P$ as seen from the source is ${\rm d}\Omega = |\cos\theta_P|\,{\rm d}A_P/d^2$, where $\cos\theta_P = (-\mathbf{\hat{d}})\cdot\mathbf{\hat{n}}_P$ is the incidence angle at the receiver.  Combining these:
\begin{equation}
  {\rm d}P_{S\to P}
  = I\,
    \underbrace{|\mathbf{\hat{d}}\cdot\mathbf{\hat{n}}_S|}_{\cos\theta_S}\;
    \underbrace{|(-\mathbf{\hat{d}})\cdot\mathbf{\hat{n}}_P|}_{\cos\theta_P}\;
    \frac{{\rm d}A_S\,{\rm d}A_P}{d^2}.
  \label{eq:app_vf}
\end{equation}
\citet{Fukue1992} considers a somewhat related geometry to that which we considered here, flat inner disk (the source, with normal
$\mathbf{\hat{e}}_z$) irradiating a flared outer disk element which is at cylindrical position $(r, 0, H)$, where $H(r)$ is the local disk height. The unit vector from the outer element toward the source is $\mathbf{\hat d} = (-r, 0, -H)/R$ with $R = \sqrt{r^2 + H^2}$, and the
outward normal to the flared surface is
$\mathbf{\hat{n}} = (-H', 0, 1)/\sqrt{1 + (H')^2}$, where $H' = {\rm d}H/{\rm d}r$. 

\citeauthor{Fukue1992}'s equation~(5) writes the irradiating flux
as
\begin{equation}
  Q_{\rm irr}^{{\rm eq.}\, (5)} \propto
    \underbrace{(-\mathbf{\hat d}\cdot\mathbf{\hat e}_z)}_{\cos\theta_S}\;
    \underbrace{(\mathbf{\hat d}\cdot\mathbf{\hat n})}_{\cos\theta_P}\;
    \underbrace{(\mathbf{\hat n}\cdot\mathbf{\hat e}_z)}_{\text{third factor}},
  \label{eq:app_fukue5}
\end{equation}
a three-cosine kernel.  However, the same paper then
evaluates the geometry explicitly and arrives at their equation~(9):
\begin{equation}
  Q_{\rm irr}^{{\rm eq.}\, (9)} \propto
    \frac{H(rH' - H)}{R^2\sqrt{1 + (H')^2}}.
  \label{eq:app_fukue9}
\end{equation}
This is however the result one would have reached by neglecting the third dot product, and using the two cosine formulation. 

Computing the two cosine factors directly:
\begin{equation}
  -\mathbf{\hat d}\cdot\mathbf{\hat e}_z = \frac{H}{R},
  \qquad
  \mathbf{\hat d}\cdot \mathbf{\hat n}
  = \frac{rH' - H}{R\sqrt{1 + (H')^2}}.
\end{equation}
Their product is
\begin{equation}
  (-\mathbf{\hat d}\cdot\mathbf{\hat e}_z)(\mathbf{\hat d}\cdot\mathbf{\hat n})
  = \frac{H(rH' - H)}{R^2\sqrt{1 + (H')^2}},
  \label{eq:app_two_cos}
\end{equation}
which is {identical} to equation~(\ref{eq:app_fukue9}).

Adding the third factor
$\mathbf{\hat n}\cdot \mathbf{\hat e}_z = 1/\sqrt{1 + (H')^2}$ gives instead
\begin{equation}
(-\mathbf{\hat d}\cdot\mathbf{\hat e}_z)\;
(\mathbf{\hat d}\cdot\mathbf{\hat n})\;
(\mathbf{\hat n}\cdot\mathbf{\hat e}_z)
  = \frac{H(rH' - H)}{R^2[1 + (H')^2]},
  \label{eq:app_three_cos}
\end{equation}
which differs from equation~(\ref{eq:app_fukue9}) by
$1/\sqrt{1 + (H')^2}$.

Fukue's equations~(5) and~(9) are mutually inconsistent, which we believe is a simple typographical error. The derived result (equation~9) is correct while the proposed kernel (equation~5) contains a spurious third factor.

\end{document}